%% file: WDSearch.tex
\documentclass[iop,revtex4]{emulateapj}  % version 5.0 or prior
\usepackage[latin1]{inputenc}
\usepackage{amsmath}
\usepackage{amsfonts}
\usepackage{amssymb}
\usepackage{graphicx}
\usepackage{color}
\usepackage{rotating}
\usepackage{pdflscape}

\newcommand{\Msun}{\mbox{M$_\odot$}}
\def\Halpha{\mbox{H\hspace{0.1ex}$\alpha$}}

\def\CaII{\ion{Ca}{2}}

\begin{document}

%\title{A spectroscopic search for White Dwarf companions to 101 nearby M dwarfs}{\thanks{Based on observations collected in service mode using the Very Large Telescope (VLT) under program IDs 095\_D-0949(A) and 096\_D-0963(A), PI: Paul Vreeswijk, with X-shooter installed at VLT Unit 2, Kueyen, operated by the European Southern Observatory (ESO) on Cerro Paranal in Chile}

\title{A spectroscopic search for White Dwarf companions to 101 nearby M dwarfs\footnote{Based on observations collected in service mode using the Very Large Telescope (VLT) under program IDs 095\_D-0949(A) and 096\_D-0963(A)}}

%\let\thefootnote\relax\footnotetext{A footnote without numbering}


%\author{Ira Bar\altaffilmark{}, Paul Vreeswijk\altaffilmark{}, Avishay Gal-Yam\altaffilmark{} and Eran O. Ofek\altaffilmark{}}
%\affil{Benoziyo Center for Astrophysics, Weizmann Institute of Science, 76100 Rehovot, Israel}

\author{Ira Bar\altaffilmark{1}, Paul Vreeswijk\altaffilmark{1}, Avishay Gal-Yam\altaffilmark{1}, Eran O. Ofek\altaffilmark{1} and Gijs Nelemans\altaffilmark{2}}

\altaffiltext{1}{Benoziyo Center for Astrophysics, Faculty of Physics, The Weizmann Institute for Science, Rehovot 76100, Israel}
\altaffiltext{2}{Department of Astrophysics/IMAPP, Radboud University, PO Box 9010, NL-6500 GL Nijmegen, The Netherlands}

\email{ira7bar@gmail.com}


%\altaffiltext{1}{Visiting Astronomer, Cerro Tololo Inter-American Observatory.
%CTIO is operated by AURA, Inc.\ under contract to the National Science
%\altaffiltext{2}{Society of Fellows, Harvard University.}
%\altaffiltext{3}{present address: Center for Astrophysics,
%    60 Garden Street, Cambridge, MA 02138}
%\altaffiltext{4}{Visiting Programmer, Space Telescope Science Institute}
%\altaffiltext{5}{Patron, Alonso's Bar and Grill}


\begin{abstract}
Recent studies of the stellar population in the solar neighborhood (\textless20\,pc) suggest that there are undetected white dwarfs (WDs) in multiple systems with main sequence companions. 
Detecting these hidden stars and obtaining a more complete census of nearby WDs is important for our understanding of binary and galactic evolution, as well as the study of explosive phenomena.
In an attempt to uncover these hidden WDs, we present intermediate resolution spectroscopy over the wavelength range 3000-25000\,\AA\ of 101 nearby M dwarfs (dMs), observed with the Very Large Telescope X-Shooter spectrograph. For each star we search for a hot component superimposed on the dM spectrum.
X-Shooter has excellent blue sensitivity and thus can reveal a faint hot WD despite the brightness of its red companion. Visual examination shows no clear evidence of a WD in any of the spectra. We place upper limits on the effective temperatures of WDs that may still be hiding by fitting dM templates to the spectra, and modeling WD spectra. On average our survey is sensitive to WDs hotter than about 5300\,K. 
This suggests that the frequency of WD companions of $T_{\textrm{eff}}\gtrsim5300\,\textrm{K}$ with separation of order $\lesssim50\,$AU among the local dM population is \textless3\% at the 95\% confidence level.
The reduced spectra are made available on via WISeREP\footnote{http://wiserep.weizmann.ac.il} repository.      

\end{abstract}



\section{INTRODUCTION}

White dwarf stars are an important ingredient of stellar populations. As the end state of over 97\% of stars \citep{Fontaine-97WD}, they play a crucial role in understanding stellar and galactic evolution. For example, relating the luminosity function of these stars to their cooling sequences can yield estimates of their age, and thus the age of the galactic disk and the universe (e.g. \citealt{winget-wds}, \citealt{garcia-wds}). 

Characterizing WDs in binary and higher multiplicity systems is paramount for many fields of research such as explosive phenomena and binary evolution. For example, the origins of type Ia supernova explosions is still an open question. There is evidence that WDs are the progenitors of these explosions \citep{nugent-progenitor}, but the trigger of the explosion is still a puzzle. The two common models suggest a WD accreting material from a main sequence or red giant companion, or two binary WDs merging and exploding (see \citealt{Howell-Ia} and \citealt{maoz-sn} for reviews). Both models struggle to ignite an explosive detonation in simulations \citep{dong-Ia}. A different model suggests an explosion due to a collision of two WDs in a triple system (\citealt{Katz-collision}, \citealt{Kushnir-collisions}). This scenario easily produces explosions, but it is not clear whether enough such systems exist to account for the observed supernova rates. Improved statistics of WDs in multiple systems can constrain these models. Furthermore, this can test binary evolution and population synthesis models, which still suffer from many uncertainties (see \citealt{toonen-ce} and \citealt{ivanova-CE-review}).

WDs are compact faint objects and are thus difficult to detect, especially in multiple systems in which brighter, main sequence companions are present. Therefore, efforts have been made to obtain a complete census of WDs in the nearby solar neighborhood, from which statistics about galactic populations can be inferred. Holberg and collaborators made two such attempts for the local volume within 20\,pc (\citealt{holberg-2002}, \citealt{holberg-2008}) reaching an estimated completeness of 80\%, suggesting there are still $\sim$33 undetected WDs left in this volume. A recent study \citep{holberg-2016} has further improved this, reaching an estimated 86\% completeness. In contrast, \cite{Katz-missing-WDS} claims that the completeness fraction of \cite{holberg-2008} is over-estimated and the actual value is smaller.

So are there "missing" WDs? The answer may hide in multiple systems. According to \cite{holberg-2016}, in the local neighborhood 74\% of WDs are single stars and only 26\% are in binary or higher multiplicity systems. This is in contrast with the progenitors of these WDs - main sequence stars of K type and earlier (up to the minimum mass for supernova) - which show multiplicity rates of $\sim45\%$ and higher (e.g. \citealt{Raghavan-multiplicity}, \citealt{mason-multiplicity}, \citealt{derosa-multiplicity}). 

\cite{Ferrario-missingwds} suggests that this discrepancy may be due to an observational bias - WDs in binaries are simply too faint compared to their companions, and are thus not detected. \cite{Katz-missing-WDS} provided observational evidence supporting this claim, by using a WD luminosity-cooling age relation to derive the theoretically expected distribution of absolute visual magnitudes. They showed that the observed single WDs in the \cite{holberg-2008} 20\,pc sample roughly follow this distribution, while the number of WDs in binaries drops compared to the expectation, for magnitudes 12 and fainter. This gap between expected and observed WDs in binaries implies that there are $\sim$100 such nearby missing WDs, hiding in the light of main sequence companions. This is a significant number, considering that the local 20\,pc volume contains $\sim$1900 non-WD stars in total.

We present an attempt to unveil some of those WD companions by obtaining spectra of 101 nearby M dwarfs using the Very Large Telescope X-Shooter spectrograph. The targets were selected based on their strong Near-UV (NUV) excess as measured by the GALEX survey \citep{galex}, an excess which may arise from the contribution of a hidden WD to the spectrum, or from magnetic activity of the dM. We attempt to detect WD companions by examining the spectra for characteristic features of WDs. As no such evidence of WD presence is visible, we then put upper limits on the effective temperatures of WDs that may still be hiding below our detection threshold.

This paper is organized as follows: In Sec. \ref{sec:data} we present the observations. The results and analysis are given in Sec. \ref{sec:results}. In Sec. \ref{sec:discussion} we discuss and summarize the main ideas.

\section{DATA}\label{sec:data}

% Sample selection and observation
\subsection{Target Selection}

Our initial sample consists of dMs within 20\,pc of the sun, taken from the Gliese Catalog of Nearby Stars, 3rd edition \citep{1995yCat.5070....0G}. We have selected dMs since earlier types would be too bright and blue to allow detection of a faint blue companion.
We have limited our selection to dMs with absolute V magnitudes of 10-16\,mag. The faint limit was set by the faintest single WD in both the \cite{holberg-2008} and Gliese catalogs. Thus, a WD companion to a dM fainter than that would have to be brighter than the dM companion and would dominate the spectrum. In that case, the star would not be classified as a dM in the first place. The bright limit was set to ensure selecting dwarf stars and not sub-giants, as the Gliese catalog does not always include the luminosity class of the M stars.  
In addition, we have discarded targets which have nearby objects in the catalog with separations of less than 30\arcsec.
To increase our chances of detection, we cross-correlated our target list with the GALEX survey, choosing targets which show NUV excess. 
This excess can be caused by the contribution of a hidden WD companion to the blue side of the spectrum. \cite{Jones-UV-dM} showed that dM-WD binaries exhibit more NUV emission than single dMs, but also showed that this emission can also be the result of magnetic activity of dMs. 

\begin{figure}
\includegraphics[width=\columnwidth]{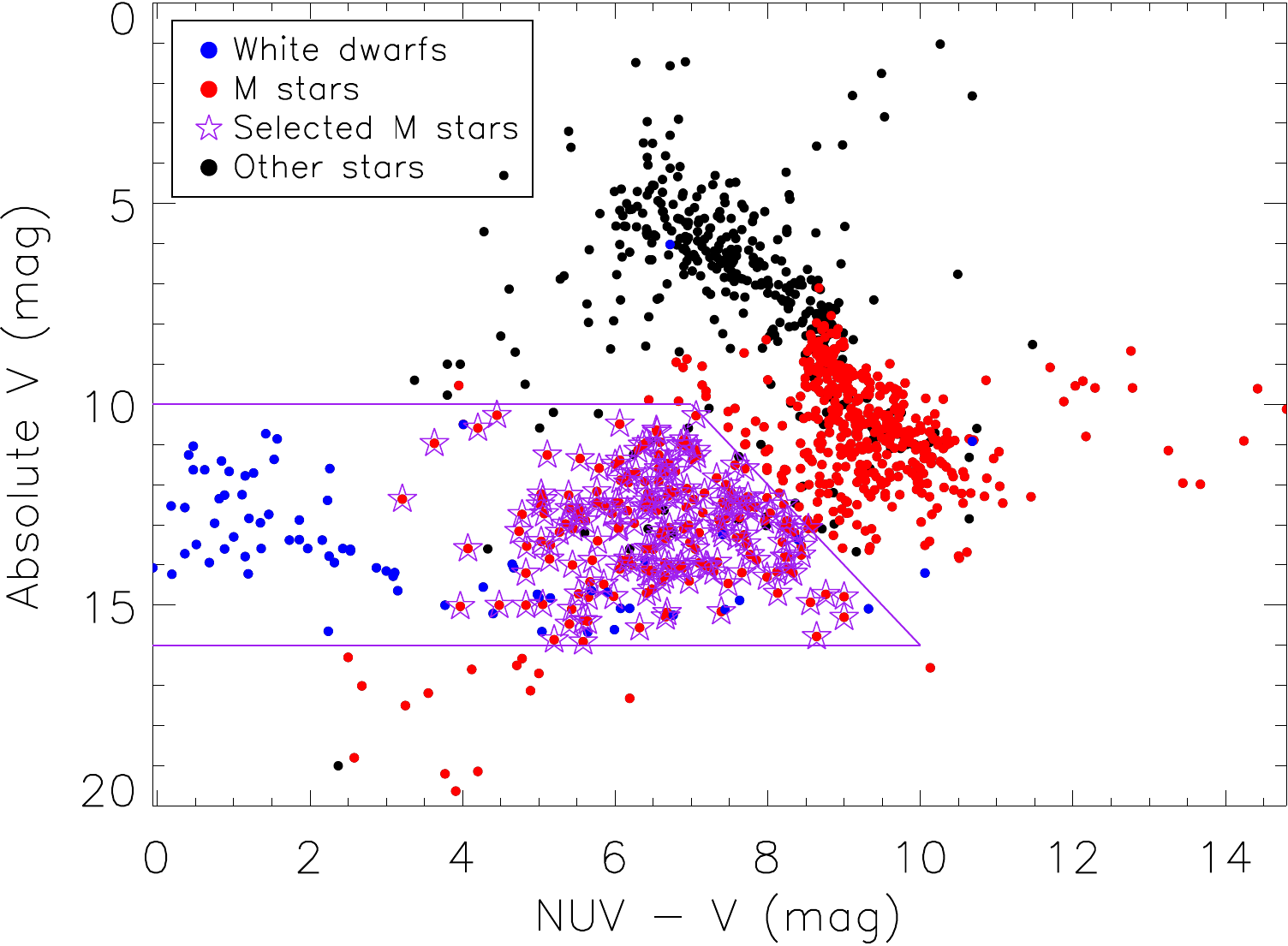}
\caption{$V$ / $NUV-V$ magnitude - color diagram of the stars in the Gliese catalog within 20\,pc of the sun. M stars are shown in red, single WDs (including additions from the \citealt{holberg-2008} catalog) are shown in blue and other stars are shown in black. The purple line illustrates our color cut and purple stars are the 138 M stars out of which our targets were selected.
\label{fig:colorcuts}}
\end{figure}

Cross matching between the Gliese and GALEX catalogs was done using tools in the astronomy \& astrophysics package for Matlab \citep{2014ascl.soft07005O}. 
Figure \ref{fig:colorcuts} illustrates our color cut which left us with a total of 138 dMs.   

\subsection{Observations and Data Reduction}

The observations were performed with the X-Shooter spectrograph mounted at the {\it Kueyen} unit of the Very Large Telescope (UT2), operated by the European Southern Observatory (ESO) on Cerro Paranal in Chile (Program IDs 095\_D-0949(A) and 096\_D-0963(A), PI: Paul Vreeswijk). X-Shooter records spectra with three separate arms simultaneously - UVB, VIS and NIR. This allows a wide wavelength range (3000-25000\,\AA) while maintaining good sensitivity throughout this range \citep{x-shooter}. In particular, the instrument is very blue-sensitive down to the atmospheric cutoff around 3000\,\AA. Thus, we benefit from the ability to characterize the dM spectrum in the red part while allowing possible detection of WD contribution in the blue part.
% show SNR plot from ETC?
Since out targets are very bright, our observation plan was submitted to ESO as a "filler" program, to be executed during poor seeing conditions (\textgreater1.5\arcsec). We obtained service-mode observations with VLT/X-Shooter for 60 (41) targets during ESO Period 95 (96), securing spectra of a total of 101 dMs. Corrected locations of all targets were calculated using proper motions from the \cite{Stauffer-gliese2mass} 2MASS-Gliese cross-correlation catalog.
Exposure times were typically 120\,s for UVB, 90\,s for VIS and four separate exposures of 30\,s each for NIR. In some cases, exposure times were increased for faint targets or decreased to prevent saturation of the detector. Moreover, saturation was avoided for bright targets ($M_V\sim10$) by using a 1x1 binning read-out mode instead of 1x2. All of the observations were made in Stare mode using the widest 5.0\arcsec\ slit, to increase the chances of including a WD companion in the spectrum (up to separation of 40\,AU for a 16\,pc target, the distance at half the 20\,pc volume). The resolving power of X-Shooter for the 5.0\arcsec\ slit was measured to be about 5000, 9000 and 5000 for the UVB, VIS and NIR arms, respectively. This was done by examining several unresolved lines in the spectra and dividing their wavelengths by their full width at half maximum (FWHM).  

Spectra were reduced using ESO's Reflex pipeline version 2.8 \citep{Reflex}. A built-in "optimal extraction" algorithm \citep{Horne-Optimal} was used to increase Signal-to-noise ratio (SNR). For the UVB and VIS arms, optimal extraction yielded a mean improvement of 5 and 4 percent in SNR, respectively, compared to standard extraction. On the other hand, for the NIR arm a mean 4 percent decrease in SNR was noted, with artifacts introduced to some of the spectra. Thus, we have decided to use optimal extraction for the UVB and VIS arms only, and standard extraction for the NIR arm.  

We removed very noisy parts in the UVB and VIS arms below 3200 and 5500\,\AA, respectively. The three spectral pieces were then stitched together using overlap regions as reference. Both the UVB and the VIS spectra include a feature at $\sim$5500-5800\,\AA\ due to the dichroic splitter of X-Shooter \citep{XSL}. Thus, we excluded this feature from the overlap regions.

The pipeline produces a flux-calibrated spectrum for each target using a spectroscopic standard star that was observed on the same night. This procedure assumes photometric nights. Therefore, we performed absolute calibration for each object using JHK band photometry from the 2MASS catalog \citep{2mass}, since these are the only bands that are reported for all of our targets in the SIMBAD database \citep{simbad}. We calculated JHK synthetic photometry for each spectrum and rescaled the flux to match the photometric data.

The log of spectroscopic observations is presented in Table \ref{tab:observations} in the appendix, along with plots of the reduced spectra. These are also available on the WISeREP\footnote{http://wiserep.weizmann.ac.il} repository \citep{WISeREP} and are searchable via object name or by type "M dwarf".

\section{Results and Analysis}\label{sec:results}

\subsection{Activity and Multiplicity}

\begin{figure*} 
\centering
\includegraphics[width=\textwidth]{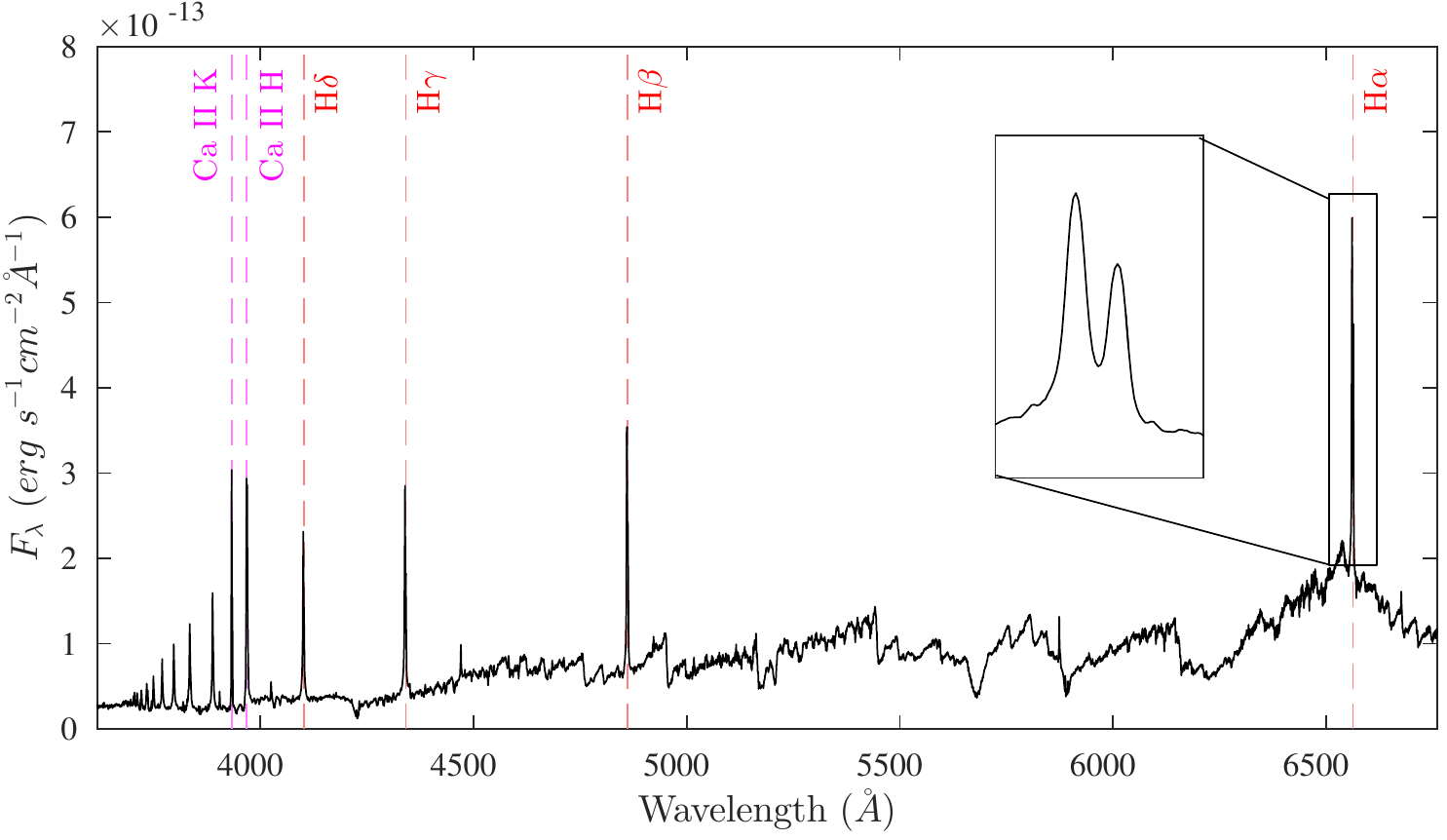}
\caption{Example emission lines for GJ2069. In red - the first four Balmer series lines. In pink - \CaII\ H/K lines. The double \Halpha\ line is shown in the inset.
\label{fig:gj2069_lines}\vspace{5pt}}
\end{figure*}

Out of the 101 observed targets, 65 show strong emission lines that are indicative of magnetic activity: the hydrogen Balmer series and the \CaII\ H and K lines \citep[see][]{Reid-dM-book}. An example of these lines for target GJ2069 is shown in Fig. \ref{fig:gj2069_lines}. The remainder of the targets show little or no emission lines. Five spectra are showing double emission lines (also shown in Fig. \ref{fig:gj2069_lines}), which suggests that these are in fact binary stars which are not listed in the Gliese catalog. A search of binary star catalogs (\citealt{mason-WDS} and \citealt{NN3129-binary}) shows that indeed these are known binary or higher multiplicity systems, and revealed 21 additional multiples in our sample with separations smaller than half of our slit width (targets were positioned in the center of the slit) that were not filtered out initially. The separations of these systems are listed in Table \ref{tab:results} and we give special care to these in our analysis later on.  

\subsection{White Dwarf Spectral Features}
According to the \cite{Kleinman-WDs-SDSS} SDSS survey, over 90\% of WDs are classified as having either hydrogen or helium atmospheres, indicating a spectral type of DA or DB, respectively. Thus, we first visually examine the spectrum of each target in search of distinct broad absorption features typical of these spectroscopic classes. For type DA these would be the hydrogen Balmer lines at 6563\,\AA, 4861\,\AA, 4341\,\AA\ and 3970\,\AA. For type DB the dominant lines are from neutral helium at 4026\,\AA, 4471\,\AA\ and 5875\,\AA\ \citep{Bergeron-DB-models}. Some WDs show only a continuum spectrum - type DC. According to \cite{Giammichele-DC-WDs}, those can be hydrogen atmosphere WDs of $T_{\textrm{eff}}\leq5000\,K $ or helium atmosphere WDs of $T_{\textrm{eff}}\leq12000\,K $. The search for this type and other more exotic types was done by looking for a clear rise in the blue part of each target. No evidence for DA or DB features, nor of a clear UV rise, was found in any of the spectra. 

\subsection{Limits on the Temperature of Unseen WDs}
Next, we place upper limits on the effective temperature of WDs that may still be hiding below our detection threshold. We do so by modeling spectra of WDs with a mass of $0.6\,\Msun$ as black-bodies of varying effective temperatures. We use this mass value as it is the peak of the narrow observed WD mass distribution (see \citealt{Bergeron-models}, \citealt{2007MNRAS.375.1315K}). Using these models and spectral templates for each dM, we determine at which temperature the WD models can be rejected.

As spectral dM templates, we tried to use the \cite{1998PASP..110..863P} spectral library, which did not provide good enough fits to our data due to the low resolution ($\sim$500\,\AA) and to a limited number of spectral subtypes. We have also tried using the PHOENIX synthetic spectral models \citep{phoenix}, which did not describe our data well in the UVB and NIR parts. Eventually, we have decided to use our own data as templates, under the assumption that most of our targets do not hide WD companions. Indeed, the self-template method yields better fits for the vast majority of our targets (see electronic Figs. \ref{fig:Results_NN3017} to \ref{fig:Results_NN4362}), compared with the alternative methods.

The best fitting template for each target was determined as the one with the lowest Residual Sum of Squares (RSS) score:
\begin{equation}\label{eqA:chi2}
\textrm{RSS}=\sum_i(f_{S,i}-f_{T,i})^2
\end{equation}
where $ f_{S,i}, f_{T,i}$ are the flux values of the current target and of each template, respectively. Before calculating the RSS score, the spectra were rescaled to each other such that the flux integral is equal to one. 
We excluded from the fit the blue part below 6300\,\AA\ to minimize possible WD contamination, while keeping the prominent dM features that extend to redder wavelengths.
We also excluded wavelength ranges of known telluric features at: 6340-6420\,\AA, 6840-6960\,\AA, 7147-7323\,\AA, 7575-7705\,\AA, 8130-8365\,\AA, 8939-9240\,\AA, 9280-9830\,\AA, 10810-11710\,\AA, 12670-12710\,\AA, 13000-15030\,\AA, 17350-19810\,\AA, 19950-20350\,\AA\ and 20480-20820\,\AA\ \citep{Moehler-tellurics}. As we only have 101 spectra to work with, dividing them into magnetically active and non-active would limit our ability to find good templates. Thus, we included both active and non-active stars in the template bank for all targets while excluding the Balmer \Halpha\ line from the fit range. As shown, for example, in Fig. \ref{fig:Results_GJ1031}, nice fits are produced also when using active templates for non-active targets and vice versa.

As white dwarf synthetic spectra were not available to us, we chose to model WDs as black-bodies of varying effective temperatures and with radii that fit a typical mass of $0.6\,\Msun$. The radius for each $T_{\textrm{eff}}$ value was calculated using $R^2=GM/g$ and the surface gravity ($\log g$) values from the publicly available WD color model grids of \cite{Bergeron-photometry}, \cite{kowalski-wd-models}, \cite{tremblay-wd-models}  and \cite{Bergeron-DB-models}.
Absolute flux values of the models were calculated using these radii and the distances to each target, obtained from the Gliese catalog.
 
Next, we determine the hottest WD that may be hiding in the data for each target. 
The dM fits are not perfect and exhibit correlated residuals (i.e. spectral regions which are systematically lower or higher between each target and its template).
Such correlated noise make it very difficult to make quantitative statements. 
We attempt to explain the observed spectrum using a combination of the template spectrum and an approximate WD model, and to ask when the temperature of the WD produces a "noticeable" effect on the composed spectrum.
In each such fit, the WD flux is completely determined by its radius (assuming a $0.6\,\Msun$ WD), its distance, and assuming a black-body emitter.
Therefore, we use several methods to put limits on the WD temperature that can be hidden in each system.
We note that since the WD luminosity is very sensitive to its effective temperature (i.e. $\propto T_{\textrm{eff}}^{4}$), any reasonable estimator will yield similar results regardless of the details of the test. 

In the first method, we reject WD models which yield an RSS score double than that of the template alone. We add the flux of the best fit template to the black-body models for a grid of effective temperatures between 1500\,K and 20000\,K, and calculate the RSS score for each temperature. We then compare these scores with the RSS of the fit with no WD. The lowest temperature for which the RSS score is more than double the score without the model is set as our upper limit. 
This test is arbitrary, but due to the extreme sensitivity of the luminosity on $T_{\textrm{eff}}$, the obtained limits are similar for different criteria. For example, changing the threshold to 3 times the RSS yields an average increase in limit temperature of only 400\,K. 
Our temperature grid follows that of the color models - from 1500 to 5500\,K in steps of 250\,K and then to 15000\,K in steps of 500\,K. As opposed to the template fitting, we now calculate the RSS score for wavelengths 3200-10000\,\AA, where WD contribution would be dominant. In addition to the \Halpha\ line and tellurics, we also remove from the fit range the rest of the Balmer emission lines, the \CaII\ lines and the X-Shooter dichroic feature. 

In the second method, we reject WD models which exceed the fit residuals envelope. The envelope is defined as the 99th percentile of the flux for the absolute value of the fit residuals. 
In this case, the limit is set as the lowest temperature for which the black-body model flux exceeds the envelope. In other words, the rejected WD model is the coldest one which is not consistent with the residuals of the fit.

\begin{figure*}[ht] 
 \centering
 \includegraphics[width=\textwidth]{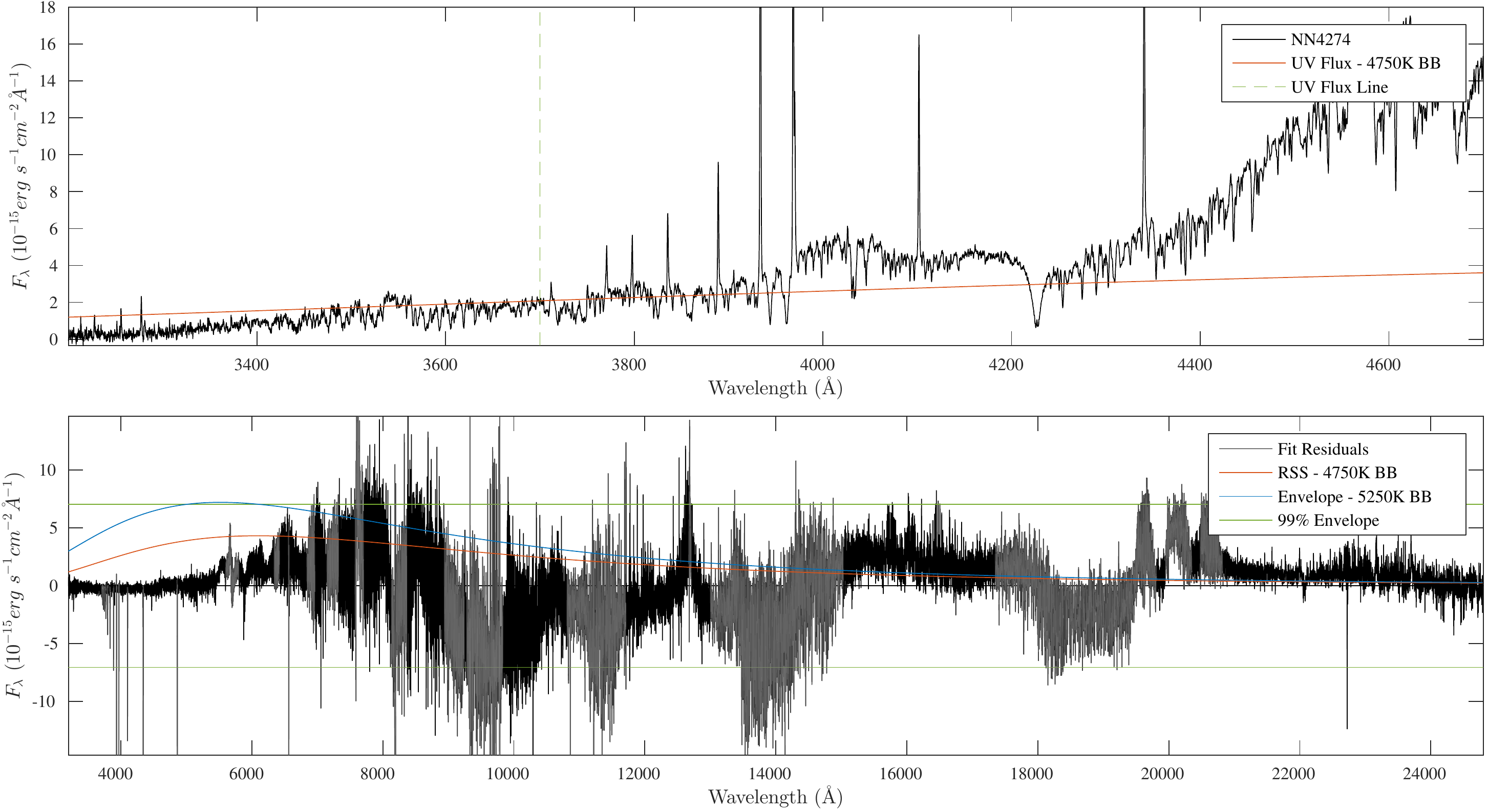}
 \caption{Example of limit calculations for NN4274. Top panel: The UV part of the spectrum is shown in black. In orange - the rejected black-body model from the UV flux method. In green - The 3700\,\AA\ line defining the UV flux region.
 Bottom panel: The fit residuals from the best fit template GJ1154 are shown in black. In orange and blue are the rejected black-body models from the RSS and envelope methods, respectively. In green - the 99\% flux envelope.
 \label{fig:limit_example}}
 \end{figure*}

The third method is robust and template-independent. Here, we compare the integrated flux of the very blue part of the dM spectra to that of our black-body models. The limit is set as the temperature of the coldest model for which the total UV flux is greater than that of the dM. 
This was done for wavelengths of 3200-3700\,\AA, as this range features only a weak continuum from the dMs. In addition, this range is bluer than the Balmer series, thus avoiding the typical absorption features of WDs and ensuring that black-body is a good approximation for WD spectra. An example of the three methods is given in Fig. \ref{fig:limit_example}.

To verify that the black-body model describes WDs well enough for our purposes, we repeat the analysis using spectra derived from photometry of the color models. For each effective temperature, we calculate flux values from the UBVRI band photometry using: 
\begin{equation}
F_{\lambda}=F_{\lambda,0}\cdot10^{(-0.4M_\lambda)}\cdot(10/D)^2
\end{equation} 
Where $M_\lambda$ are the UBVRI magnitudes, $F_0$ are vega zero point fluxes calculated in \cite{Bergeron-photometry} and D is the distance to each target, which is used to normalize the flux as the magnitudes were calculated for 10\,pc, i.e. absolute magnitudes. The wavelength of each flux value is taken as the effective wavelength of each band, also reported in \cite{Bergeron-photometry}. We then interpolate between these points to obtain a model spectrum. We do this for the hydrogen-atmosphere and the helium-atmosphere color models.
Both hydrogen and helium models yield the same limits as the black-body model up to 500\,K differences in effective temperature, showing that indeed using the black-body approximation is good enough for our analysis.

As noted before, 26 of our targets are listed in binary catalogs as binaries or higher multiplicity systems with separations of less than 2.5\arcsec\ (close binaries, hereafter), which is within half the X-Shooter slit width that we have used.
Five of those display double emission lines in their spectra. The spectra of the rest of these close binaries may or may not contain more than one star. Thus, they are problematic for our limit analysis and must be analyzed with care. As a first precaution, we have taken these targets out of the template bank. Second, for these cases we only report the limits obtained using the UV flux method, which is independent of template.

The limits obtained and best fit templates for each target are listed in Table \ref{tab:results}. Electronic Figs. \ref{fig:Results_NN3017} to \ref{fig:Results_NN4362} display, for each target:
\begin{enumerate}
  \item{The spectrum.}
 \item{The best fit template, rescaled to the flux of the target.}
 \item{The residuals of the fit and the rejected black-body models used to derive the limits.}
 \end{enumerate}
Figures \ref{fig:Results_NN3010} and following display the spectra of the close-binary targets, for which no template analysis was performed.

\clearpage

\LongTables
\begin{deluxetable}{llccccl} % l is left

\tablecolumns{7}
\tablewidth{0pt}
\tabletypesize{\small}
\tablecaption{Obtained results of spectral analysis}
\tablehead{
\colhead{Name}               &
\colhead{Template}            &
\colhead{RSS}	&
\colhead{Env.}	&
\colhead{UV Flux}	&
\colhead{Act.\tablenotemark{a}}	&
\colhead{Sep.\tablenotemark{b}}	\\
\colhead{}               &
\colhead{}            &
\colhead{[K]}	&
\colhead{[K]}	&
\colhead{[K]}	&
\colhead{}	&
\colhead{[\arcsec]}		
}
\startdata

NN3010	&		&		&		&	5250	&	V	&	0.1	\\
NN3017	&	GJ2021	&	5250	&	6500	&	5250	&	V	&		\\
NN3027	&	NN3682	&	5250	&	6500	&	5500	&	V	&		\\
NN3033	&	NN3518	&	5000	&	6500	&	4750	&	V	&		\\
NN3034	&	GJ1204	&	5000	&	6500	&	5000	&		&		\\
NN3039	&		&		&		&	6000	&	V	&	0.7	\\
GJ1019	&	NN3227	&	6000	&	6000	&	5000	&		&		\\
NN3056	&	NN3142	&	4500	&	6000	&	5250	&		&		\\
GJ1024	&	GJ1031	&	6000	&	7000	&	5500	&		&		\\
GJ1029	&	NN3937	&	4750	&	6000	&	3750	&		&		\\
GJ1031	&	NN3225	&	5000	&	5500	&	5250	&	V	&		\\
GJ2021	&	GJ1031	&	4500	&	5500	&	4750	&		&		\\
NN3076	&		&		&		&	4250	&	V	&	0.3	\\
NN3119	&	NN4292	&	4000	&	5000	&	4250	&	V	&		\\
Gl83.1	&	NN3225	&	4000	&	5000	&	4750	&	V	&		\\
NN3129	&		&		&		&	5500	&	D\tablenotemark{c}	&		\\
NN3142	&	NN3149	&	5250	&	7000	&	5500	&	V	&		\\
NN3143	&	Gl828	&	6000	&	6500	&	5500	&	V	&		\\
NN3149	&	GJ1031	&	6000	&	7000	&	6500	&	V	&		\\
NN3148	&	GJ1284	&	6500	&	8000	&	6500	&	V	&		\\
Gl102	&	GJ1031	&	5250	&	6500	&	5250	&		&		\\
GJ1055	&	NN3253	&	3750	&	5000	&	4250	&		&		\\
NN3224	&		&		&		&	4500	&	D	&	0.3+2	\\
NN3225	&	Gl83.1	&	4500	&	5250	&	4000	&	V	&		\\
NN3227	&	Gl729	&	7500	&	7000	&	6000	&	V	&		\\
NN3237	&	NN3253	&	6500	&	6500	&	5500	&	V	&		\\
NN3253	&	NN3237	&	4500	&	5500	&	4500	&		&		\\
NN3261	&		&		&		&	6500	&	V	&	0.8	\\
NN3296	&	NN3225	&	5500	&	6000	&	5000	&	V	&		\\
NN3304	&		&		&		&	5500	&	V	&	1.1	\\
NN3322	&		&		&		&	6500	&	V	&	1.4	\\
NN3326	&	NN3967	&	6000	&	6500	&	5250	&	V	&		\\
NN3332	&		&		&		&	8000	&	V	&	0.8	\\
NN3344	&	Gl207.1	&	10000	&	9500	&	8500	&		&		\\
Gl207.1	&	GJ1284	&	10500	&	9500	&	8000	&	V	&		\\
GJ1083	&		&		&		&	4000	&	V	&	0.5	\\
NN3405	&	Gl828	&	6500	&	7500	&	6000	&		&		\\
GJ1093	&	GJ1286	&	4000	&	5250	&	3750	&		&		\\
NN3423	&	Gl729	&	5500	&	6500	&	5500	&	V	&		\\
GJ1096	&	NN3225	&	5000	&	6000	&	4500	&	V	&		\\
Gl268.3	&		&		&		&	6500	&		&	0.1	\\
NN3454	&		&		&		&	4750	&	V	&	0.3	\\
NN3463	&	GJ1103	&	4250	&	5250	&	4750	&		&		\\
GJ1103	&	NN3463	&	5500	&	5500	&	5000	&		&		\\
NN3466	&		&		&		&	5500	&		&	1	\\
GJ1108	&		&		&		&	7500	&	D	&	0.3+14	\\
Gl300	&		&		&		&	5250	&		&	2	\\
NN3503	&	NN3518	&	4500	&	5500	&	4250	&		&		\\
GJ2069	&		&		&		&	7000	&	D	&	1+10+22	\\
Gl324	&	NN3967	&	6500	&	7000	&	5500	&		&		\\
NN3518	&	NN3033	&	5000	&	6000	&	5000	&	V	&		\\
GJ1116	&		&		&		&	3750	&	V	&	1.8	\\
NN3543	&	NN3344	&	8500	&	8000	&	6500	&	Ca	&		\\
NN3549	&	NN3937	&	4000	&	5250	&	4250	&	V	&		\\
Gl347	&	GJ1031	&	5250	&	6000	&	4500	&		&		\\
Gl359	&	Gl347	&	4500	&	6000	&	4500	&		&		\\
NN3571	&	Gl347	&	4750	&	6000	&	4750	&	V	&		\\
NN3572	&	GJ1186	&	8500	&	8000	&	6500	&		&		\\
NN3573	&	NN3571	&	7000	&	7500	&	5000	&		&		\\
NN3590	&	GJ1186	&	6000	&	7500	&	5000	&		&		\\
NN3647	&	GJ1031	&	6000	&	8000	&	6500	&		&		\\
NN3654	&	Gl781.1	&	7000	&	7500	&	5500	&		&		\\
NN3657	&	GJ1031	&	4250	&	6000	&	4250	&		&		\\
NN3668	&	NN3463	&	4500	&	6000	&	4750	&		&		\\
NN3682	&	NN3780	&	6000	&	6500	&	6000	&	V	&		\\
NN3685	&	NN3149	&	6500	&	7000	&	6000	&	V	&		\\
GJ1154	&	NN3149	&	4250	&	5500	&	4750	&	V	&		\\
NN3780	&	NN3682	&	6000	&	6500	&	6000	&	V	&		\\
GJ1179	&	NN3682	&	4750	&	5250	&	3750	&	V	&		\\
NN3808	&	GJ2021	&	6000	&	6500	&	5000	&	V	&		\\
Gl540.2	&	NN4292	&	5500	&	6500	&	5250	&	V	&		\\
NN3856	&	NN3780	&	5250	&	6500	&	5500	&	V	&		\\
GJ1186	&	NN4292	&	5250	&	5500	&	4500	&		&		\\
NN3900	&	Gl828	&	5500	&	7000	&	6000	&		&		\\
NN3937	&	GJ1029	&	4750	&	5500	&	4250	&		&		\\
GJ1204	&	GJ1284	&	6000	&	6500	&	5250	&	V	&		\\
NN3967	&	NN3326	&	4500	&	6000	&	5000	&	V	&		\\
NN3981	&		&		&		&	5500	&	V	&	0.6	\\
GJ1210	&		&		&		&	4750	&	V	&	0.4	\\
NN4032	&	NN3227	&	8500	&	8000	&	6500	&		&		\\
NN4071	&	NN3423	&	6000	&	7000	&	5500	&	V	&		\\
Gl729	&	NN3227	&	4750	&	6000	&	5250	&	V	&		\\
Gl781.1	&	NN4279	&	7500	&	8000	&	7000	&	V	&		\\
Gl791.2	&		&		&		&	5000	&	V	&	0+0.2	\\
Gl828	&	NN3143	&	4750	&	6500	&	5500	&		&		\\
NN4201	&		&		&		&	5500	&	V	&	0.8	\\
Gl836	&	NN4071	&	6000	&	6500	&	6000	&		&		\\
NN4215	&	NN4292	&	4500	&	5500	&	4500	&		&		\\
NN4231	&		&		&		&	7500	&	V	&	0.2+0.7	\\
NN4239	&	NN3657	&	5250	&	6000	&	4500	&		&		\\
NN4274	&	GJ1154	&	4750	&	5250	&	4750	&	V	&		\\
NN4279	&	Gl781.1	&	5000	&	6000	&	5500	&	V	&		\\
NN4282	&		&		&		&	7500	&	V	&	1.5	\\
NN4292	&	NN3119	&	5500	&	5500	&	4500	&	V	&		\\
NN4302	&	Gl781.1	&	6000	&	6500	&	5500	&		&		\\
NN4326	&		&		&		&	6500	&	V	&	0.1	\\
GJ1284	&	Gl83.1	&	9500	&	9000	&	7500	&	V	&		\\
GJ1286	&	Gl207.1	&	5000	&	5250	&	4000	&		&		\\
NN4360	&		&		&		&	4500	&	V	&	0.6	\\
NN4362	&	GJ1284	&	7500	&	8000	&	7000	&	V	&		\\
NN4378	&		&		&		&	7000	&	D	&	0.6+20

\enddata
\tablecomments{RSS, Env. and UV Flux are the limits obtained using the three methods. Targets without reported templates and RSS/Envelope limits are those defined as close binaries.
\footnotetext[a]{Activity - B+Ca stands for Balmer and \CaII\ emission lines, Ca for \CaII\ lines only, D for double emission lines (Balmer and \CaII).}
\footnotetext[b]{Companion separations from the WDS catalog, listed for targets where at least one companion is closer than 2.5\arcsec.}
\footnotetext[c]{Listed as binary in \cite{NN3129-binary}, no separation reported.}}

\label{tab:results}
\end{deluxetable}

\begin{figure*}[] 
\centering
\includegraphics[width=\textwidth]{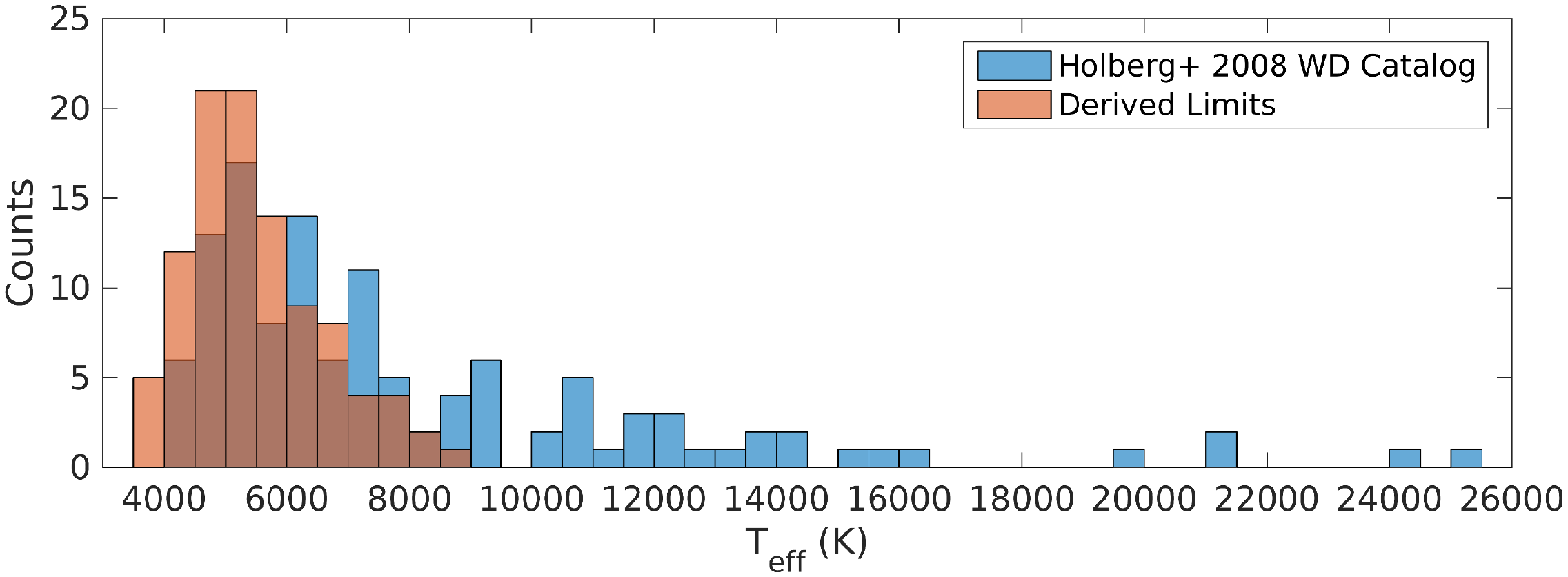}
\caption{The effective temperature distributions of the limits derived in this study and of the WDs in the \cite{holberg-2008} catalog.
\label{fig:histograms}}
\end{figure*}

\section{DISCUSSION}\label{sec:discussion}
Motivated by the evidence for missing WDs in multiple systems, we have used X-Shooter to record spectra of 101 dMs which show NUV excess. We have examined these spectra for evidence of WD spectral features. When those were not found, we have used the spectra themselves as templates, modeled WDs as black-bodies and obtained upper limits for the effective temperatures of WDs that may still be hidden, using three different methods. It is important to note that the limits obtained are only within our slit width of 5\arcsec. Since our targets are positioned roughly at the center of the slit, this corresponds to a companion at 40\,AU for a target at 16\,pc (the distance at half the 20\,pc volume). Though our slit is only 5\arcsec\ wide, it is 11\arcsec\ long, which could allow a companion at larger separations. Thus, we take order $\sim$50\,AU as our typical separation limit. 

Figure \ref{fig:histograms} shows the effective temperature distribution of our derived limits, selected as the tightest limit for each target among the three methods. 
Also shown is the effective temperature distribution of WDs in the \cite{holberg-2008} catalog. According to these, our limit analysis is sensitive to $\sim$75\% of WDs in the local volume. The average obtained limit for our 101 dM sample is 5300\,K. Thus, the frequency of dM-WD binaries with WD hotter than that is smaller than 3\% at the 95\% confidence level.  

The Gliese catalog lists $\sim$1900 non-WD stars with distances below 20\,pc. Assuming that 100 WD companions are missing from that volume \citep{Katz-missing-WDS} 
and that WDs do not prefer specific class of companions, then 5\% of the stars
should have WD companions.
As we have detected no WDs in our 101 dM sample, this hypothesis can be rejected at the 99.3\% confidence level.
However, our survey is sensitive to WDs hotter than $\sim$5300\,K, corresponding to $\sim$75\% of the known local population. In addition, our slit width limits our survey to separations of order $\lesssim50$\,AU, which corresponds to 55\% of dM-WD pairs \citep{farihi-wddm}. Taking both into account, the expectancy is to detect only two WDs in our sample, and this hypothesis can be rejected at the 86\% confidence level. The missing WDs may be colder than our detection limit or outside our slit, thus undetected. Another possibility is that the missing WDs are not hidden in the shadow of dMs, but rather are companions to stars of earlier types, as proposed in \cite{Ferrario-missingwds}. Or else, perhaps not as many WDs are missing from the local volume as claimed. 

It is interesting to note that selecting targets according to NUV excess produced a strong bias towards active dMs. For example, for spectral types M4 and M5, which make up the majority of our sample, we noted a 62\% activity fraction, as opposed to $\sim$35\% for the SDSS survey of \cite{west-activity}. Studies of magnetic activity of dMs can thus benefit from our data set.

As mentioned before, our spectra are available online for public use. In our observations, we covered the entire wavelength range of active and non-active dMs at medium resolution, including several binaries with resolved double emission lines. We hope that this data will be useful for future research.

\acknowledgements

We would like to thank Barak Zackay for many insightful discussions.\\
This work used the astronomy \& astrophysics package for Matlab \citep{2014ascl.soft07005O}\\
This work has made use of the SIMBAD data base, operated at CDS,
Strasbourg, France.\\
This work makes use of data obtained from the Weizmann interactive supernova data repository - http://wiserep.weizmann.ac.il\\
This work makes use of data products from the 2MASS,
which is a joint project of the University of Massachusetts and
the Infrared Processing and Analysis Center/California Institute of
Technology, funded by NASA and the NSF. \\
This work has made use of the Washington Double Star Catalog maintained at the U.S. Naval Observatory.\\
This work has made use of the tables for evolutionary sequences of white dwarf atmospheres of \cite{Bergeron-photometry}, \cite{kowalski-wd-models}, \cite{tremblay-wd-models}  and \cite{Bergeron-DB-models} - http://www.astro.umontreal.ca/bergeron/CoolingModels \\
This work has made use of observations made with the NASA Galaxy Evolution Explorer. GALEX is operated for NASA by the California Institute of Technology under NASA contract NAS5-98034.\\
We acknowledge support from the Weizmann Institute, in particular the Kimmel Award.

\clearpage
\appendix
The appendix contains Table \ref{tab:observations} of the observation log, Figs. \ref{fig:Results_NN3017} to \ref{fig:Results_NN4362} which contain the spectra, the best fit templates, the fit residuals and black-body models of the derived limits. Excluded fit ranges of tellurics, emission lines and dichroic feature are colored in gray. Also included are Figs. \ref{fig:Results_NN3010} and following, which display the spectra of the close binary targets.

\clearpage
\LongTables
\begin{landscape}

\centering
\begin{deluxetable}{llllllllllccc} 

\tablecolumns{13}
\tablewidth{0pt}
\tablecaption{Observed M dwarfs}
\tablehead{
\colhead{Name}               &
\colhead{RA}    &
\colhead{DEC}           &
\colhead{Spectral}        &
\colhead{$M_V$}            &
\colhead{Distance}            &
\colhead{Observation}	&
\colhead{Observation}	&
\colhead{Airmass}	&
\colhead{Seeing}	&
\colhead{UVB Exposure}	&
\colhead{VIS Exposure}	&
\colhead{NIR Exposure} \\
\colhead{}               &
\colhead{}    &
\colhead{}           &
\colhead{Type}        &
\colhead{}            &
\colhead{[pc]}            &
\colhead{Date}	&
\colhead{UT Time}	&
\colhead{[\arcsec]}	&
\colhead{[\arcsec]}	&
\colhead{[s]}	&
\colhead{[s]}	&
\colhead{[s]}	
}
\startdata
NN3010	&	00:08:53.95	&	+20:50:22.4	&	M5	&	13.54	&	10.64	&	2015-07-16	&	09:54:56	&	1.452	&	0.9	&	120	&	90	&	4x30	\\
NN3017	&	00:15:36.88	&	-29:46:02.7	&	M4	&	14.31	&	17.86	&	2015-08-13	&	05:25:54	&	1.138	&	1.5	&	120	&	90	&	4x30	\\
NN3027	&	00:18:54.11 	&	+27:48:47.4	&	M4	&	13.86	&	19.23	&	2015-10-30	&	00:41:26	&	1.914	&	1.2	&	120	&	90	&	4x30	\\
NN3033	&	00:24:35.78	&	+30:02:30.2	&	M5	&	14.54	&	18.87	&	2015-10-30	&	01:03:07	&	1.937	&	1.1	&	120	&	90	&	4x30	\\
NN3034	&	00:25:20.42	&	+22:53:04.0	&	M4	&	14.30	&	18.87	&	2015-06-20	&	10:06:16	&	1.566	&	1.1	&	120	&	90	&	4x30	\\
NN3039	&	00:32:34.97	&	+07:29:25.3	&	M4	&	12.70	&	11.63	&	2015-07-25	&	08:58:25	&	1.182	&	1.1	&	120	&	90	&	4x30	\\
GJ1019	&	00:43:35.52 	&	+26:28:25.1	&	M4	&	14.52	&	19.27	&	2015-10-14	&	05:33:38	&	1.886	&	1.1	&	120	&	90	&	4x30	\\
NN3056	&	00:47:08.22	&	-23:30:31.1	&	M3	&	14.40	&	18.52	&	2015-08-13	&	05:47:17	&	1.169	&	1.6	&	120	&	90	&	4x30	\\
GJ1024	&	00:56:39.19	&	+17:27:30.3	&	M4	&	13.71	&	17.42	&	2015-07-25	&	08:49:12	&	1.371	&	1.1	&	120	&	90	&	4x30	\\
GJ1029	&	01:05:39.95 	&	+28:29:31.3	&	M5	&	14.80	&	12.56	&	2015-10-14	&	06:01:32	&	1.919	&	1.2	&	120	&	90	&	4x30	\\
GJ1031	&	01:08:19.12	&	-28:48:23.6	&	M3	&	13.42	&	13.16	&	2015-08-13	&	05:33:08	&	1.277	&	1.2	&	120	&	90	&	4x30	\\
GJ2021	&	01:09:19.02	&	-24:4x30:28.3	&	M4	&	14.52	&	15.38	&	2015-08-13	&	05:40:14	&	1.262	&	1.3	&	120	&	90	&	4x30	\\
NN3076	&	01:11:25.65	&	+15:26:18.5	&	M5	&	14.36	&	8.47	&	2015-07-25	&	07:50:46	&	1.510	&	1.2	&	120	&	90	&	4x30	\\
NN3119	&	01:51:04.76	&	-06:07:10.0	&	M4.5	&	14.60	&	10.00	&	2015-07-25	&	09:23:14	&	1.089	&	1.0	&	120	&	90	&	4x30	\\
Gl83.1	&	02:00:14.15	&	+13:02:38.3	&	M4	&	12.28	&	4.47	&	2015-07-25	&	08:03:38	&	1.631	&	1.2	&	120	&	90	&	4x30	\\
NN3129	&	02:02:44.89	&	+13:34:4x30.9	&	M4.5	&	14.27	&	17.24	&	2015-07-25	&	08:19:09	&	1.570	&	1.2	&	120	&	90	&	4x30	\\
NN3142	&	02:12:55.31	&	+00:00:17.3	&	M4	&	13.50	&	16.67	&	2015-07-25	&	09:33:35	&	1.155	&	0.9	&	120	&	90	&	4x30	\\
NN3143	&	02:15:34.75 	&	+33:57:34.9	&	M3.5	&	13.58	&	17.24	&	2015-10-13	&	05:53:42	&	1.929	&	1.0	&	120	&	90	&	4x30	\\
NN3149	&	02:16:35.90	&	-30:58:05.4	&	M4	&	13.00	&	11.36	&	2015-08-13	&	06:10:58	&	1.409	&	1.0	&	120	&	90	&	4x30	\\
NN3148	&	02:16:41.99	&	-4x30:59:15.8	&	M3	&	12.00	&	11.36	&	2015-08-13	&	06:03:07	&	1.453	&	1.0	&	120	&	90	&	4x30	\\
Gl102	&	02:33:37.24	&	+24:55:27.3	&	M3.5	&	12.96	&	9.77	&	2015-07-25	&	09:50:28	&	1.651	&	0.8	&	120	&	90	&	4x30	\\
GJ1055	&	03:09:00.59 	&	+10:01:16.4	&	M5	&	14.85	&	11.88	&	2015-12-15	&	04:41:59	&	1.536	&	0.8	&	120	&	90	&	4x30	\\
NN3224	&	03:25:42.18 	&	+05:51:50.4	&	M4.5	&	14.70	&	12.99	&	2015-12-15	&	04:50:56	&	1.417	&	0.8	&	120	&	90	&	4x30	\\
NN3225	&	03:26:45.08	&	+19:14:39.3	&	M4.5	&	14.96	&	16.67	&	2015-10-11	&	09:23:10	&	1.855	&	1.0	&	120	&	90	&	4x30	\\
NN3227	&	03:28:49.89 	&	+26:29:10.3	&	M4	&	13.40	&	18.18	&	2015-10-11	&	08:01:56	&	1.689	&	0.4	&	120	&	90	&	4x30	\\
NN3237	&	03:36:41.04 	&	+03:29:17.5	&	M5	&	13.86	&	14.29	&	2015-12-15	&	04:59:38	&	1.369	&	0.8	&	120	&	90	&	4x30	\\
NN3253	&	03:52:42.29 	&	+17:00:55.0	&	M5	&	13.70	&	9.62	&	2015-10-14	&	06:27:09	&	1.360	&	0.8	&	120	&	90	&	4x30	\\
NN3261	&	04:05:38.91 	&	+05:44:40.3	&	M3.5	&	12.89	&	15.87	&	2015-12-15	&	05:07:15	&	1.325	&	0.8	&	120	&	90	&	4x30	\\
NN3296	&	04:33:34.62 	&	+20:44:41.7	&	M5	&	14.60	&	15.63	&	2015-10-14	&	07:22:20	&	1.432	&	1.2	&	120	&	90	&	4x30	\\
NN3304	&	04:38:13.13 	&	+28:12:58.7	&	M4.5	&	12.53	&	10.00	&	2015-10-13	&	07:31:23	&	1.665	&	1.0	&	120	&	90	&	4x30	\\
NN3322	&	05:01:58.86 	&	+09:58:57.7	&	M3.5	&	11.47	&	7.04	&	2015-12-17	&	05:47:09	&	1.367	&	0.7	&	120	&	90	&	4x30	\\
NN3326	&	05:04:14.67 	&	+11:03:27.3	&	M5	&	13.75	&	13.70	&	2015-12-17	&	05:58:45	&	1.419	&	0.6	&	120	&	90	&	4x30	\\
NN3332	&	05:06:49.54 	&	-21:35:04.8	&	M3	&	11.66	&	12.05	&	2015-10-21	&	06:26:10	&	1.063	&	1.1	&	60	&	45	&	4x15	\\
NN3344	&	05:16:00.36 	&	-72:13:59.7	&	M2	&	11.70	&	19.23	&	2015-10-21	&	06:35:03	&	1.523	&	0.8	&	120	&	90	&	4x30	\\
Gl207.1	&	05:33:44.55 	&	+01:56:39.5	&	M3	&	11.53	&	15.08	&	2015-12-17	&	05:37:45	&	1.163	&	0.6	&	60	&	45	&	4x10	\\
GJ1083	&	05:40:25.91 	&	+24:48:02.3	&	M7	&	14.85	&	10.34	&	2015-10-11	&	08:28:00	&	1.560	&	1.3	&	120	&	90	&	4x30	\\
NN3405	&	06:42:13.46 	&	+03:35:26.5	&	M4	&	13.33	&	15.63	&	2015-12-17	&	06:29:22	&	1.159	&	0.7	&	120	&	90	&	4x30	\\
GJ1093	&	06:59:29.95 	&	+19:20:41.0	&	M5	&	14.83	&	7.76	&	2015-10-11	&	08:48:44	&	1.542	&	1.0	&	120	&	90	&	4x30	\\
NN3423	&	07:03:23.25 	&	+34:41:54.9	&	M4	&	13.17	&	13.33	&	2015-11-06	&	08:31:36	&	1.957	&	1.1	&	120	&	90	&	4x30	\\
GJ1096	&	07:16:18.07 	&	+33:09:03.8	&	M5	&	14.48	&	14.90	&	2015-11-07	&	08:40:59	&	1.873	&	1.1	&	120	&	90	&	4x30	\\
Gl268.3	&	07:16:19.93 	&	+27:08:29.8	&	M3	&	10.85	&	7.94	&	2015-10-11	&	08:59:03	&	1.842	&	0.8	&	120	&	90	&	4x10	\\
NN3454	&	07:36:25.38 	&	+07:04:38.6	&	M4.5	&	13.22	&	6.17	&	2015-12-17	&	06:37:36	&	1.174	&	0.6	&	120	&	90	&	4x30	\\
NN3463	&	07:51:51.87 	&	+05:32:51.1	&	M4.5	&	14.75	&	15.92	&	2015-12-07	&	07:59:18	&	1.165	&	0.7	&	120	&	90	&	4x30	\\
GJ1103	&	07:51:54.99 	&	-00:00:23.2	&	M4.5	&	13.50	&	8.79	&	2015-12-07	&	08:26:09	&	1.132	&	0.8	&	120	&	90	&	4x30	\\
NN3466	&	07:54:55.22 	&	-29:21:04.4	&	M4	&	13.38	&	12.50	&	2015-10-21	&	07:08:15	&	1.462	&	0.8	&	120	&	90	&	4x30	\\
GJ1108	&	08:08:55.43 	&	+32:49:02.6	&	M2.8+M3.3	&	12.12	&	17.24	&	2015-12-06	&	08:05:23	&	1.855	&	0.9	&	120	&	90	&	4x20	\\
Gl300	&	08:12:40.98 	&	-21:33:18.1	&	M3.5	&	12.10	&	5.88	&	2015-10-21	&	06:57:12	&	1.734	&	0.7	&	120	&	90	&	4x30	\\
NN3503	&	08:31:22.82 	&	-10:29:59.9	&	M4	&	15.00	&	15.38	&	2015-10-21	&	07:25:13	&	1.802	&	1.1	&	120	&	90	&	4x30	\\
GJ2069	&	08:31:37.48 	&	+19:23:37.5	&	M5	&	11.89	&	8.77	&	2015-11-19	&	07:33:41	&	1.596	&	1.0	&	120	&	90	&	4x30	\\
Gl324	&	08:52:40.41 	&	+28:18:55.5	&	M4	&	13.14	&	13.09	&	2015-12-09	&	07:47:06	&	1.683	&	0.8	&	120	&	90	&	4x30	\\
NN3518	&	08:55:19.62 	&	-23:52:14.1	&	M4	&	14.00	&	12.20	&	2015-10-21	&	07:32:49	&	1.776	&	0.9	&	120	&	90	&	4x30	\\
GJ1116	&	08:58:14.39 	&	+19:45:46.1	&	M5.5	&	14.06	&	5.23	&	2015-11-19	&	07:14:53	&	1.883	&	1.0	&	120	&	90	&	4x30	\\
NN3543	&	09:16:20.35	&	-18:37:31.3	&	M2	&	10.75	&	12.50	&	2015-04-15	&	00:09:21	&	1.008	&	0.5	&	120	&	70	&	4x10	\\
NN3549	&	09:18:41.36 	&	+26:45:46.4	&	M5	&	16.00	&	20.00	&	2015-12-09	&	08:12:52	&	1.625	&	1.0	&	120	&	90	&	4x30	\\
Gl347	&	09:28:55.54	&	-07:22:22.0	&	M4.5	&	15.00	&	16.72	&	2015-05-19	&	01:03:31	&	1.332	&	1.2	&	120	&	90	&	4x30	\\
Gl359	&	09:41:02.70	&	+22:01:21.0	&	M4.5	&	14.23	&	12.17	&	2015-04-09	&	23:55:43	&	1.557	&	0.8	&	120	&	90	&	4x30	\\
NN3571	&	09:53:54.82	&	+20:56:52.2	&	M4	&	14.05	&	10.20	&	2015-04-14	&	23:41:40	&	1.548	&	0.4	&	120	&	90	&	4x30	\\
NN3572	&	09:55:43.61 	&	+35:21:41.7	&	M3.5	&	12.73	&	17.54	&	2016-02-04	&	07:13:20	&	2.241	&	0.8	&	120	&	90	&	4x30	\\
NN3573	&	09:56:26.53	&	+22:38:57.9	&	M4	&	14.20	&	16.13	&	2015-04-12	&	01:24:52	&	1.470	&	0.7	&	120	&	90	&	4x30	\\
NN3590	&	10:15:06.93 	&	+31:25:08.7	&	M4	&	13.60	&	18.18	&	2015-12-14	&	08:15:25	&	1.909	&	0.7	&	120	&	90	&	4x30	\\
NN3647	&	11:11:51.74 	&	+32:33:11.4	&	M3.5	&	12.38	&	12.20	&	2016-02-03	&	08:53:00	&	2.250	&	0.8	&	120	&	90	&	4x30	\\
NN3654	&	11:16:37.08	&	-27:57:30.5	&	M3.5	&	13.70	&	15.63	&	2015-05-17	&	02:22:37	&	1.126	&	0.9	&	120	&	90	&	4x30	\\
NN3657	&	11:23:07.96 	&	+25:53:36.8	&	M5	&	15.14	&	17.33	&	2016-02-05	&	05:55:29	&	1.668	&	0.8	&	120	&	90	&	4x30	\\
NN3668	&	11:31:08.78	&	-14:57:41.2	&	M5	&	14.29	&	12.82	&	2015-07-15	&	23:42:57	&	1.388	&	0.7	&	120	&	90	&	4x30	\\
NN3682	&	11:43:23.43	&	+25:18:13.5	&	M4	&	13.83	&	18.87	&	2015-05-19	&	00:52:43	&	1.551	&	1.1	&	120	&	90	&	4x30	\\
NN3685	&	11:47:40.46	&	+00:15:19.7	&	M4	&	13.25	&	15.63	&	2015-07-15	&	23:50:34	&	1.520	&	0.9	&	120	&	90	&	4x30	\\
GJ1154	&	12:14:15.60	&	+00:37:22.9	&	M4.5	&	13.73	&	8.46	&	2015-07-16	&	00:00:32	&	1.431	&	1.0	&	120	&	90	&	4x30	\\
NN3780	&	13:23:37.34	&	-25:54:47.8	&	M3.5	&	12.90	&	12.66	&	2015-08-10	&	23:35:04	&	1.256	&	1.2	&	120	&	90	&	4x30	\\
GJ1179	&	13:48:11.82	&	+23:36:50.9	&	M5	&	15.32	&	11.99	&	2015-07-16	&	00:08:44	&	1.595	&	0.9	&	120	&	90	&	4x30	\\
NN3808	&	13:48:48.66	&	+04:06:00.9	&	M4	&	14.34	&	16.39	&	2015-08-13	&	23:26:41	&	1.417	&	1.3	&	120	&	90	&	4x30	\\
Gl540.2	&	14:13:04.24	&	-12:01:31.5	&	M5	&	13.86	&	11.63	&	2015-08-13	&	23:36:29	&	1.192	&	0.9	&	120	&	90	&	4x30	\\
NN3856	&	14:32:11.01	&	+16:00:49.1	&	M5	&	13.61	&	14.93	&	2015-04-21	&	06:55:55	&	1.465	&	0.8	&	120	&	90	&	4x30	\\
GJ1186	&	14:53:37.31	&	+11:34:02.2	&	M4.5	&	15.29	&	18.55	&	2015-04-21	&	07:42:35	&	1.463	&	0.9	&	120	&	90	&	4x30	\\
NN3900	&	15:19:11.00	&	-12:45:08.2	&	M4	&	12.58	&	13.33	&	2015-06-12	&	03:50:18	&	1.065	&	0.9	&	120	&	90	&	4x30	\\
NN3937	&	16:04:20.00	&	-06:16:57.8	&	M4.5	&	15.51	&	16.56	&	2015-04-21	&	08:12:07	&	1.120	&	0.7	&	120	&	90	&	4x30	\\
GJ1204	&	16:36:05.18	&	+08:48:47.7	&	M4	&	13.80	&	15.34	&	2015-06-12	&	04:52:04	&	1.234	&	0.8	&	120	&	90	&	4x30	\\
NN3967	&	16:40:06.23	&	+00:42:16.9	&	M5	&	13.69	&	11.20	&	2015-06-12	&	04:09:01	&	1.106	&	0.8	&	120	&	90	&	4x30	\\
NN3981	&	16:58:24.94	&	+13:58:11.5	&	M4	&	13.13	&	12.99	&	2015-04-02	&	09:25:52	&	1.288	&	0.8	&	120	&	90	&	4x30	\\
GJ1210	&	17:07:40.42	&	+07:22:01.7	&	M5	&	14.01	&	12.82	&	2015-06-12	&	05:02:54	&	1.192	&	0.7	&	120	&	90	&	4x30	\\
NN4032	&	17:53:00.42	&	+16:54:59.3	&	M3.5	&	12.69	&	17.54	&	2015-04-21	&	07:20:12	&	1.432	&	0.9	&	120	&	90	&	4x30	\\
NN4071	&	18:42:45.07	&	+13:54:22.0	&	M5	&	12.81	&	10.42	&	2015-04-21	&	07:32:02	&	1.491	&	0.8	&	120	&	90	&	4x30	\\
Gl729	&	18:49:50.13	&	-23:50:14.4	&	M3.5	&	10.46	&	2.93	&	2015-06-12	&	04:42:07	&	1.067	&	0.8	&	120	&	30	&	4x10	\\
Gl781.1	&	20:07:45.27	&	-31:45:24.9	&	M4	&	12.50	&	19.72	&	2015-06-17	&	10:09:34	&	1.307	&	0.8	&	120	&	90	&	4x30	\\
Gl791.2	&	20:29:49.07	&	+09:41:23.1	&	M4.5	&	13.05	&	8.76	&	2015-06-17	&	09:44:17	&	1.473	&	0.6	&	120	&	90	&	4x30	\\
Gl828	&	21:26:53.22	&	-44:48:44.6	&	M3.5	&	14.10	&	14.93	&	2015-07-13	&	04:56:12	&	1.155	&	1.1	&	360	&	300	&	4x100	\\
NN4201	&	21:32:22.36	&	+24:33:42.0	&	M4	&	12.66	&	12.35	&	2015-05-28	&	09:50:41	&	1.530	&	0.3	&	120	&	90	&	4x30	\\
Gl836	&	21:39:02.08	&	-24:09:40.8	&	M4	&	13.43	&	13.95	&	2015-07-13	&	05:11:28	&	1.097	&	0.8	&	360	&	300	&	4x60	\\
NN4215	&	21:44:08.31	&	+17:04:38.2	&	M4.5	&	14.81	&	17.54	&	2015-06-15	&	09:49:04	&	1.387	&	0.9	&	120	&	90	&	4x30	\\
NN4231	&	21:52:10.59	&	+05:37:33.7	&	M2.4	&	12.11	&	15.63	&	2015-05-22	&	09:38:37	&	1.197	&	0.8	&	120	&	90	&	4x30	\\
NN4239	&	21:56:56.63	&	-01:54:00.5	&	M5	&	14.64	&	13.33	&	2015-07-13	&	04:38:48	&	1.397	&	1.1	&	360	&	300	&	4x100	\\
NN4274	&	22:23:07.34	&	-17:36:36.2	&	M4.5	&	13.25	&	7.46	&	2015-09-09	&	04:51:51	&	1.036	&	0.9	&	120	&	90	&	4x30	\\
NN4279	&	22:27:03.07	&	+06:49:33.4	&	M3.5	&	13.22	&	13.89	&	2015-09-09	&	05:01:00	&	1.222	&	1.0	&	120	&	90	&	4x30	\\
NN4282	&	22:33:22.92	&	-09:36:53.0	&	M3	&	12.41	&	16.95	&	2015-09-09	&	05:09:45	&	1.076	&	0.8	&	120	&	90	&	4x30	\\
NN4292	&	22:43:23.71	&	+22:08:17.8	&	M5	&	15.00	&	15.87	&	2015-07-25	&	07:26:48	&	1.463	&	1.3	&	120	&	90	&	4x30	\\
NN4302	&	22:54:47.15	&	-05:28:19.8	&	M4	&	13.90	&	20.00	&	2015-08-21	&	08:55:18	&	1.574	&	0.8	&	120	&	90	&	4x30	\\
NN4326	&	23:17:28.57	&	+19:36:46.2	&	M2	&	12.10	&	12.82	&	2015-07-25	&	08:39:29	&	1.438	&	1.1	&	120	&	90	&	4x30	\\
GJ1284	&	23:30:13.82	&	-20:23:29.3	&	M2	&	11.16	&	10.87	&	2015-09-09	&	05:36:07	&	1.013	&	0.8	&	100	&	75	&	4x15	\\
GJ1286	&	23:35:11.31	&	-02:23:33.4	&	M5	&	14.69	&	7.22	&	2015-09-09	&	05:47:55	&	1.099	&	0.7	&	120	&	90	&	4x30	\\
NN4360	&	23:45:30.87	&	-16:10:27.5	&	M5	&	14.50	&	9.01	&	2015-09-09	&	05:56:39	&	1.025	&	0.6	&	120	&	90	&	4x30	\\
NN4362	&	23:48:35.42	&	-27:39:44.4	&	M2.5	&	12.40	&	18.87	&	2015-08-13	&	05:54:23	&	1.042	&	1.3	&	120	&	90	&	4x30	\\
NN4378	&	23:57:20.84	&	-12:58:47.4	&	M4	&	12.93	&	17.86	&	2015-09-09	&	06:07:33	&	1.035	&	0.6	&	120	&	90	&	4x30	\\		
\enddata
\tablecomments{Observed M dwarfs, ordered by RA. Spectral types are from SIMBAD \citep{simbad}, distances are from the Gliese catalog}
\label{tab:observations}
\end{deluxetable}

\clearpage
\end{landscape}

\input{limit_file_list}

\clearpage

\input{nonlimit_file_list}

\end{document}

%% file: limit_file_list.tex
\begin{figure}[h]
\includegraphics[width=\textwidth]{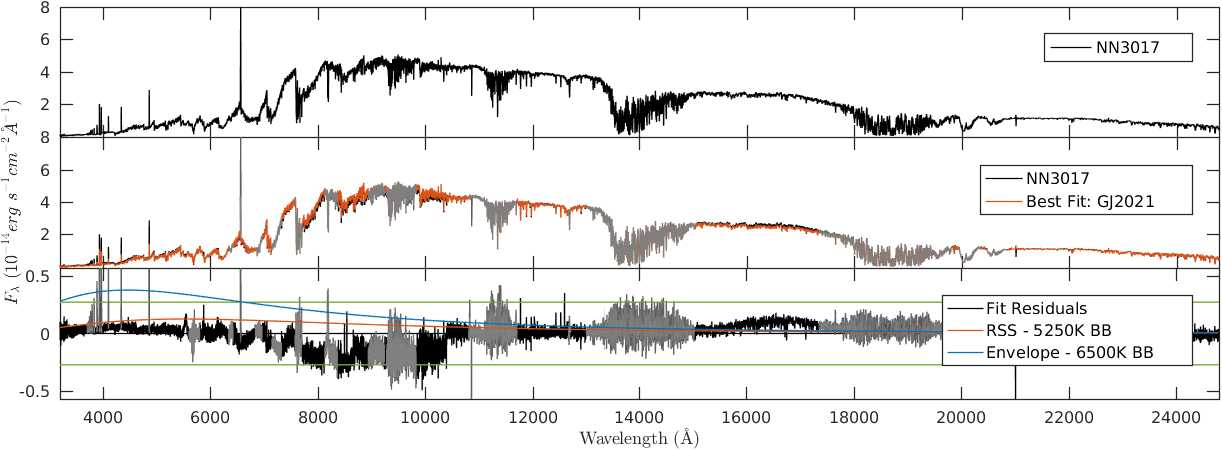}
\centering
\caption{NN3017\label{fig:Results_NN3017}}
\end{figure}

\begin{figure}[h]
\includegraphics[width=\textwidth]{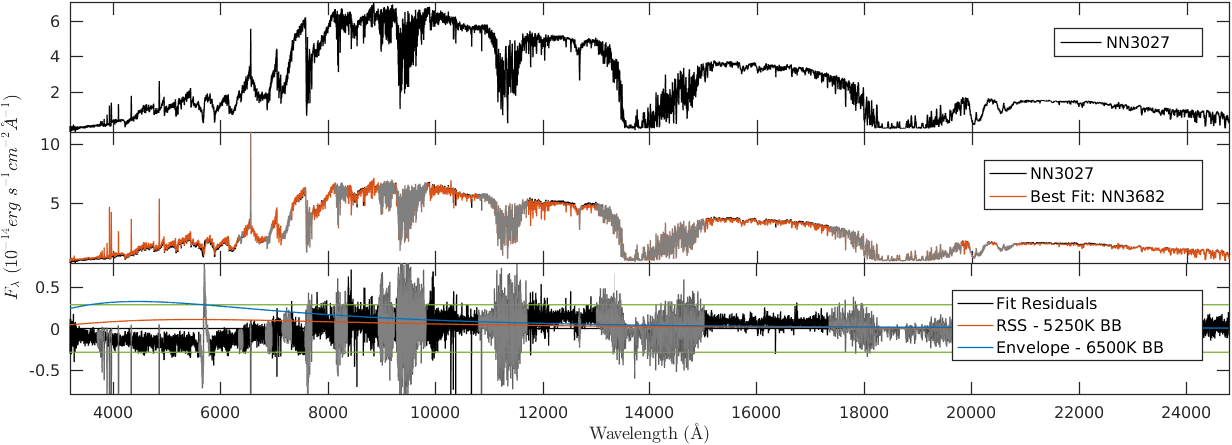}
\centering
\caption{NN3027\label{fig:Results_NN3027}}
\end{figure}

\begin{figure}[h]
\includegraphics[width=\textwidth]{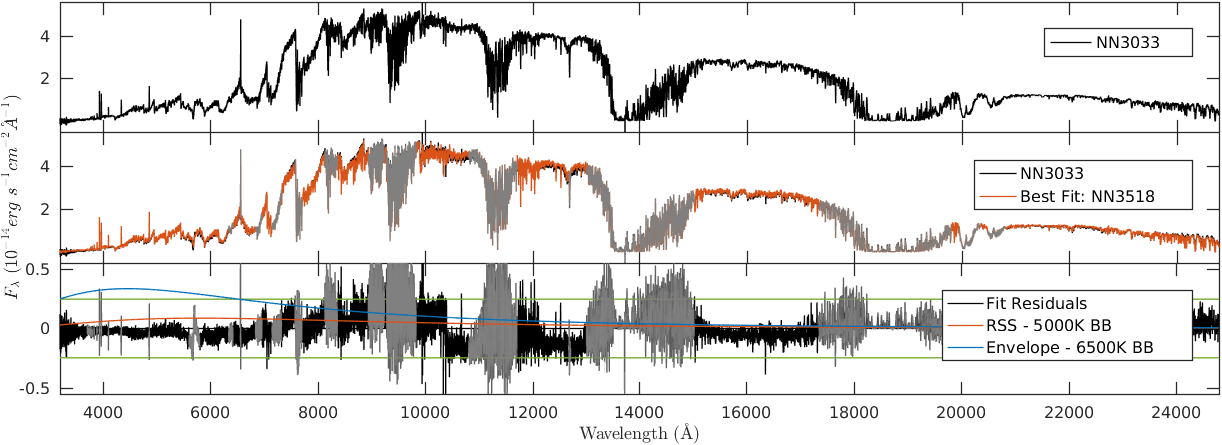}
\centering
\caption{NN3033\label{fig:Results_NN3033}}
\end{figure}

\clearpage

\begin{figure}[h]
\includegraphics[width=\textwidth]{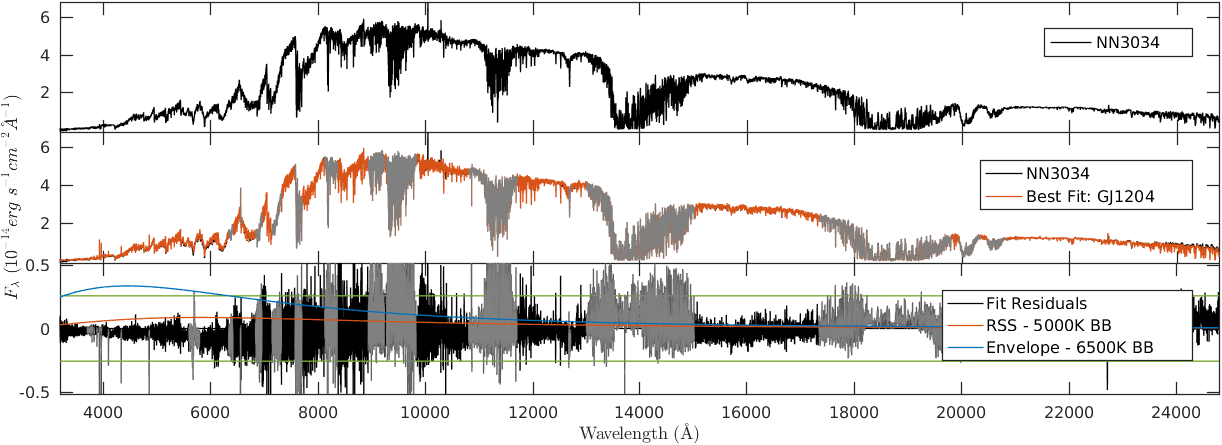}
\centering
\caption{NN3034\label{fig:Results_NN3034}}
\end{figure}

\begin{figure}[h]
\includegraphics[width=\textwidth]{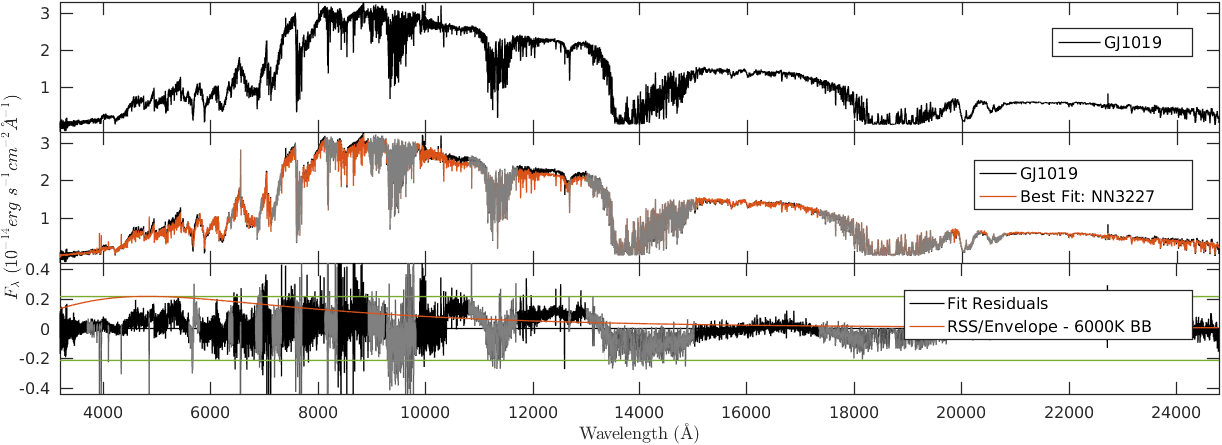}
\centering
\caption{GJ1019\label{fig:Results_GJ1019}}
\end{figure}

\begin{figure}[h]
\includegraphics[width=\textwidth]{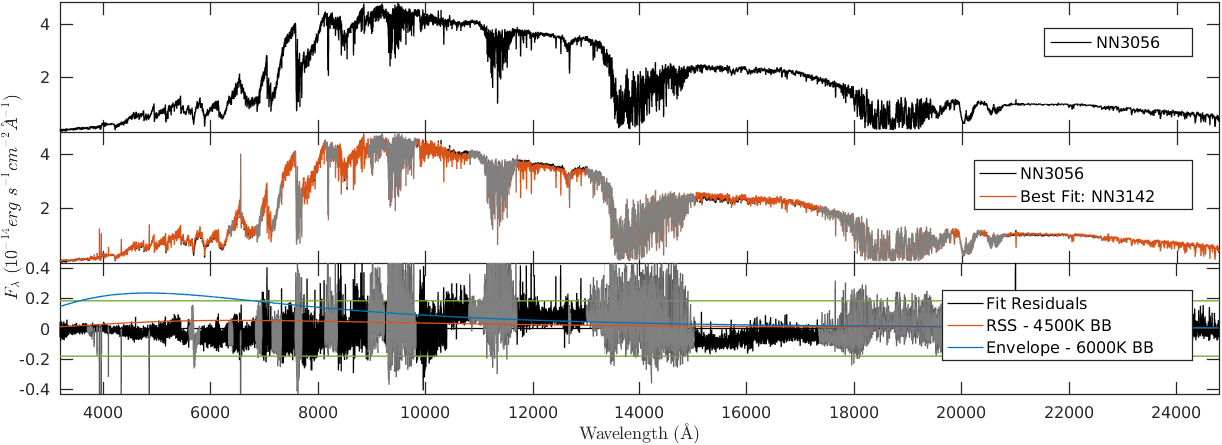}
\centering
\caption{NN3056\label{fig:Results_NN3056}}
\end{figure}

\clearpage

\begin{figure}[h]
\includegraphics[width=\textwidth]{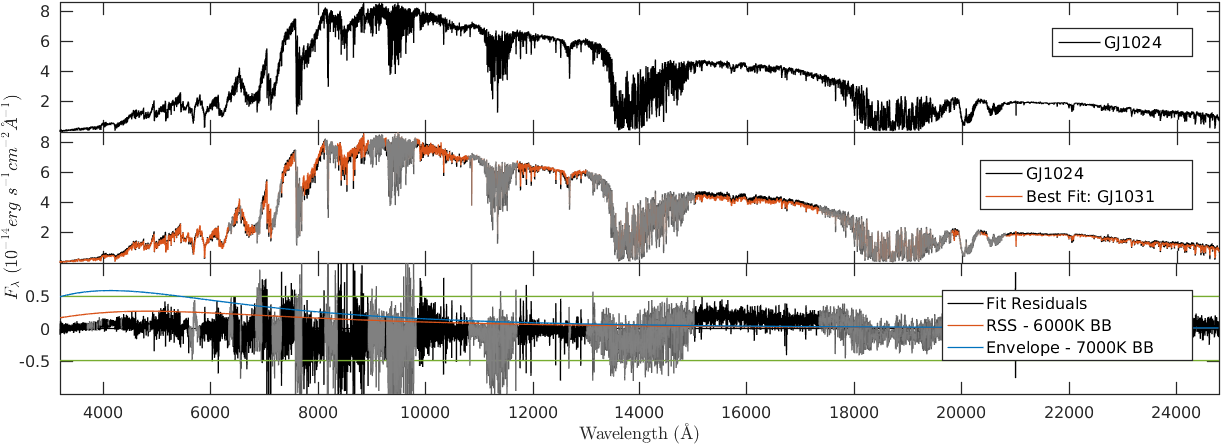}
\centering
\caption{GJ1024\label{fig:Results_GJ1024}}
\end{figure}

\begin{figure}[h]
\includegraphics[width=\textwidth]{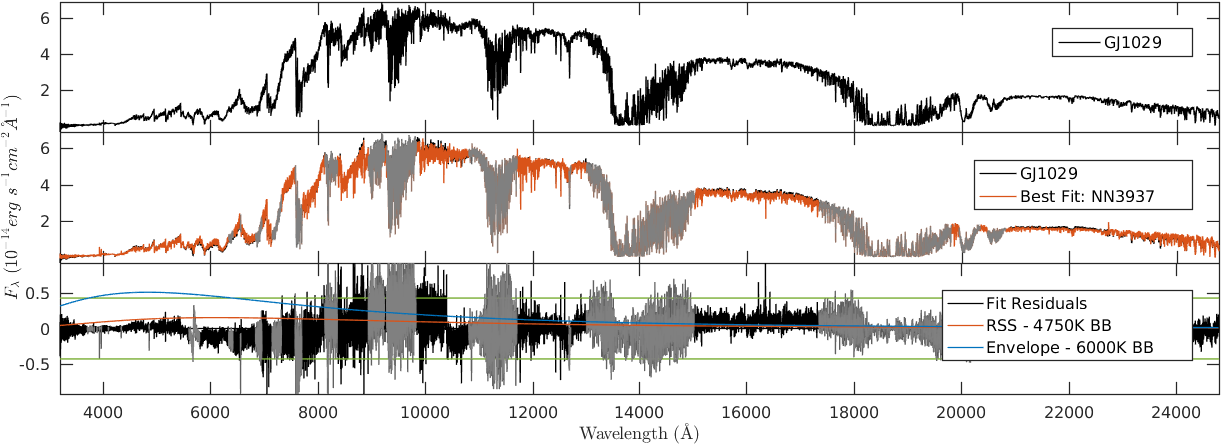}
\centering
\caption{GJ1029\label{fig:Results_GJ1029}}
\end{figure}

\begin{figure}[h]
\includegraphics[width=\textwidth]{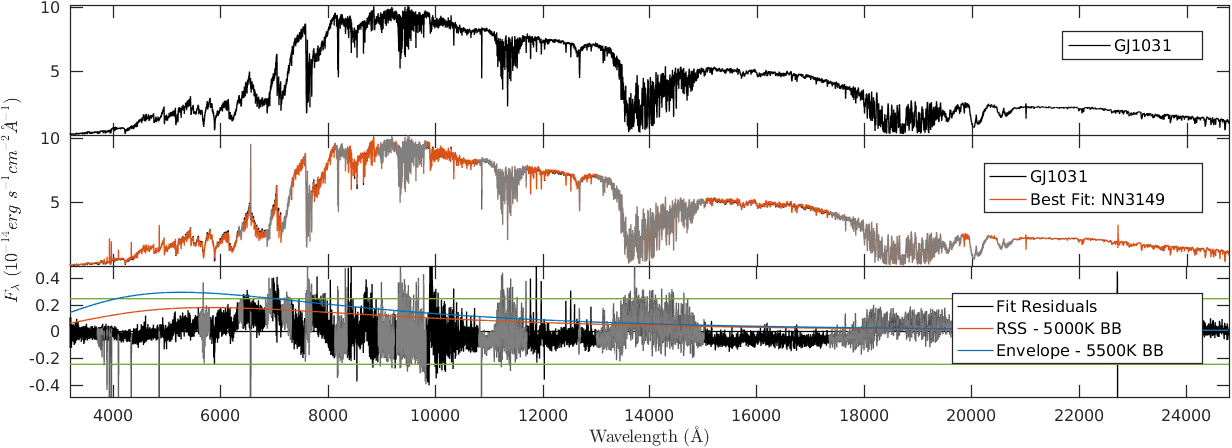}
\centering
\caption{GJ1031\label{fig:Results_GJ1031}}
\end{figure}

\clearpage

\begin{figure}[h]
\includegraphics[width=\textwidth]{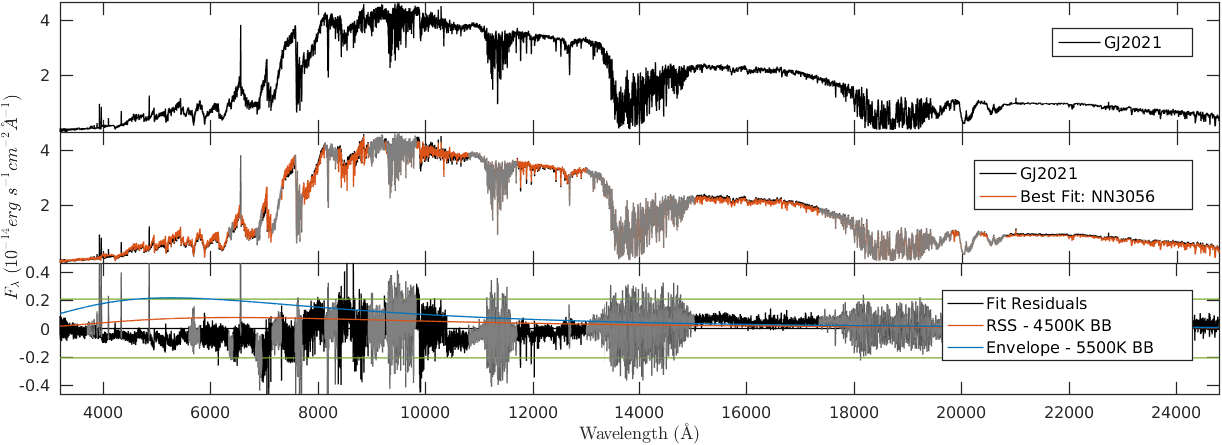}
\centering
\caption{GJ2021\label{fig:Results_GJ2021}}
\end{figure}

\begin{figure}[h]
\includegraphics[width=\textwidth]{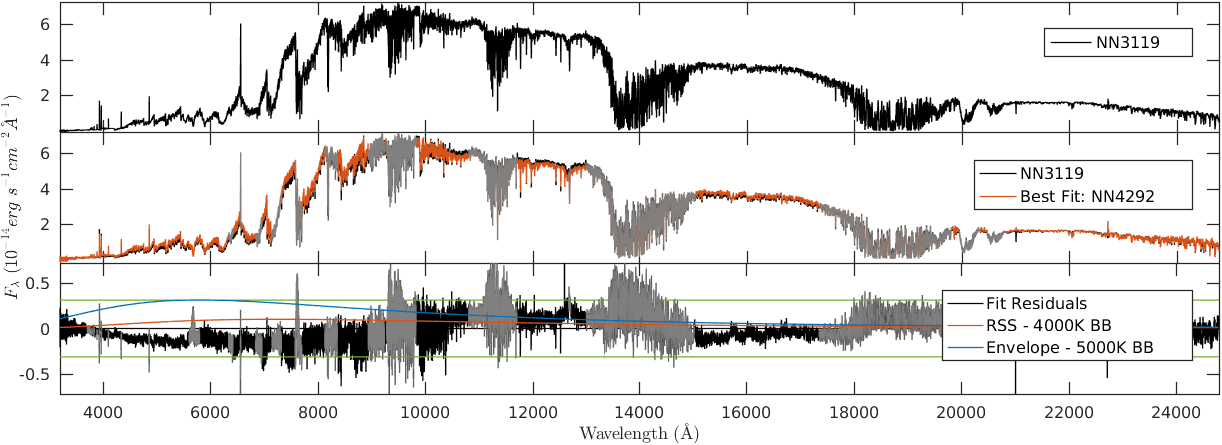}
\centering
\caption{NN3119\label{fig:Results_NN3119}}
\end{figure}

\begin{figure}[h]
\includegraphics[width=\textwidth]{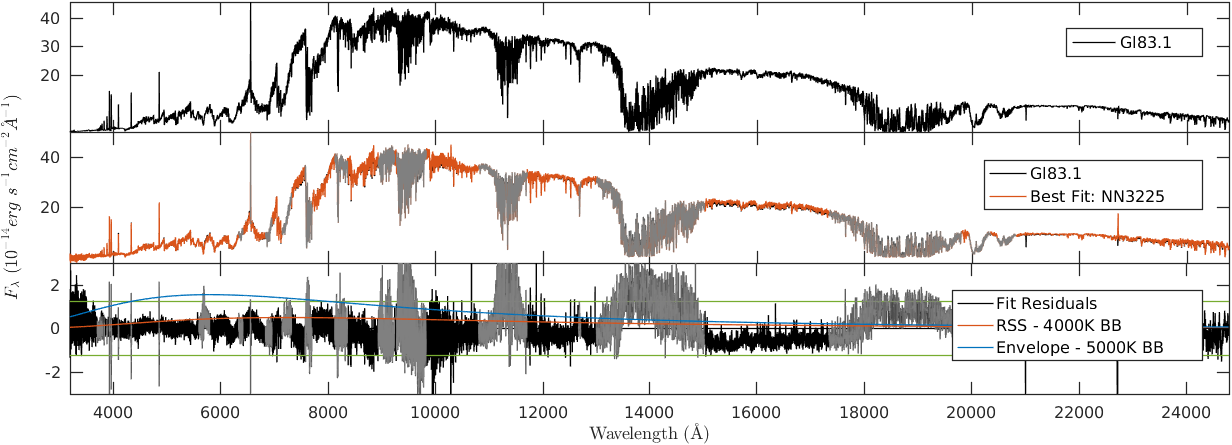}
\centering
\caption{Gl83.1\label{fig:Results_Gl83.1}}
\end{figure}

\clearpage

\begin{figure}[h]
\includegraphics[width=\textwidth]{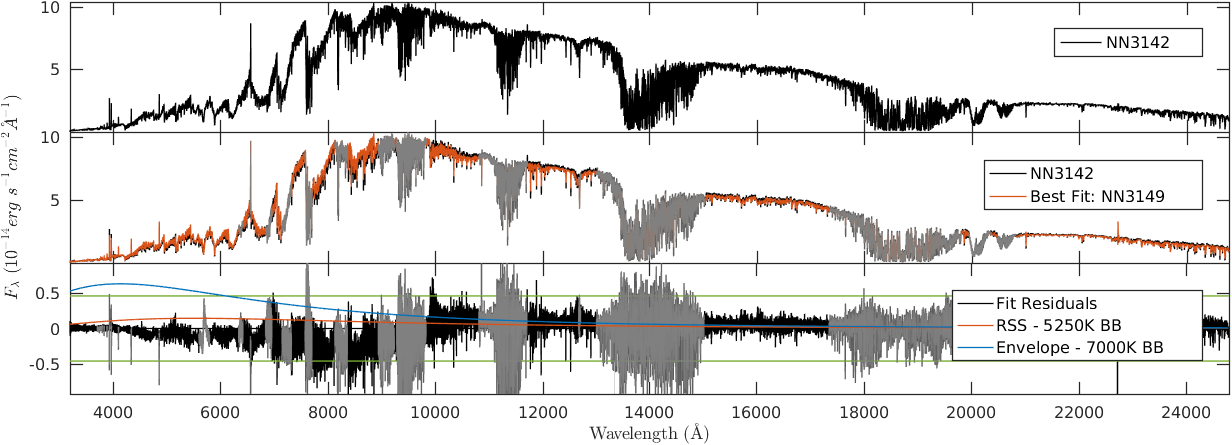}
\centering
\caption{NN3142\label{fig:Results_NN3142}}
\end{figure}

\begin{figure}[h]
\includegraphics[width=\textwidth]{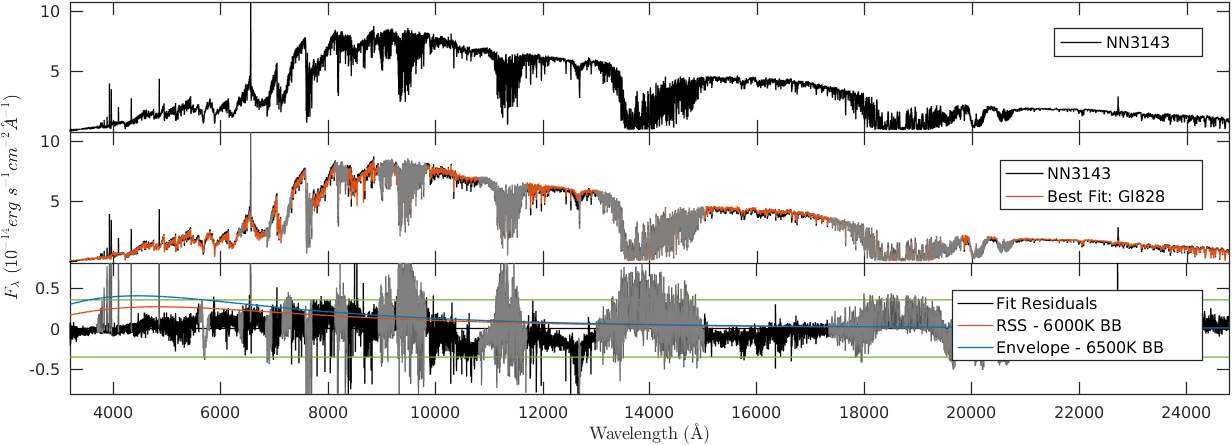}
\centering
\caption{NN3143\label{fig:Results_NN3143}}
\end{figure}

\begin{figure}[h]
\includegraphics[width=\textwidth]{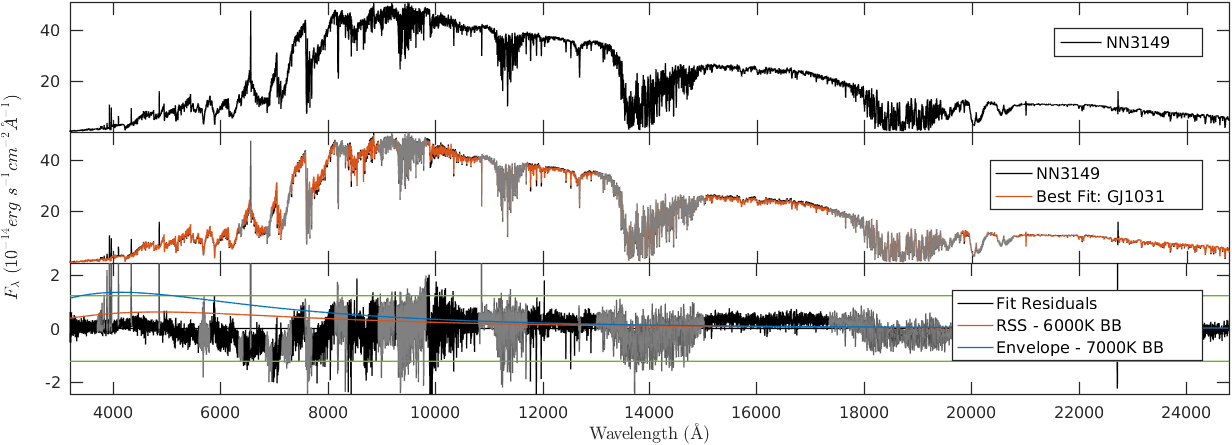}
\centering
\caption{NN3149\label{fig:Results_NN3149}}
\end{figure}

\clearpage

\begin{figure}[h]
\includegraphics[width=\textwidth]{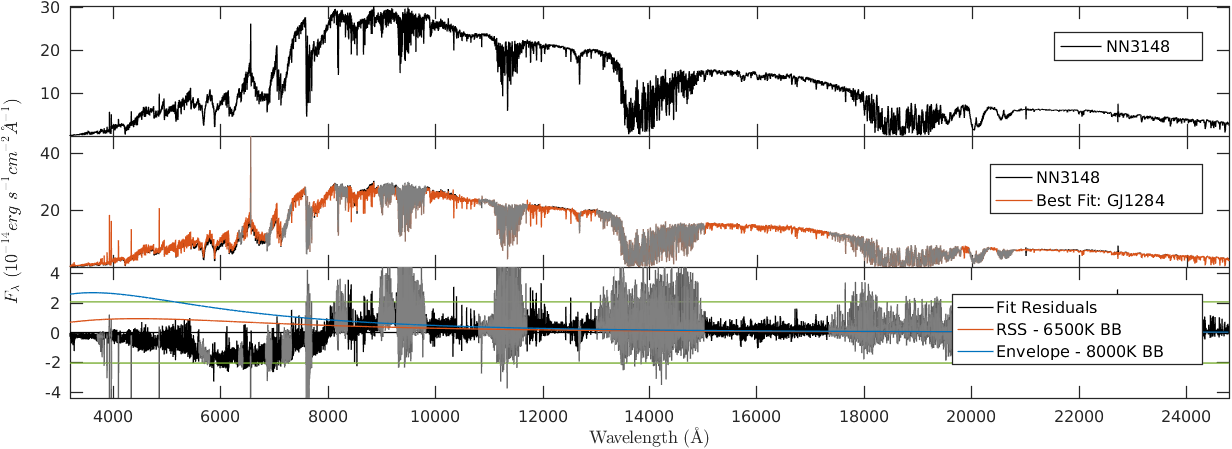}
\centering
\caption{NN3148\label{fig:Results_NN3148}}
\end{figure}

\begin{figure}[h]
\includegraphics[width=\textwidth]{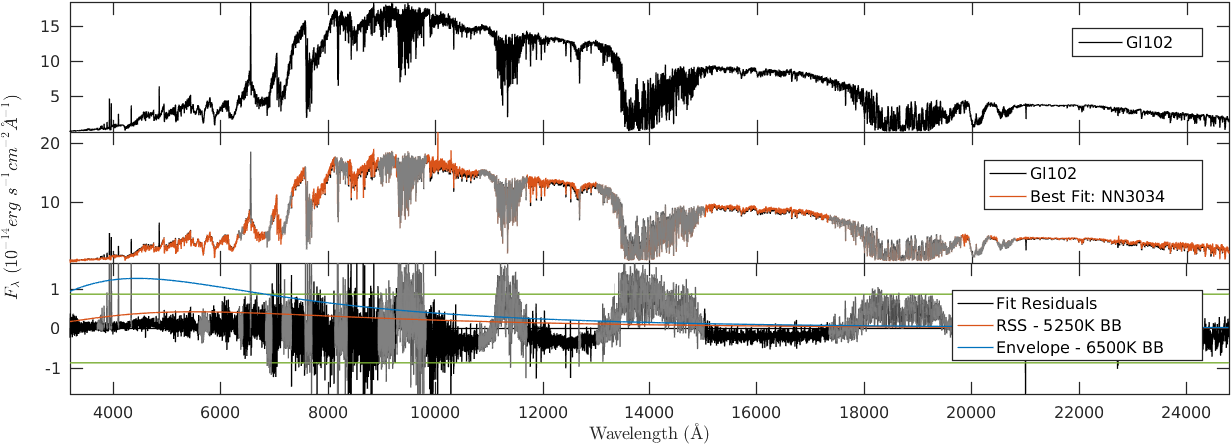}
\centering
\caption{Gl102\label{fig:Results_Gl102}}
\end{figure}

\begin{figure}[h]
\includegraphics[width=\textwidth]{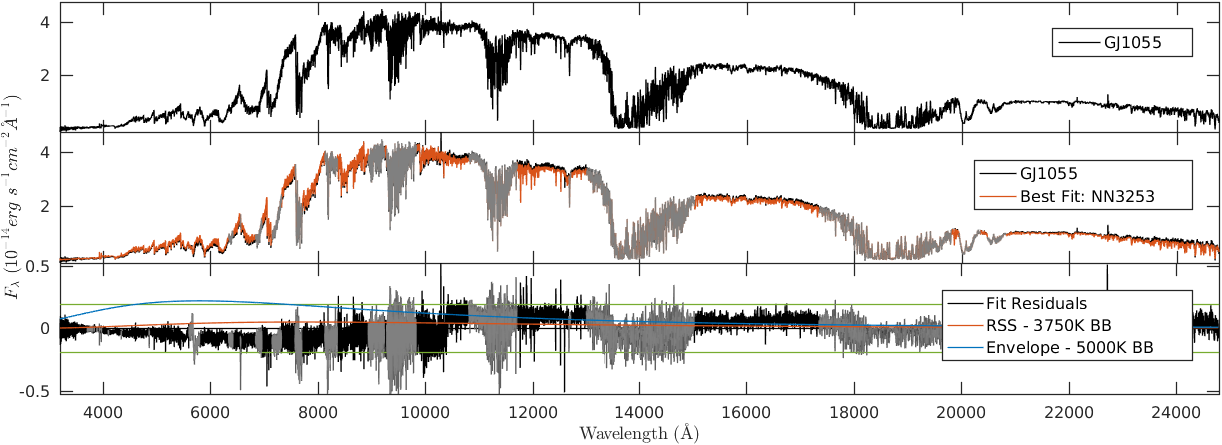}
\centering
\caption{GJ1055\label{fig:Results_GJ1055}}
\end{figure}

\clearpage

\begin{figure}[h]
\includegraphics[width=\textwidth]{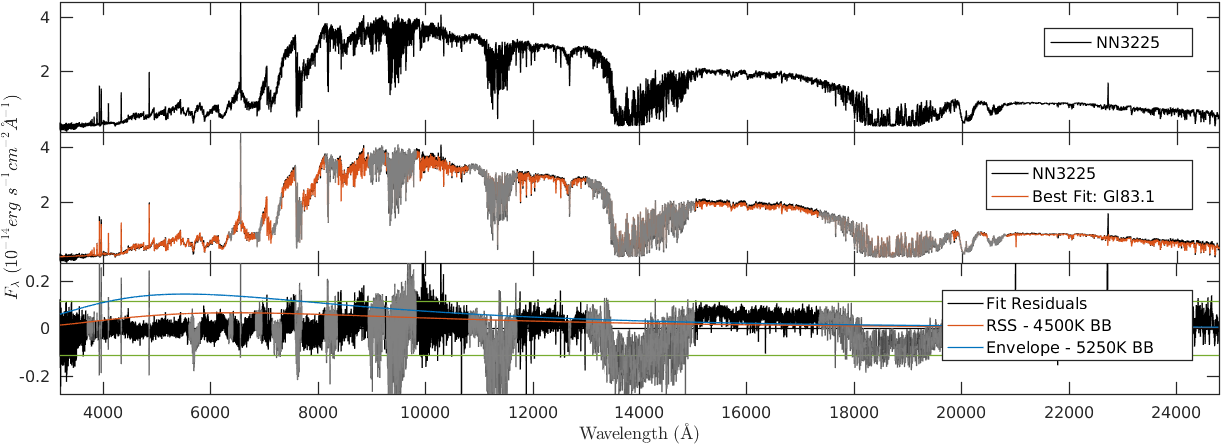}
\centering
\caption{NN3225\label{fig:Results_NN3225}}
\end{figure}

\begin{figure}[h]
\includegraphics[width=\textwidth]{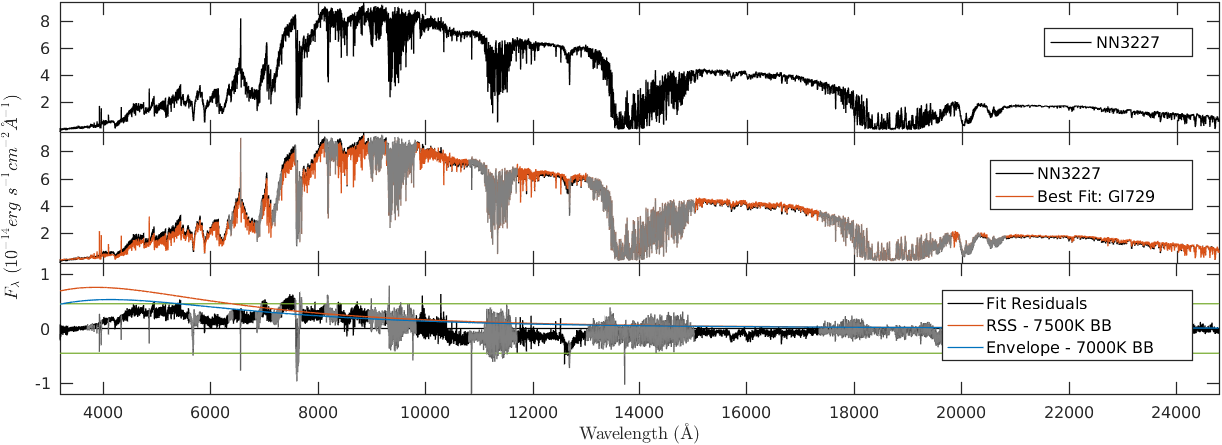}
\centering
\caption{NN3227\label{fig:Results_NN3227}}
\end{figure}

\begin{figure}[h]
\includegraphics[width=\textwidth]{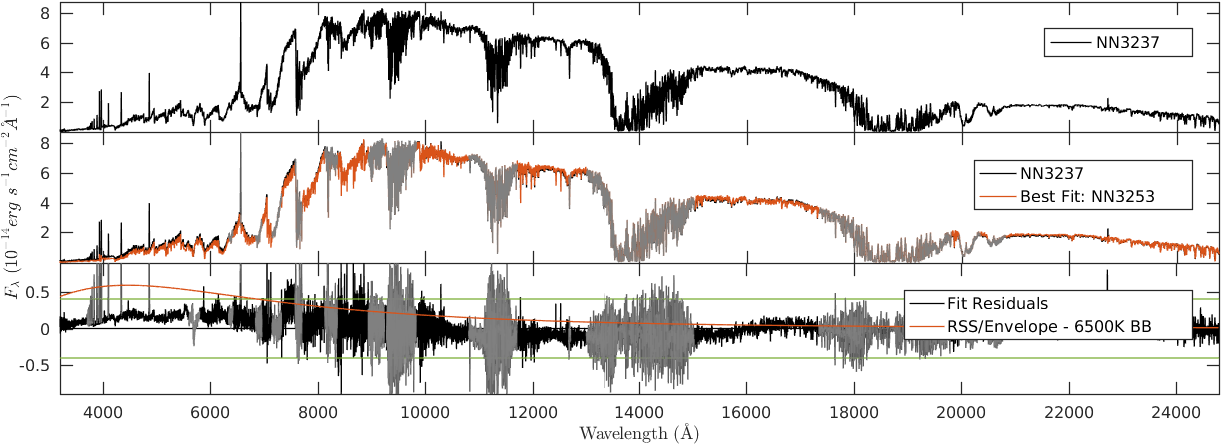}
\centering
\caption{NN3237\label{fig:Results_NN3237}}
\end{figure}

\clearpage

\begin{figure}[h]
\includegraphics[width=\textwidth]{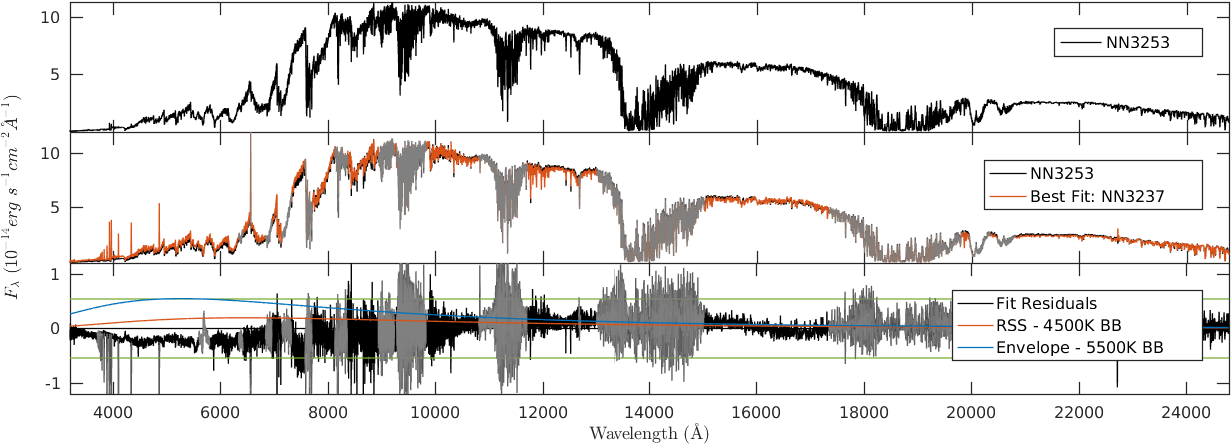}
\centering
\caption{NN3253\label{fig:Results_NN3253}}
\end{figure}

\begin{figure}[h]
\includegraphics[width=\textwidth]{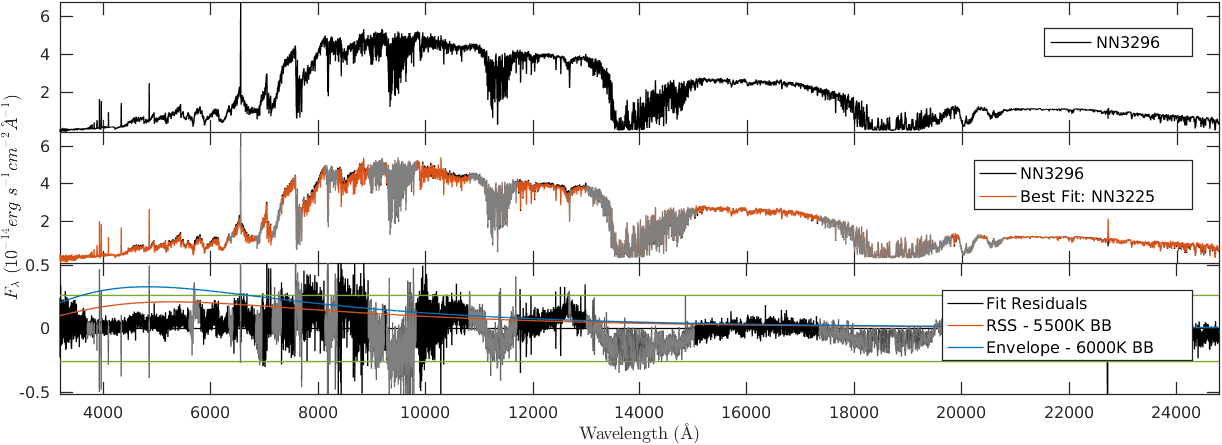}
\centering
\caption{NN3296\label{fig:Results_NN3296}}
\end{figure}

\begin{figure}[h]
\includegraphics[width=\textwidth]{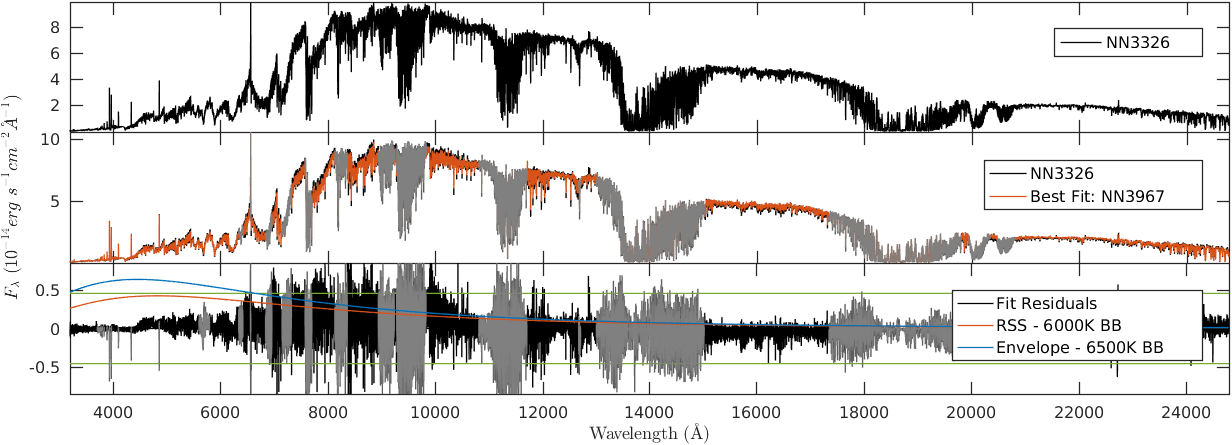}
\centering
\caption{NN3326\label{fig:Results_NN3326}}
\end{figure}

\clearpage

\begin{figure}[h]
\includegraphics[width=\textwidth]{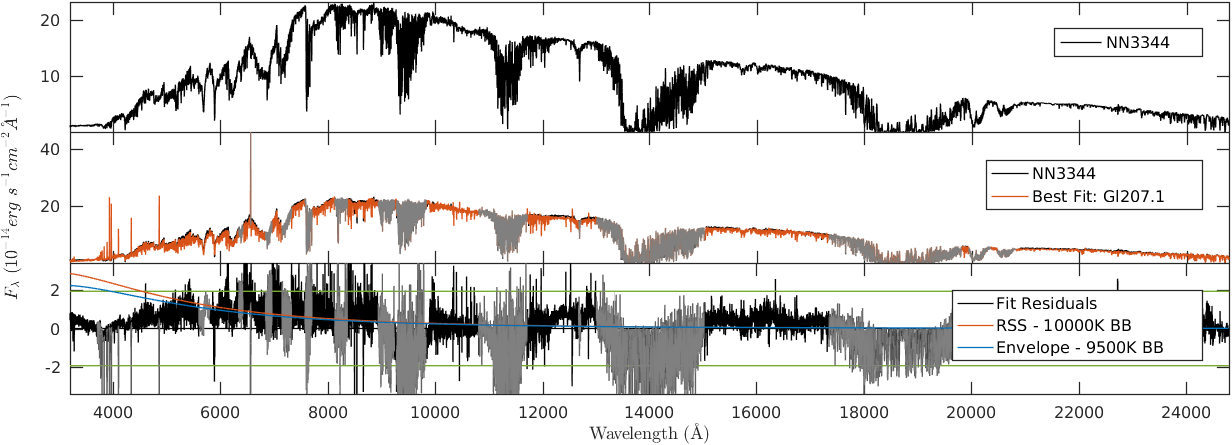}
\centering
\caption{NN3344\label{fig:Results_NN3344}}
\end{figure}

\begin{figure}[h]
\includegraphics[width=\textwidth]{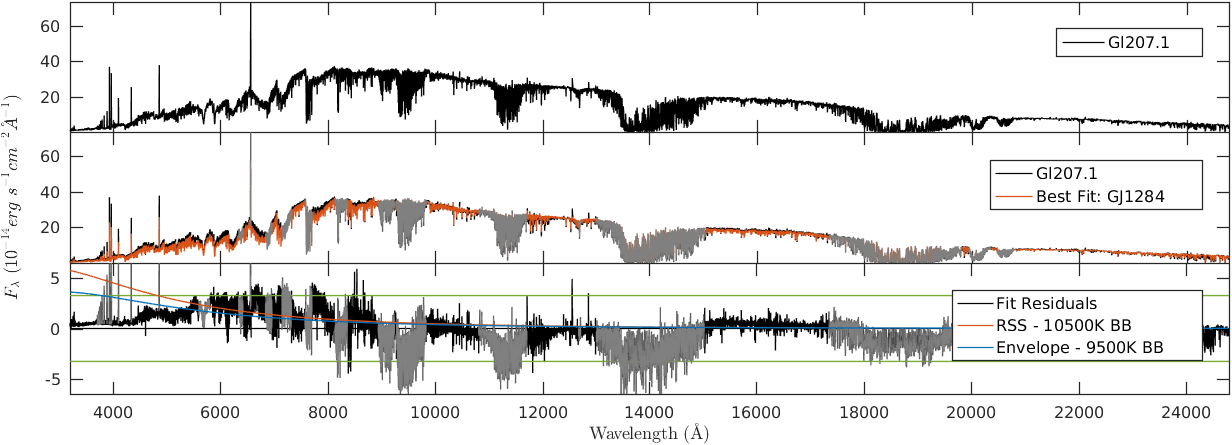}
\centering
\caption{Gl207.1\label{fig:Results_Gl207.1}}
\end{figure}

\begin{figure}[h]
\includegraphics[width=\textwidth]{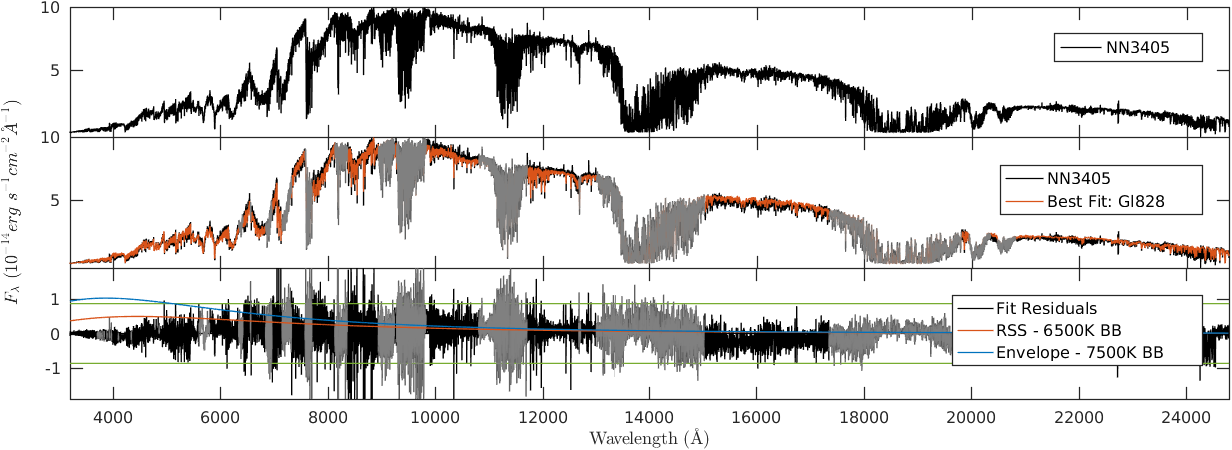}
\centering
\caption{NN3405\label{fig:Results_NN3405}}
\end{figure}

\clearpage

\begin{figure}[h]
\includegraphics[width=\textwidth]{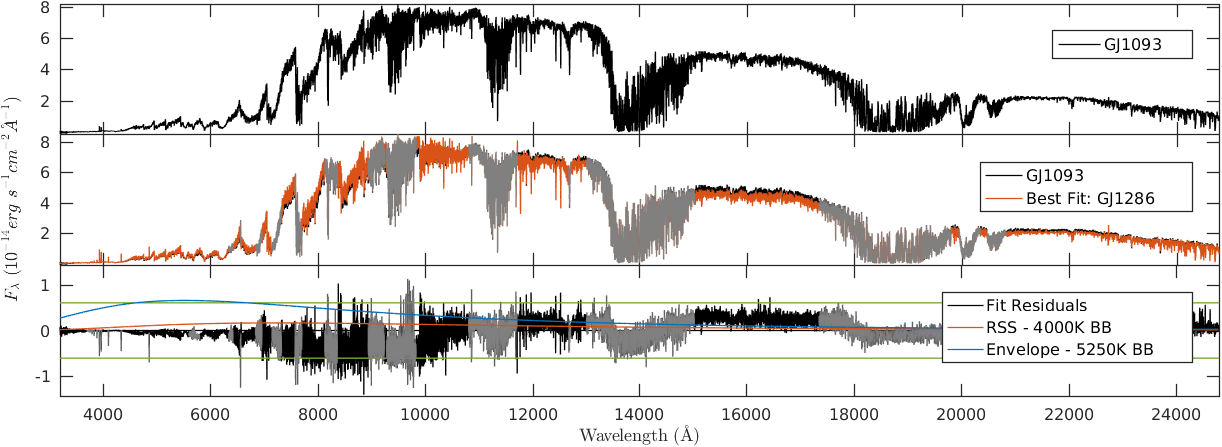}
\centering
\caption{GJ1093\label{fig:Results_GJ1093}}
\end{figure}

\begin{figure}[h]
\includegraphics[width=\textwidth]{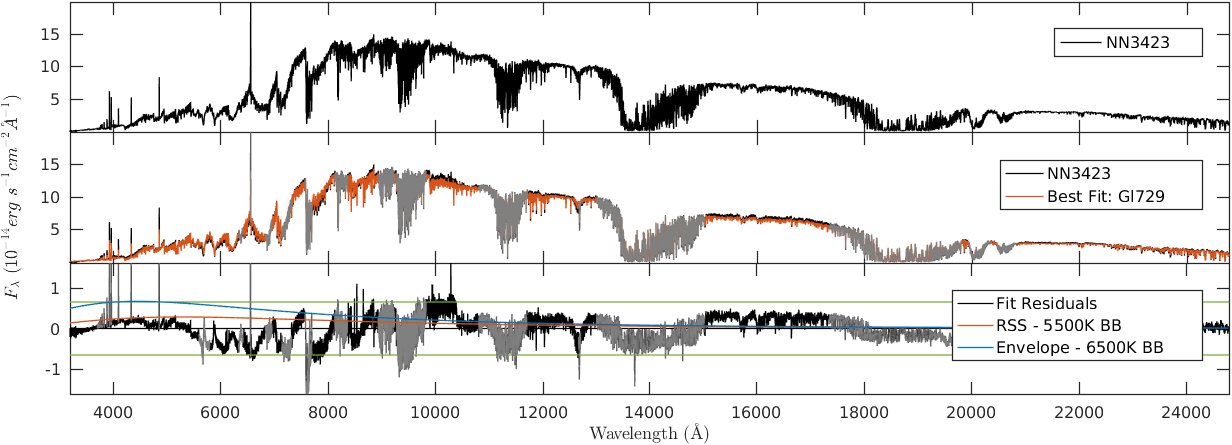}
\centering
\caption{NN3423\label{fig:Results_NN3423}}
\end{figure}

\begin{figure}[h]
\includegraphics[width=\textwidth]{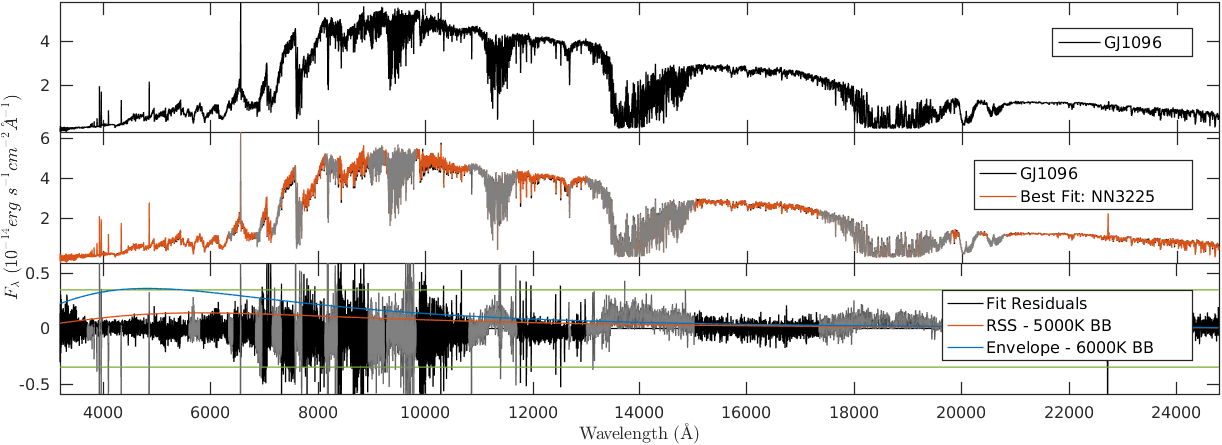}
\centering
\caption{GJ1096\label{fig:Results_GJ1096}}
\end{figure}

\clearpage

\begin{figure}[h]
\includegraphics[width=\textwidth]{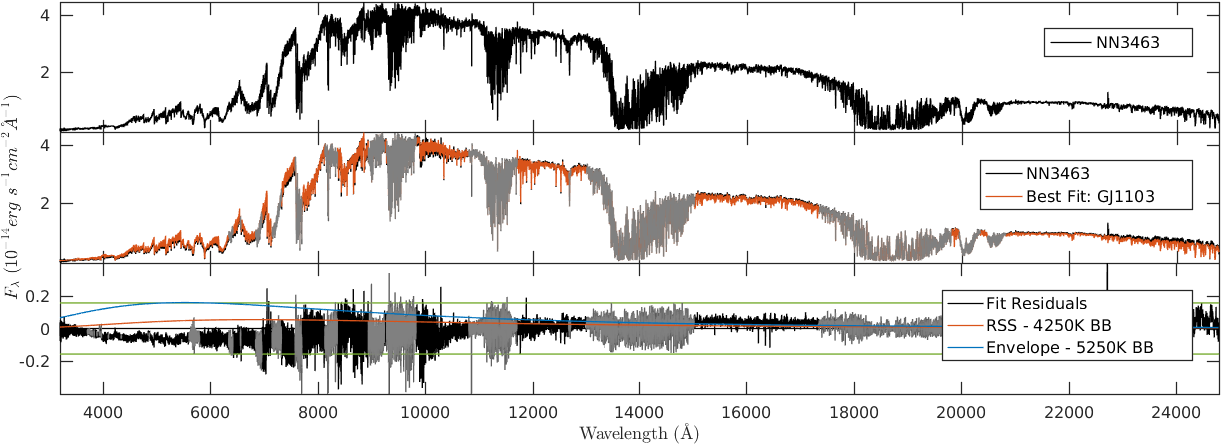}
\centering
\caption{NN3463\label{fig:Results_NN3463}}
\end{figure}

\begin{figure}[h]
\includegraphics[width=\textwidth]{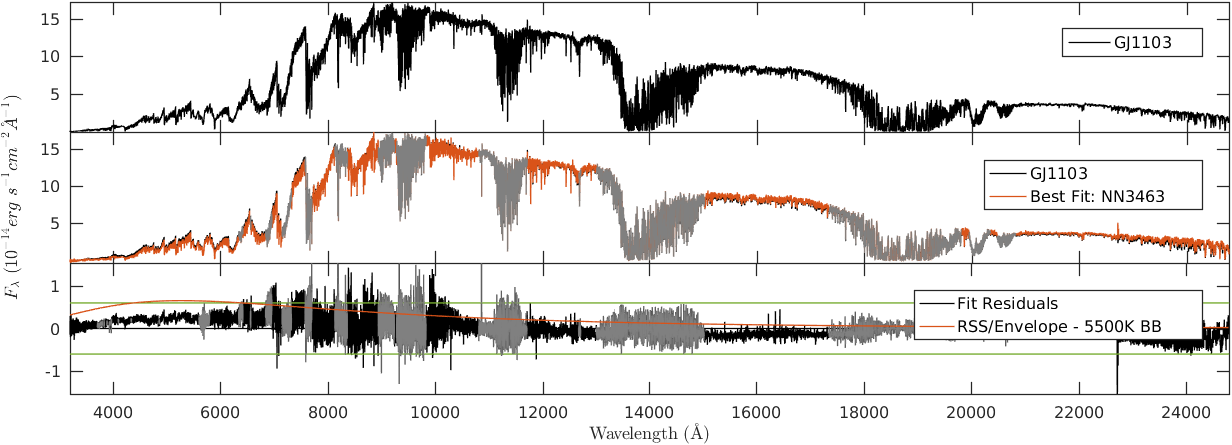}
\centering
\caption{GJ1103\label{fig:Results_GJ1103}}
\end{figure}

\begin{figure}[h]
\includegraphics[width=\textwidth]{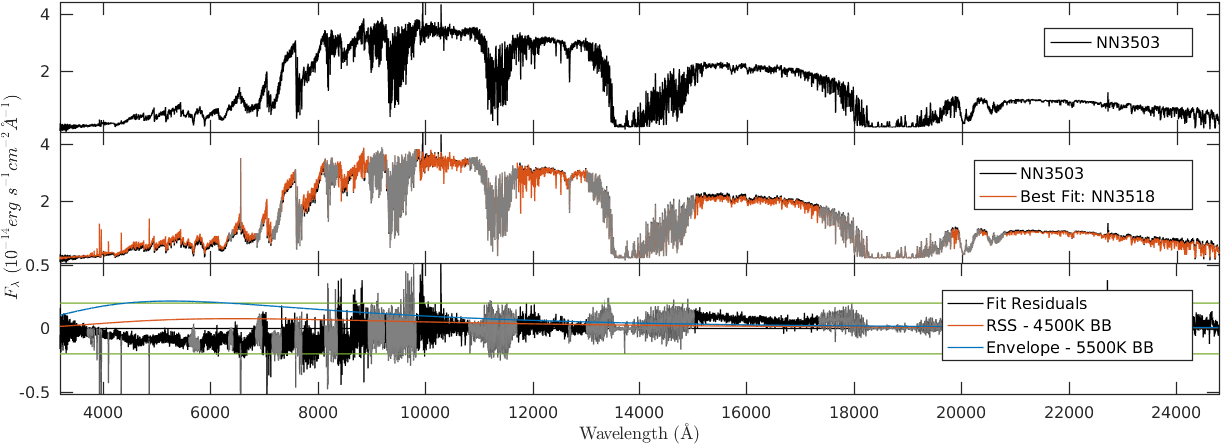}
\centering
\caption{NN3503\label{fig:Results_NN3503}}
\end{figure}

\clearpage

\begin{figure}[h]
\includegraphics[width=\textwidth]{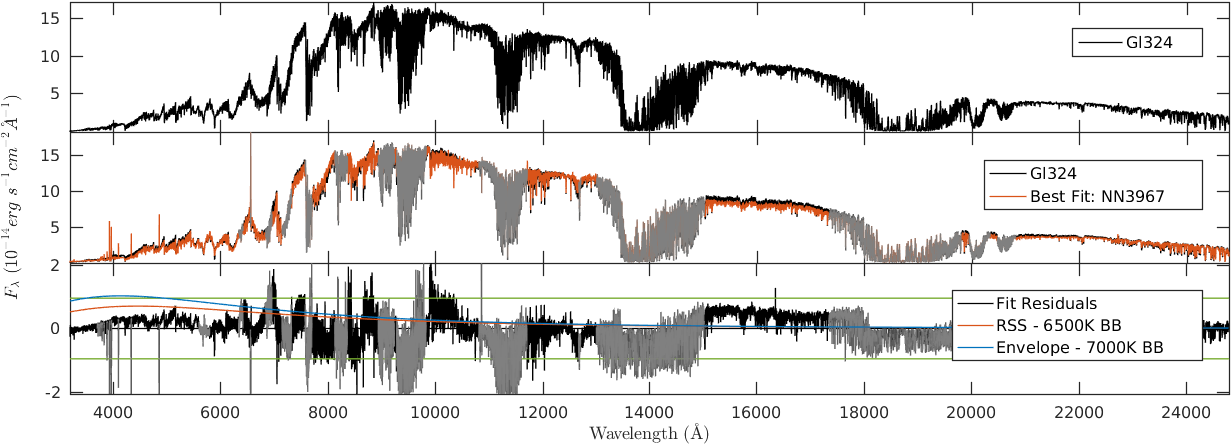}
\centering
\caption{Gl324\label{fig:Results_Gl324}}
\end{figure}

\begin{figure}[h]
\includegraphics[width=\textwidth]{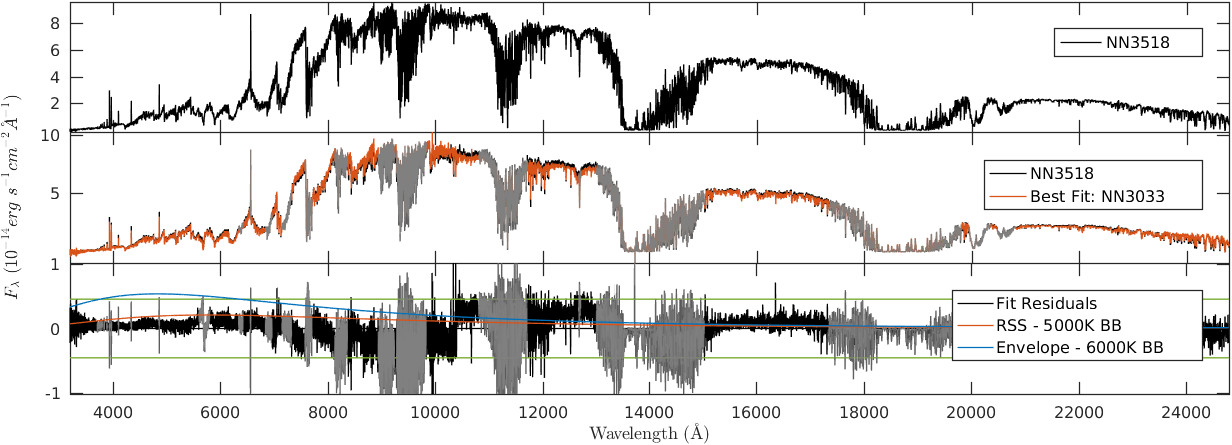}
\centering
\caption{NN3518\label{fig:Results_NN3518}}
\end{figure}

\begin{figure}[h]
\includegraphics[width=\textwidth]{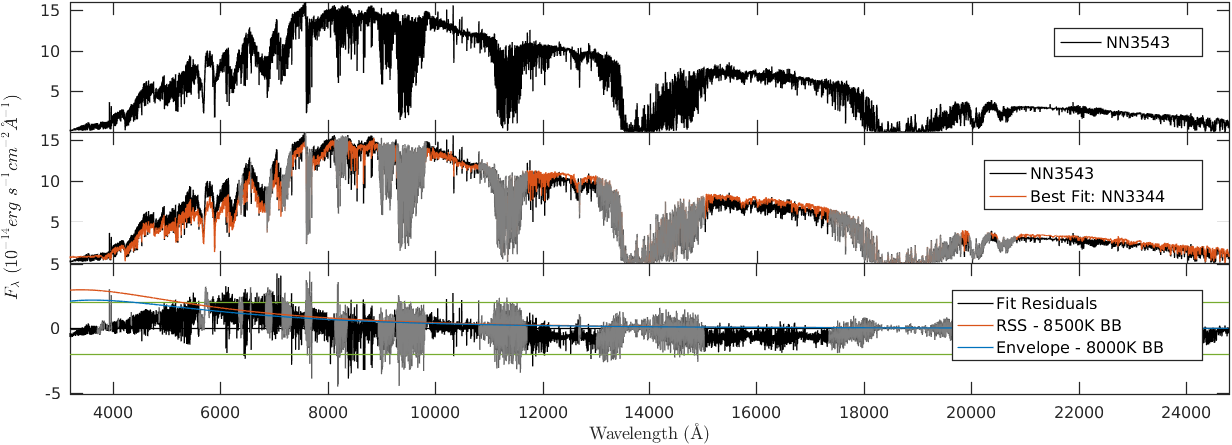}
\centering
\caption{NN3543\label{fig:Results_NN3543}}
\end{figure}

\clearpage

\begin{figure}[h]
\includegraphics[width=\textwidth]{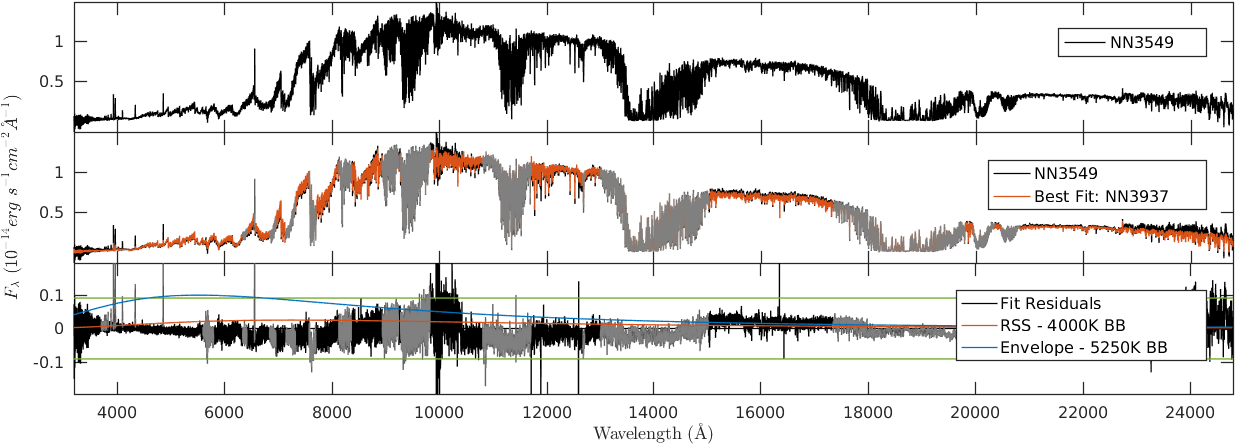}
\centering
\caption{NN3549\label{fig:Results_NN3549}}
\end{figure}

\begin{figure}[h]
\includegraphics[width=\textwidth]{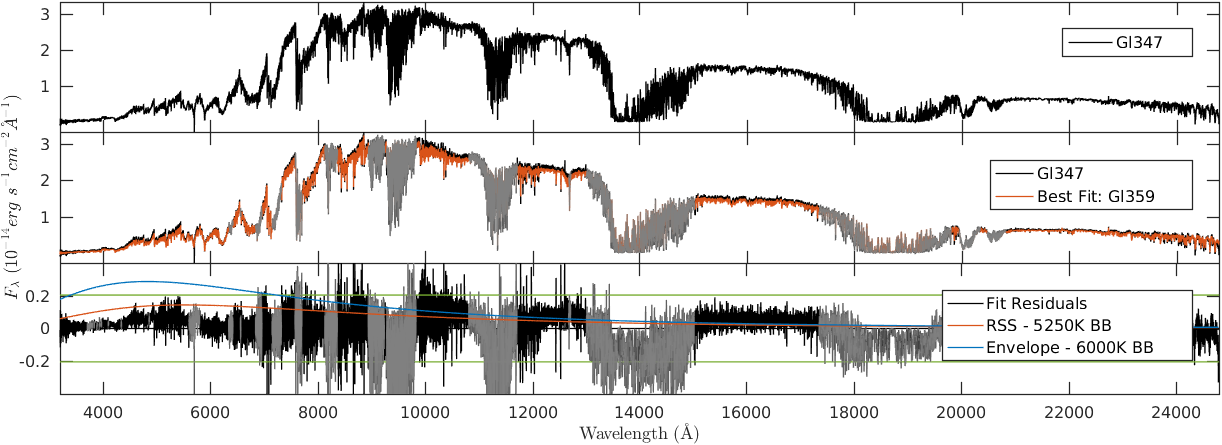}
\centering
\caption{Gl347\label{fig:Results_Gl347}}
\end{figure}

\begin{figure}[h]
\includegraphics[width=\textwidth]{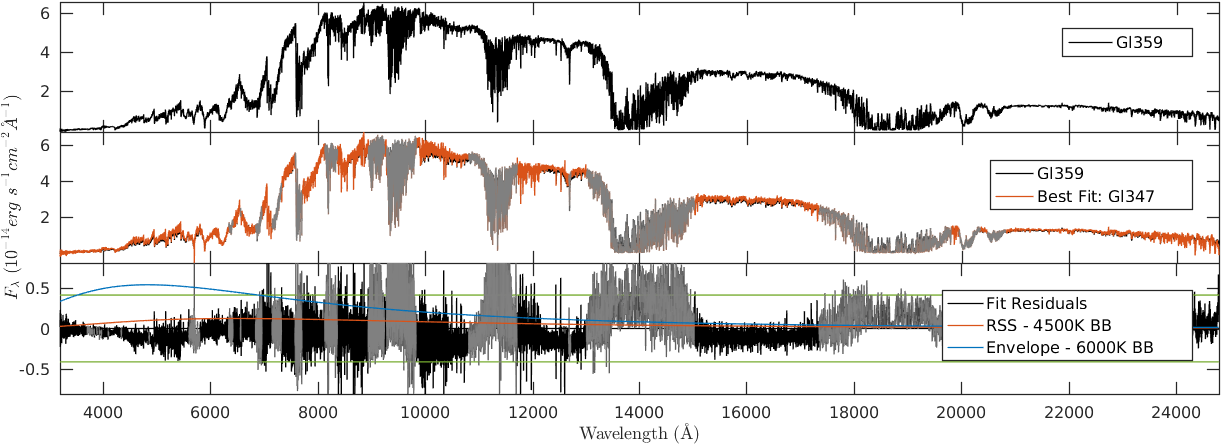}
\centering
\caption{Gl359\label{fig:Results_Gl359}}
\end{figure}

\clearpage

\begin{figure}[h]
\includegraphics[width=\textwidth]{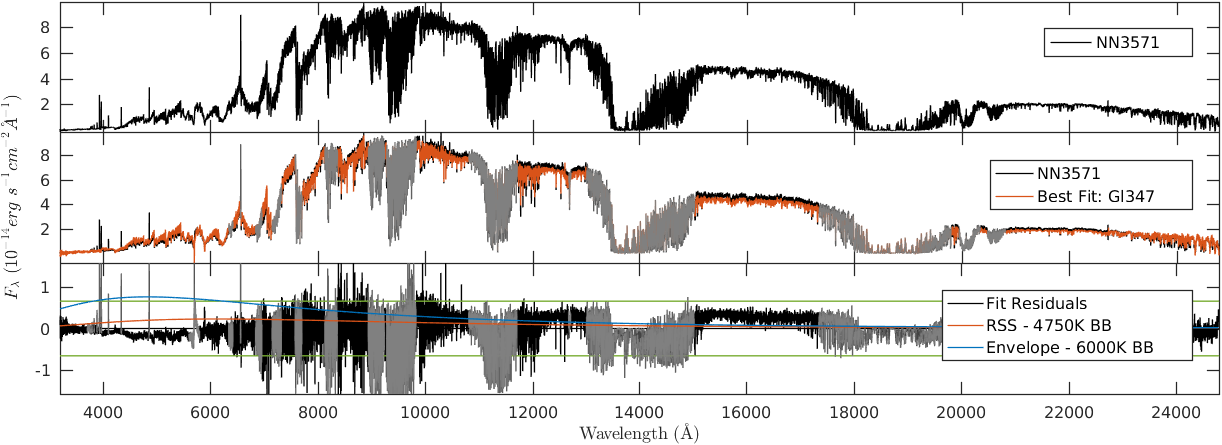}
\centering
\caption{NN3571\label{fig:Results_NN3571}}
\end{figure}

\begin{figure}[h]
\includegraphics[width=\textwidth]{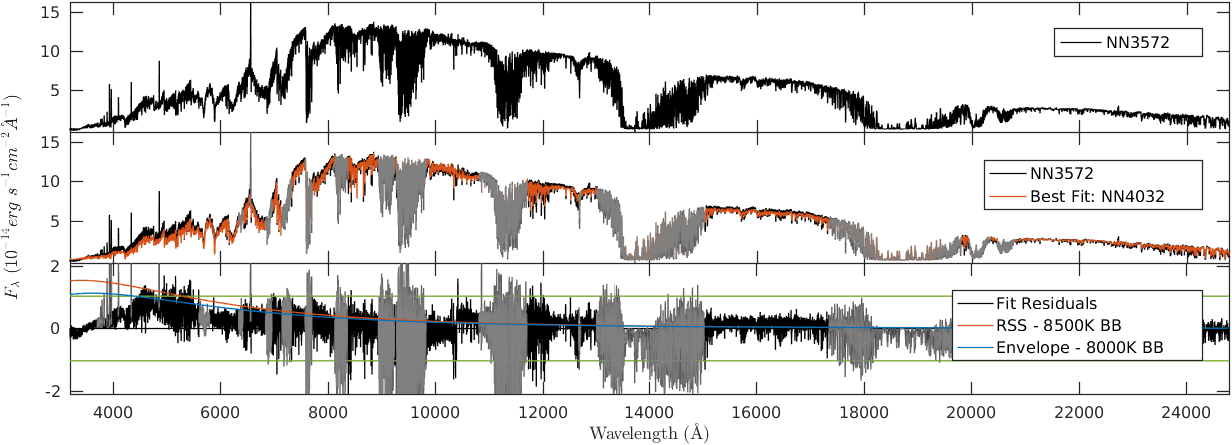}
\centering
\caption{NN3572\label{fig:Results_NN3572}}
\end{figure}

\begin{figure}[h]
\includegraphics[width=\textwidth]{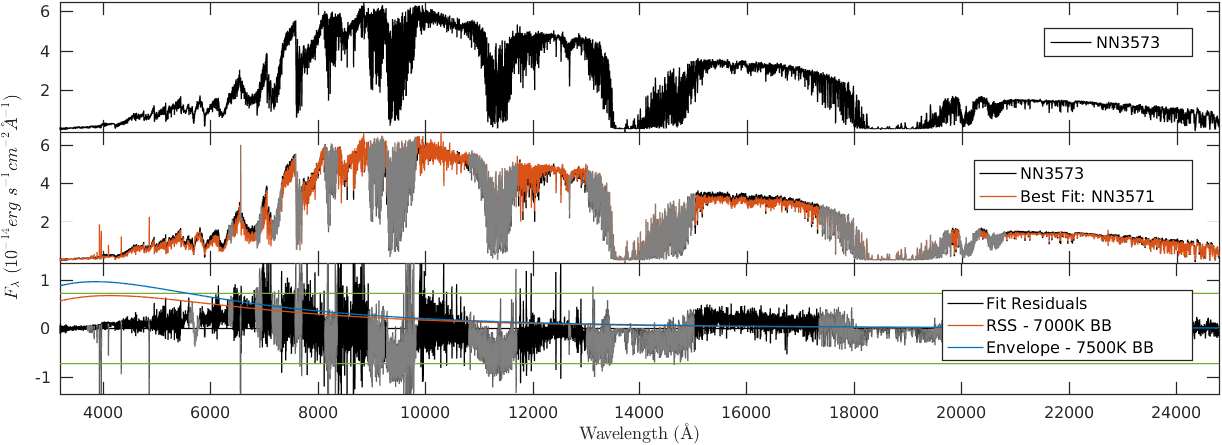}
\centering
\caption{NN3573\label{fig:Results_NN3573}}
\end{figure}

\clearpage

\begin{figure}[h]
\includegraphics[width=\textwidth]{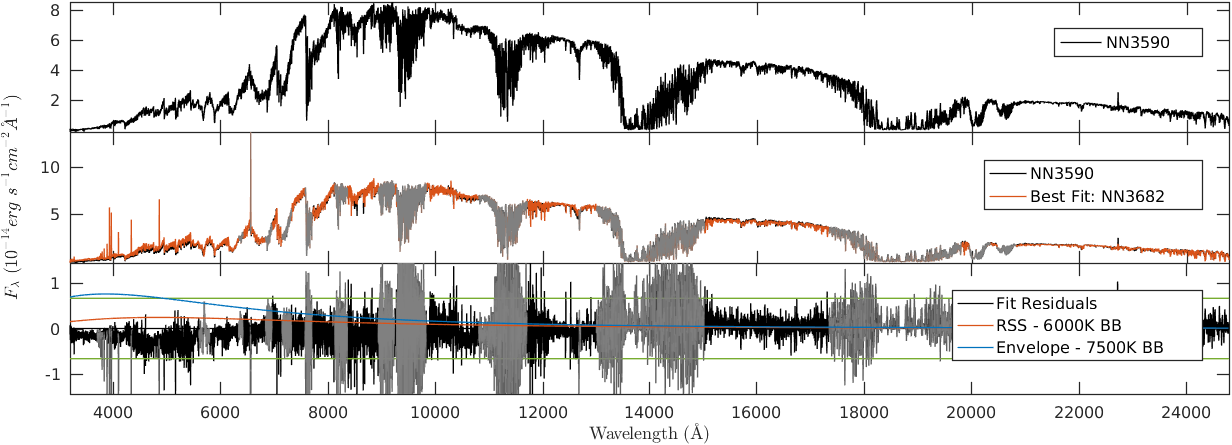}
\centering
\caption{NN3590\label{fig:Results_NN3590}}
\end{figure}

\begin{figure}[h]
\includegraphics[width=\textwidth]{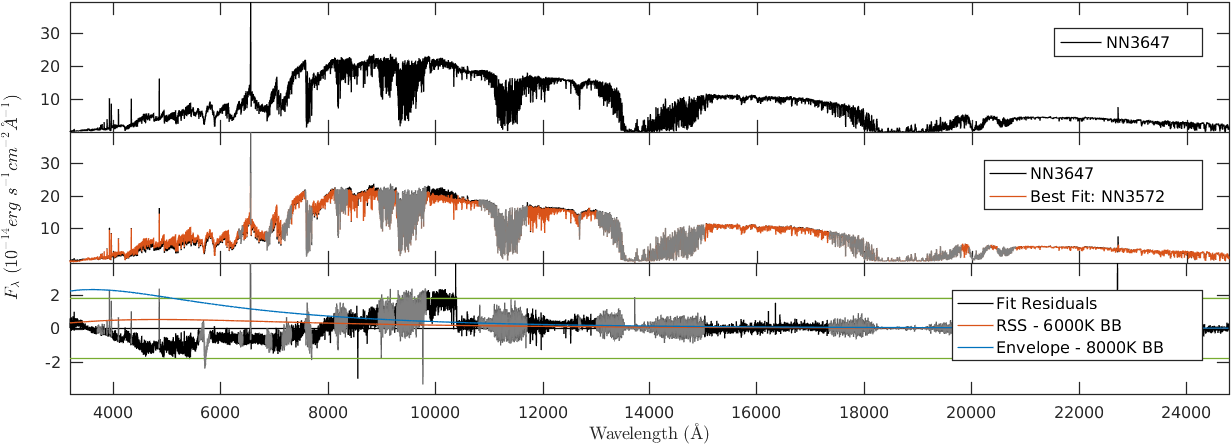}
\centering
\caption{NN3647\label{fig:Results_NN3647}}
\end{figure}

\begin{figure}[h]
\includegraphics[width=\textwidth]{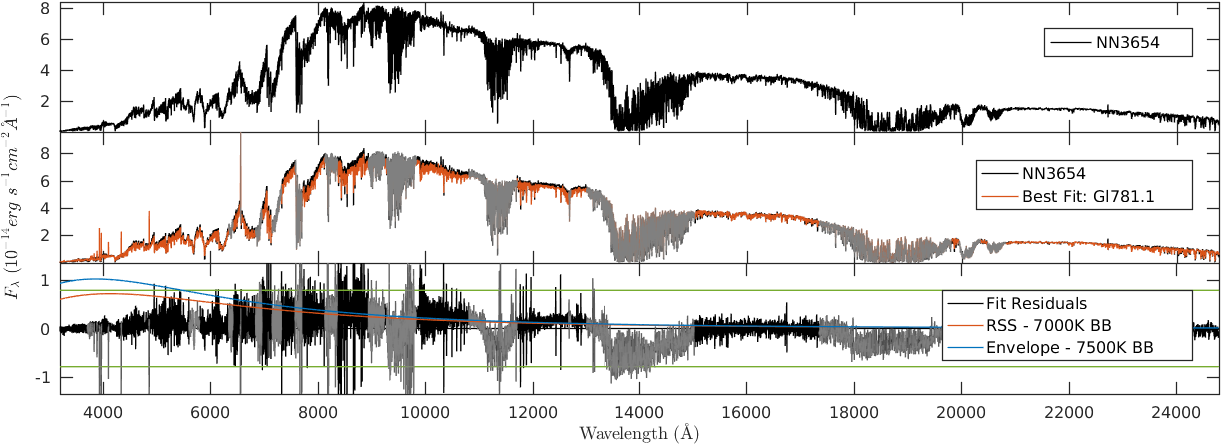}
\centering
\caption{NN3654\label{fig:Results_NN3654}}
\end{figure}

\clearpage

\begin{figure}[h]
\includegraphics[width=\textwidth]{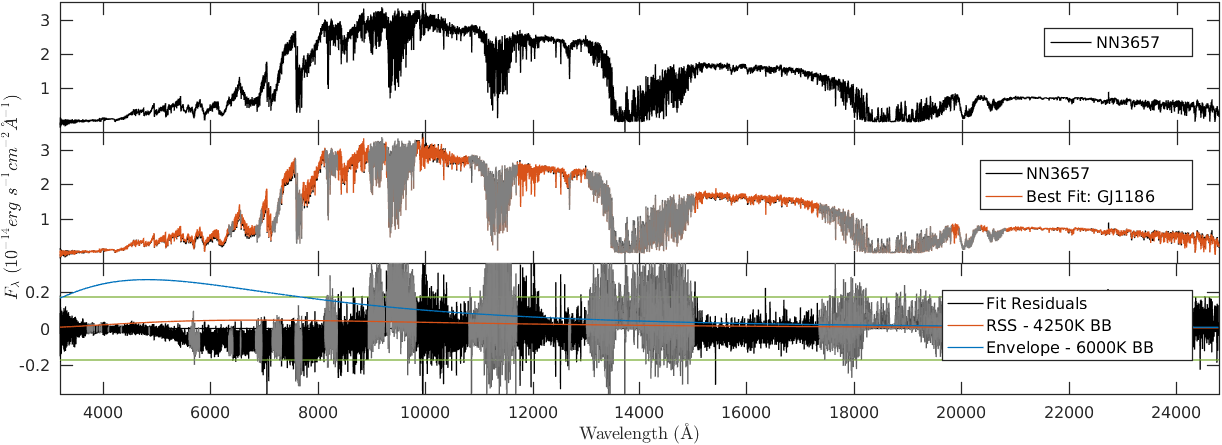}
\centering
\caption{NN3657\label{fig:Results_NN3657}}
\end{figure}

\begin{figure}[h]
\includegraphics[width=\textwidth]{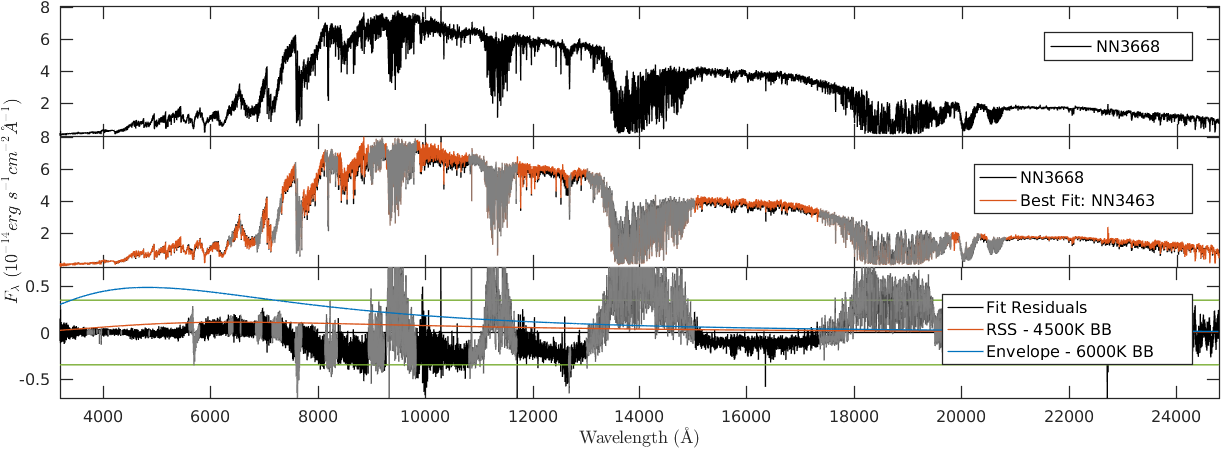}
\centering
\caption{NN3668\label{fig:Results_NN3668}}
\end{figure}

\begin{figure}[h]
\includegraphics[width=\textwidth]{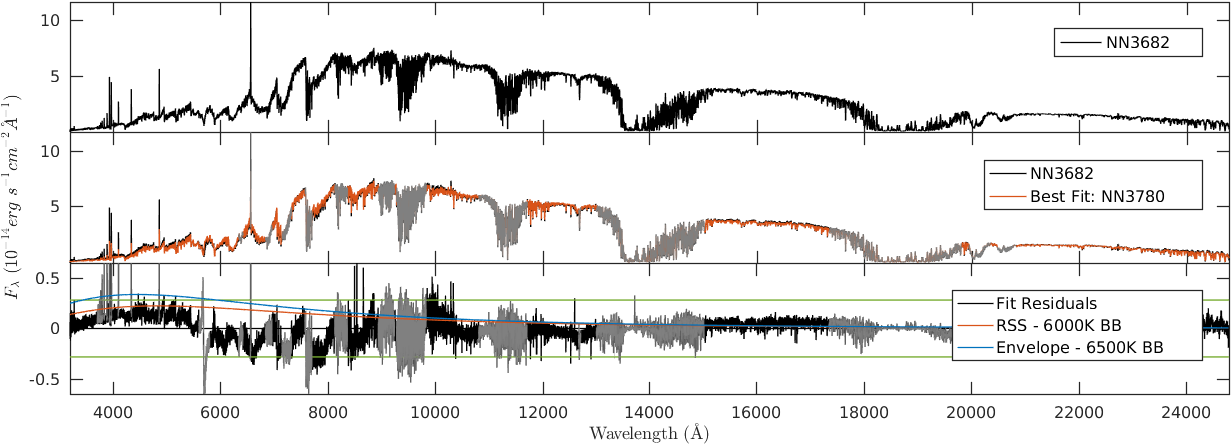}
\centering
\caption{NN3682\label{fig:Results_NN3682}}
\end{figure}

\clearpage

\begin{figure}[h]
\includegraphics[width=\textwidth]{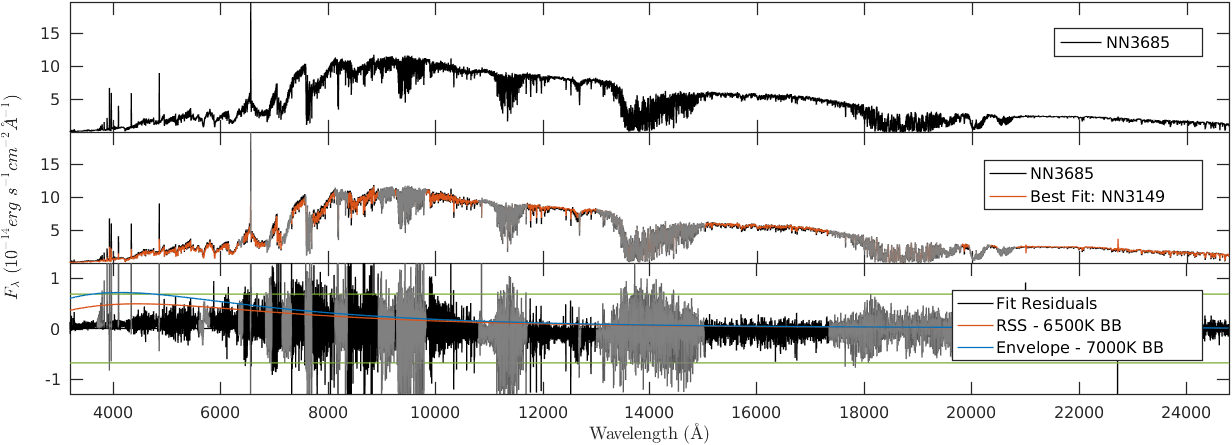}
\centering
\caption{NN3685\label{fig:Results_NN3685}}
\end{figure}

\begin{figure}[h]
\includegraphics[width=\textwidth]{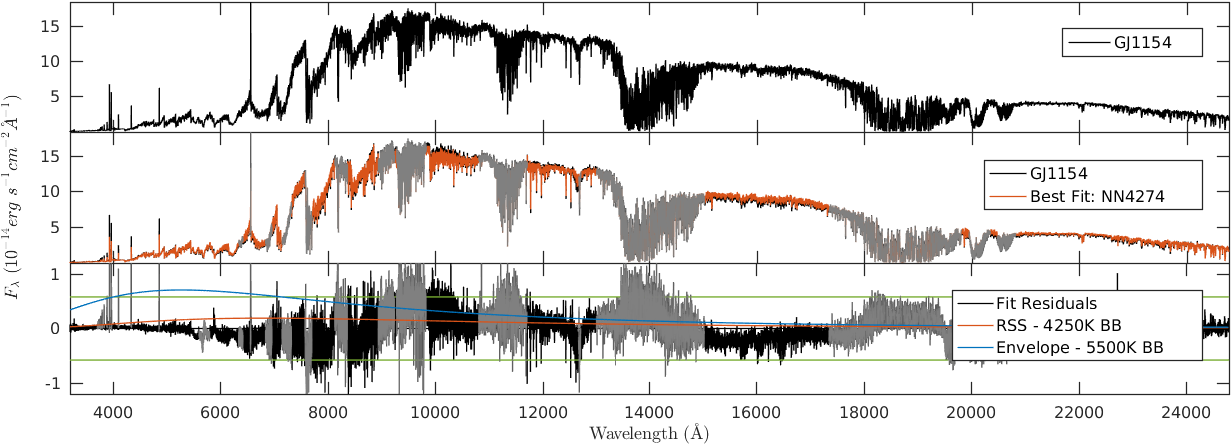}
\centering
\caption{GJ1154\label{fig:Results_GJ1154}}
\end{figure}

\begin{figure}[h]
\includegraphics[width=\textwidth]{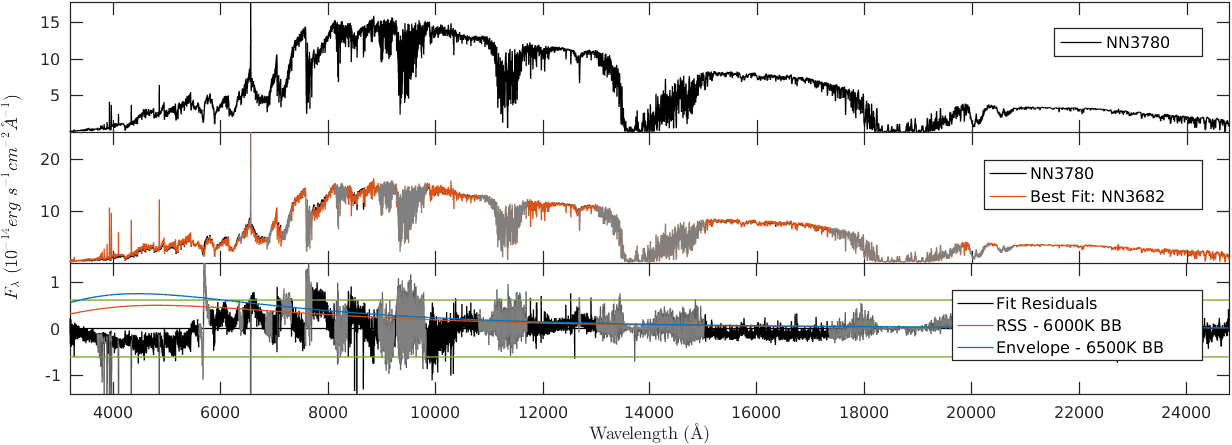}
\centering
\caption{NN3780\label{fig:Results_NN3780}}
\end{figure}

\clearpage

\begin{figure}[h]
\includegraphics[width=\textwidth]{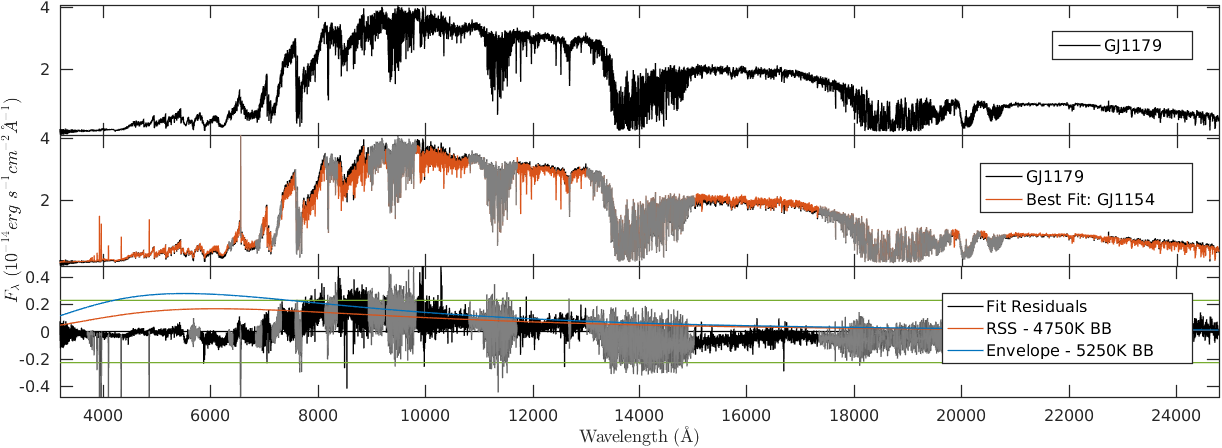}
\centering
\caption{GJ1179\label{fig:Results_GJ1179}}
\end{figure}

\begin{figure}[h]
\includegraphics[width=\textwidth]{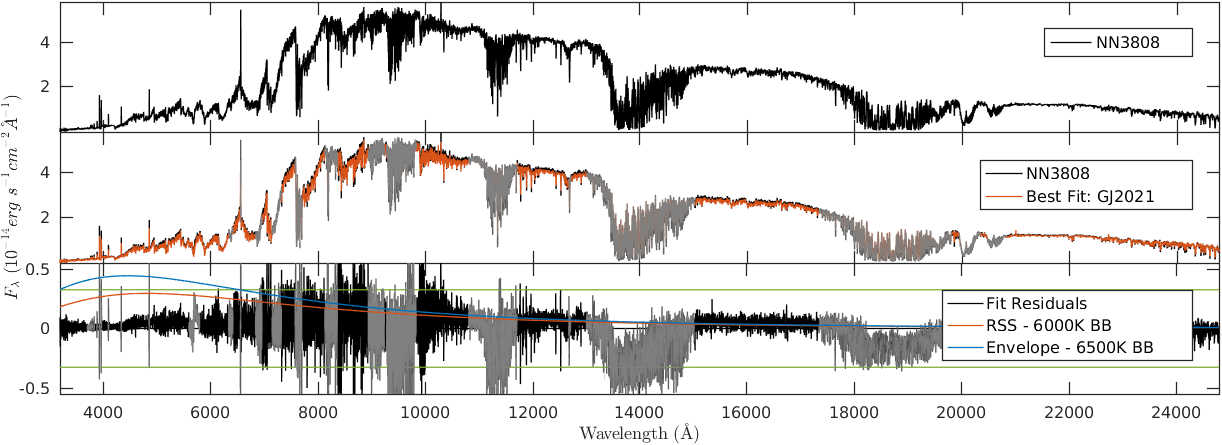}
\centering
\caption{NN3808\label{fig:Results_NN3808}}
\end{figure}

\begin{figure}[h]
\includegraphics[width=\textwidth]{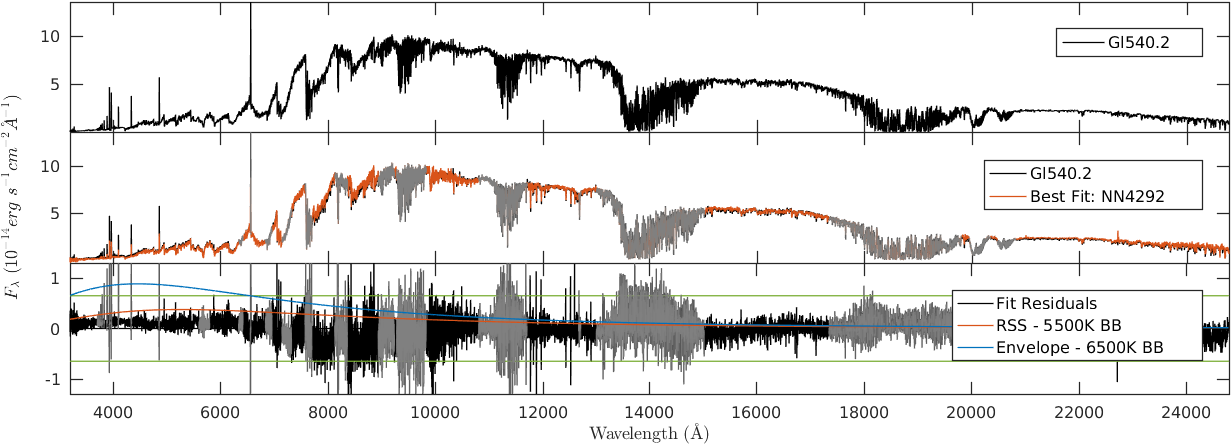}
\centering
\caption{Gl540.2\label{fig:Results_Gl540.2}}
\end{figure}

\clearpage

\begin{figure}[h]
\includegraphics[width=\textwidth]{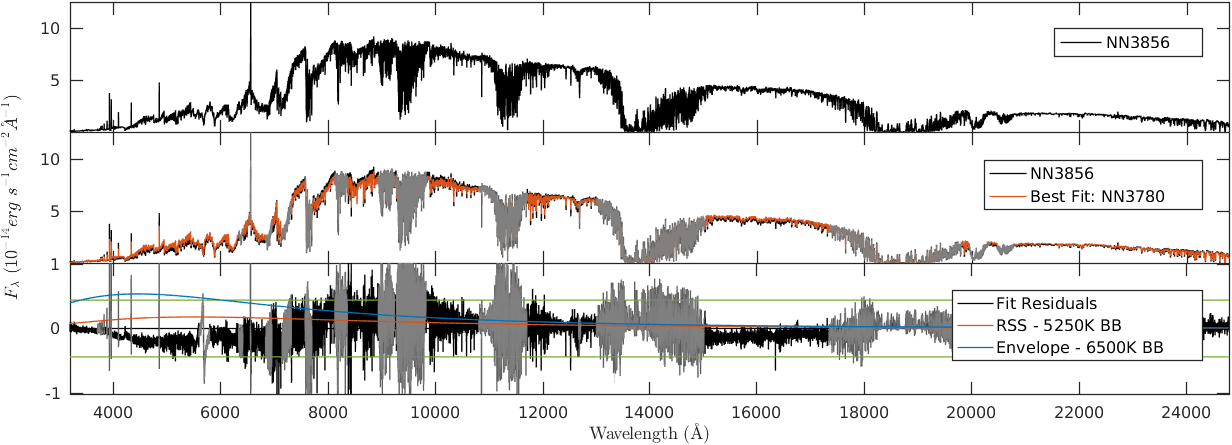}
\centering
\caption{NN3856\label{fig:Results_NN3856}}
\end{figure}

\begin{figure}[h]
\includegraphics[width=\textwidth]{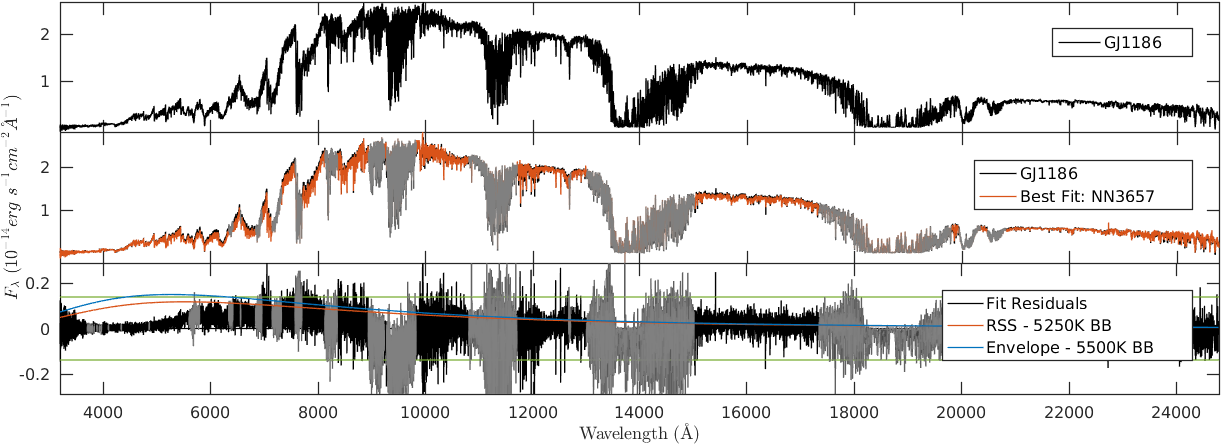}
\centering
\caption{GJ1186\label{fig:Results_GJ1186}}
\end{figure}

\begin{figure}[h]
\includegraphics[width=\textwidth]{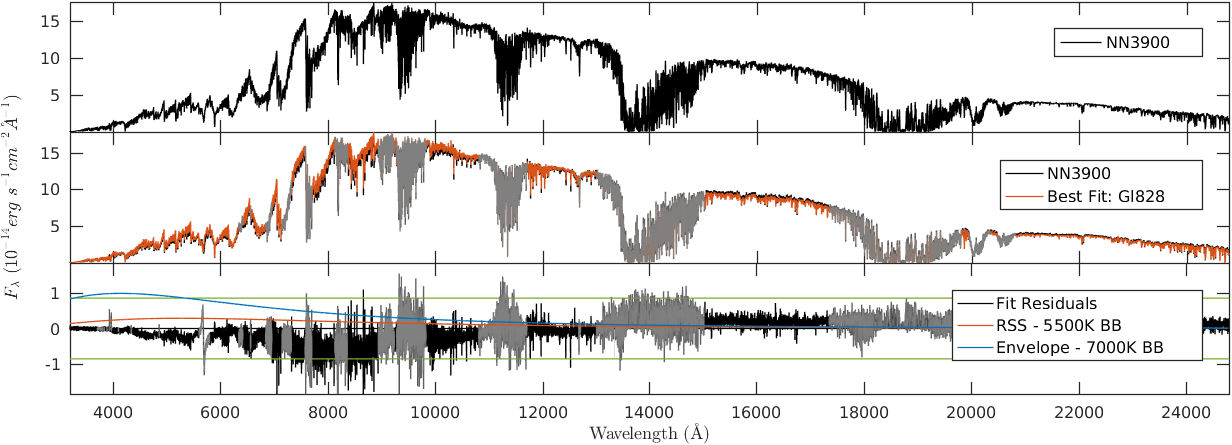}
\centering
\caption{NN3900\label{fig:Results_NN3900}}
\end{figure}

\clearpage

\begin{figure}[h]
\includegraphics[width=\textwidth]{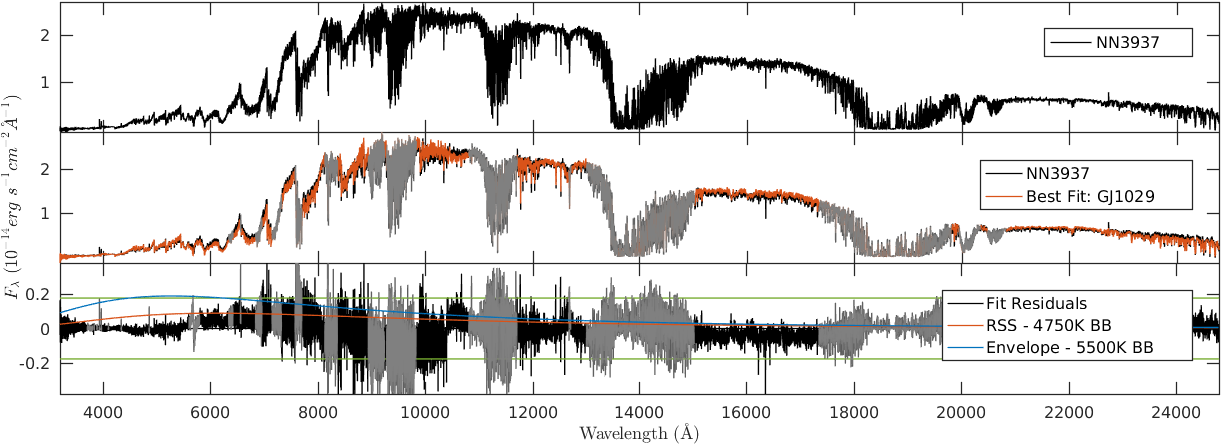}
\centering
\caption{NN3937\label{fig:Results_NN3937}}
\end{figure}

\begin{figure}[h]
\includegraphics[width=\textwidth]{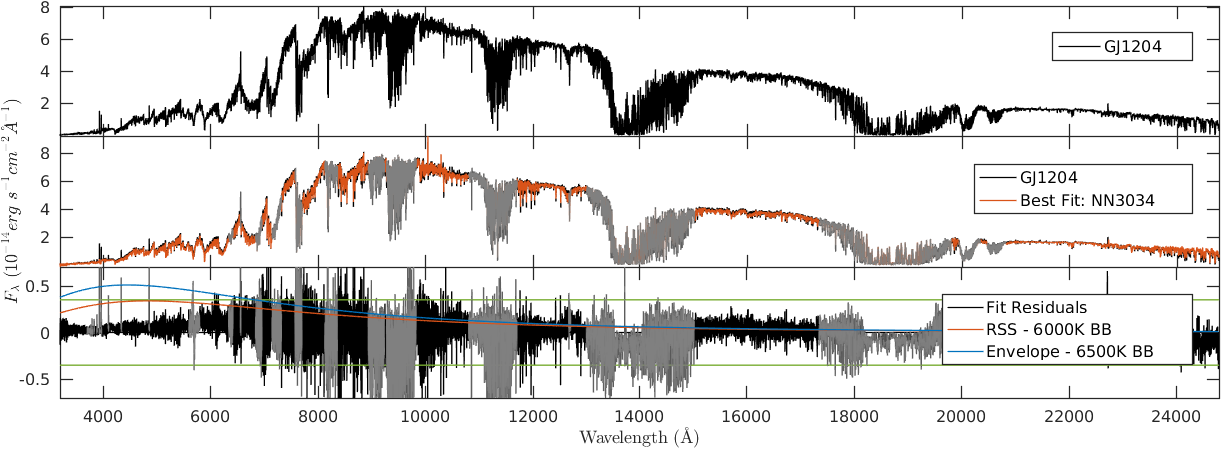}
\centering
\caption{GJ1204\label{fig:Results_GJ1204}}
\end{figure}

\begin{figure}[h]
\includegraphics[width=\textwidth]{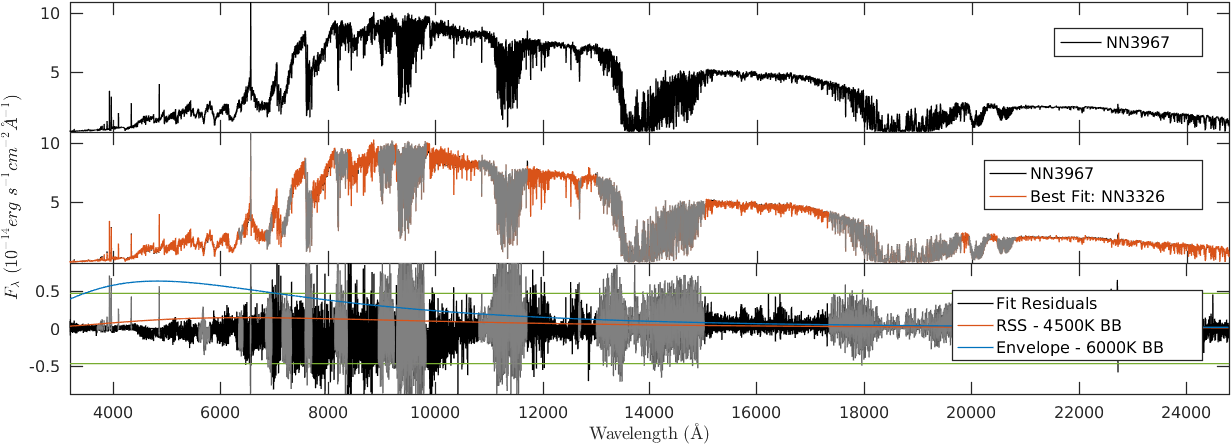}
\centering
\caption{NN3967\label{fig:Results_NN3967}}
\end{figure}

\clearpage

\begin{figure}[h]
\includegraphics[width=\textwidth]{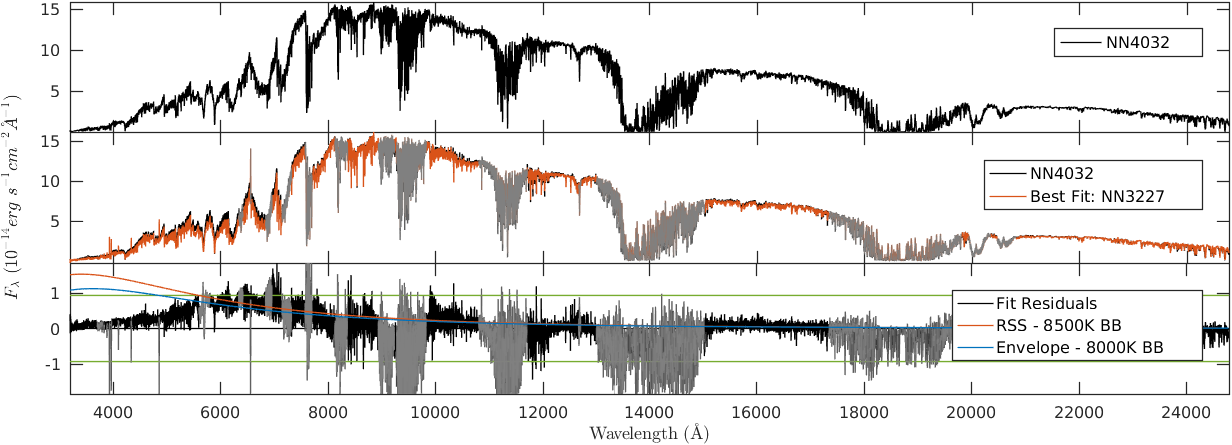}
\centering
\caption{NN4032\label{fig:Results_NN4032}}
\end{figure}

\begin{figure}[h]
\includegraphics[width=\textwidth]{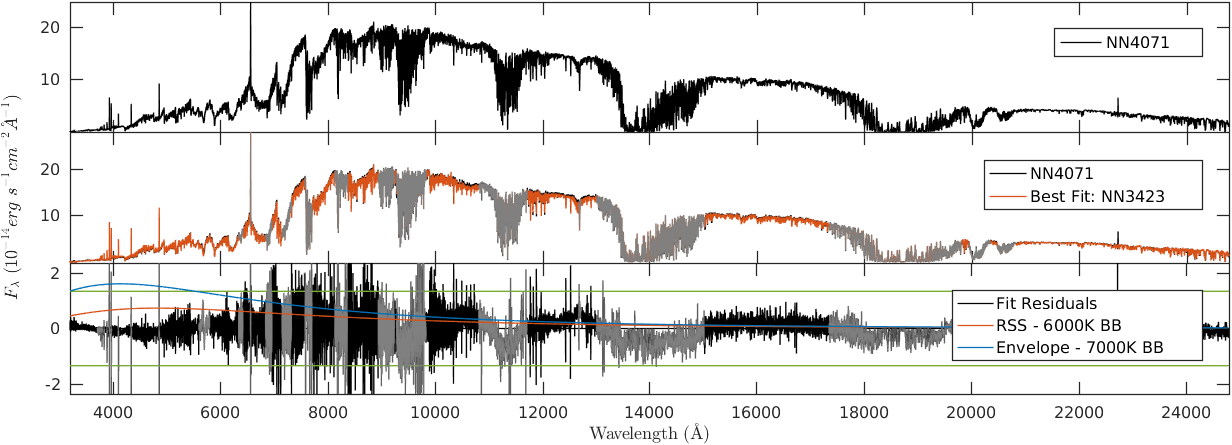}
\centering
\caption{NN4071\label{fig:Results_NN4071}}
\end{figure}

\begin{figure}[h]
\includegraphics[width=\textwidth]{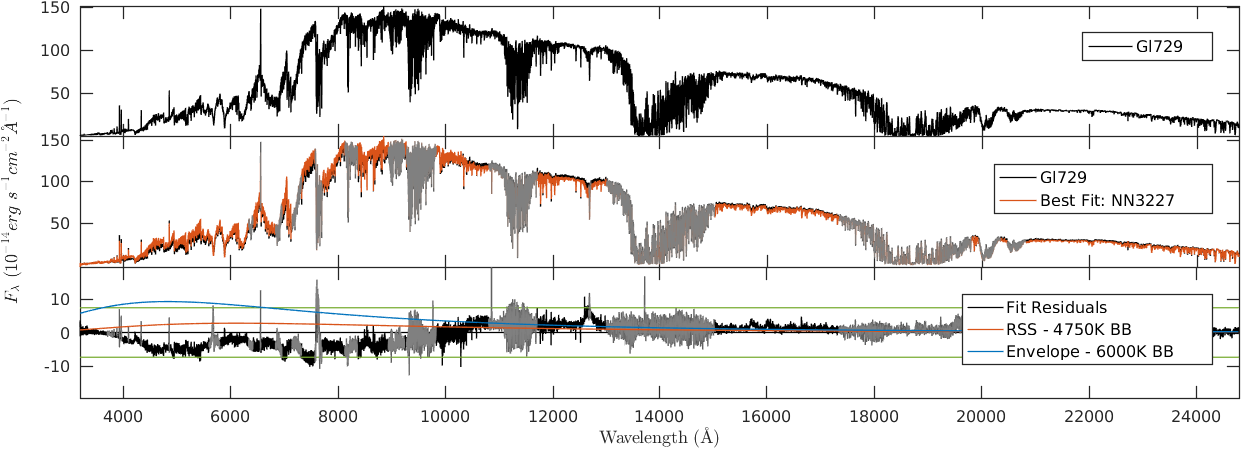}
\centering
\caption{Gl729\label{fig:Results_Gl729}}
\end{figure}

\clearpage

\begin{figure}[h]
\includegraphics[width=\textwidth]{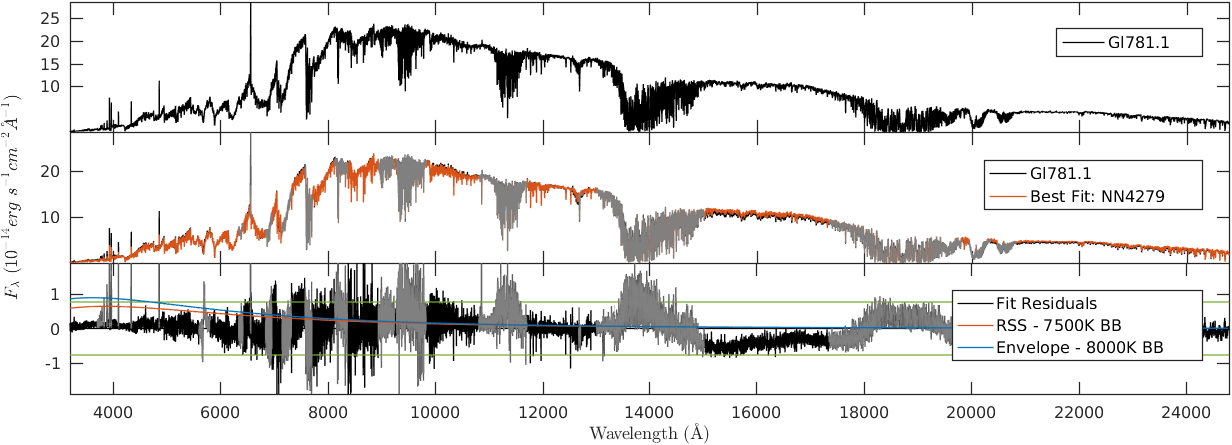}
\centering
\caption{Gl781.1\label{fig:Results_Gl781.1}}
\end{figure}

\begin{figure}[h]
\includegraphics[width=\textwidth]{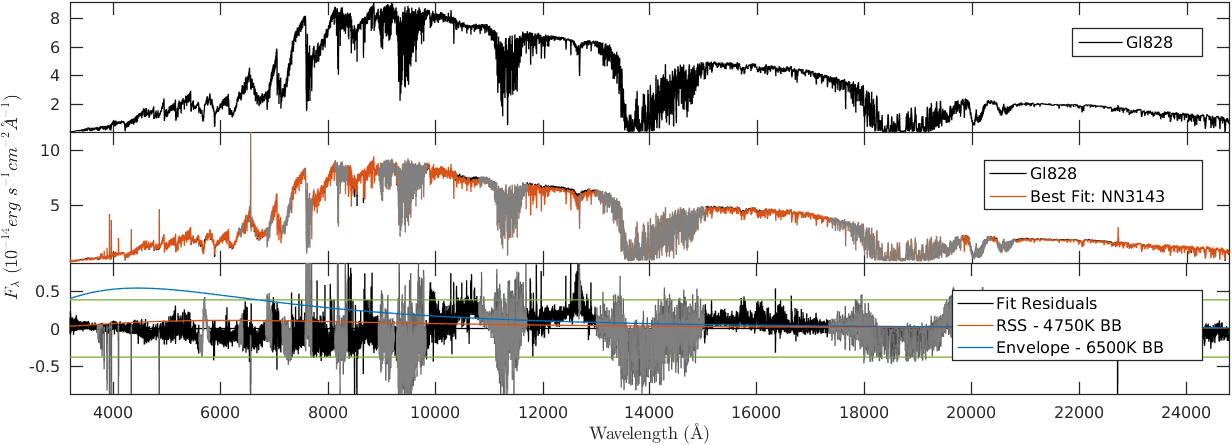}
\centering
\caption{Gl828\label{fig:Results_Gl828}}
\end{figure}

\begin{figure}[h]
\includegraphics[width=\textwidth]{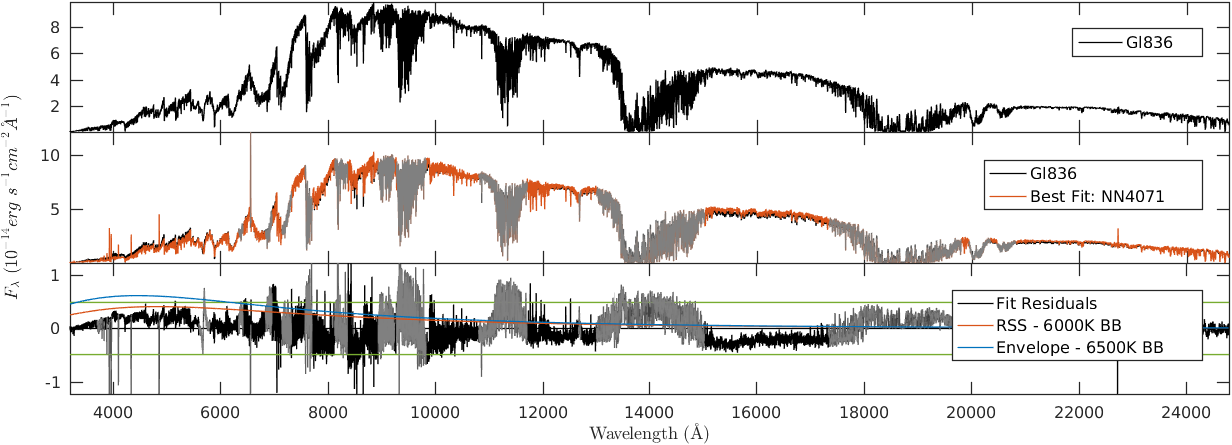}
\centering
\caption{Gl836\label{fig:Results_Gl836}}
\end{figure}

\clearpage

\begin{figure}[h]
\includegraphics[width=\textwidth]{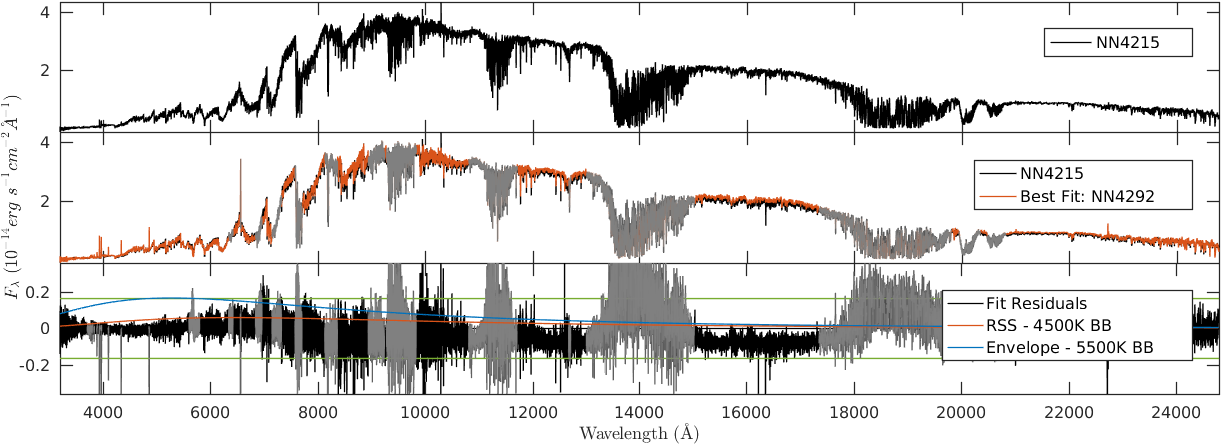}
\centering
\caption{NN4215\label{fig:Results_NN4215}}
\end{figure}

\begin{figure}[h]
\includegraphics[width=\textwidth]{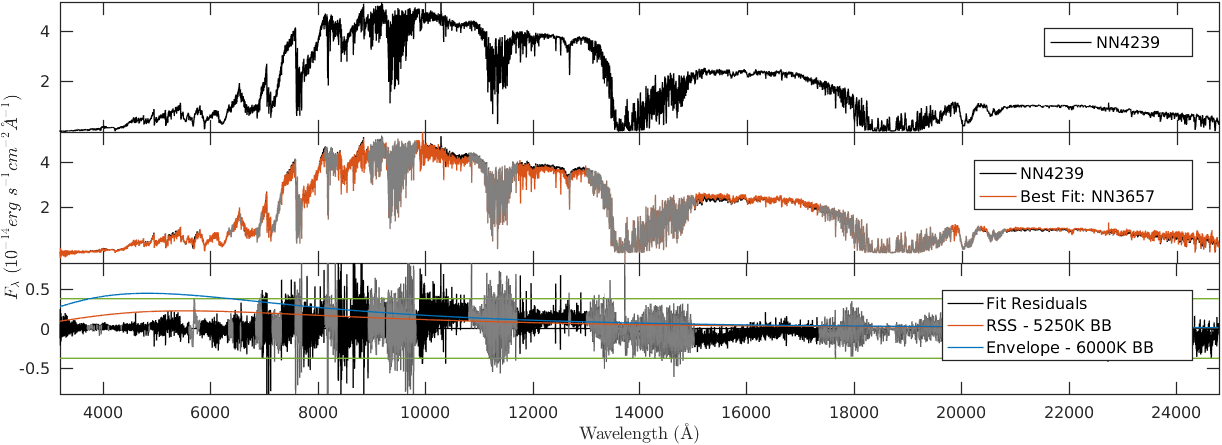}
\centering
\caption{NN4239\label{fig:Results_NN4239}}
\end{figure}

\begin{figure}[h]
\includegraphics[width=\textwidth]{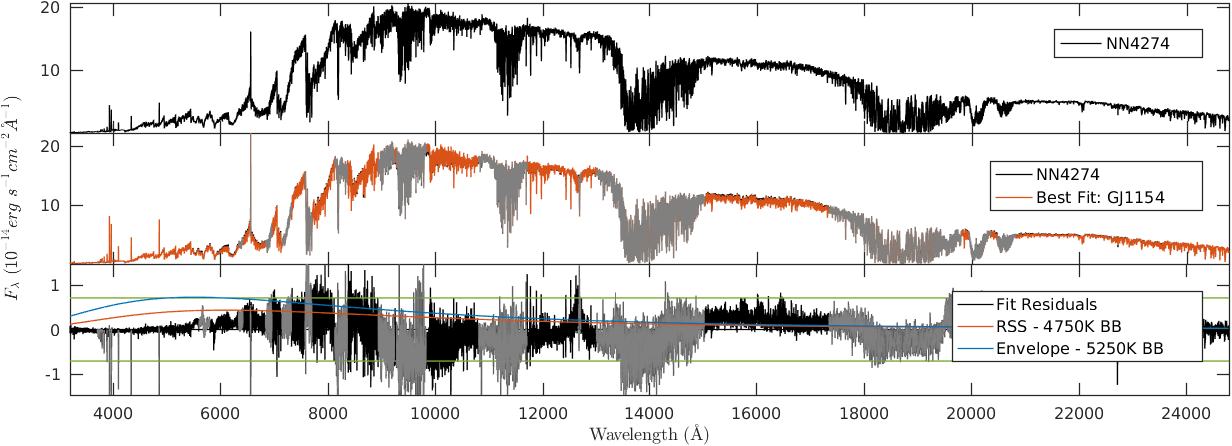}
\centering
\caption{NN4274\label{fig:Results_NN4274}}
\end{figure}

\clearpage

\begin{figure}[h]
\includegraphics[width=\textwidth]{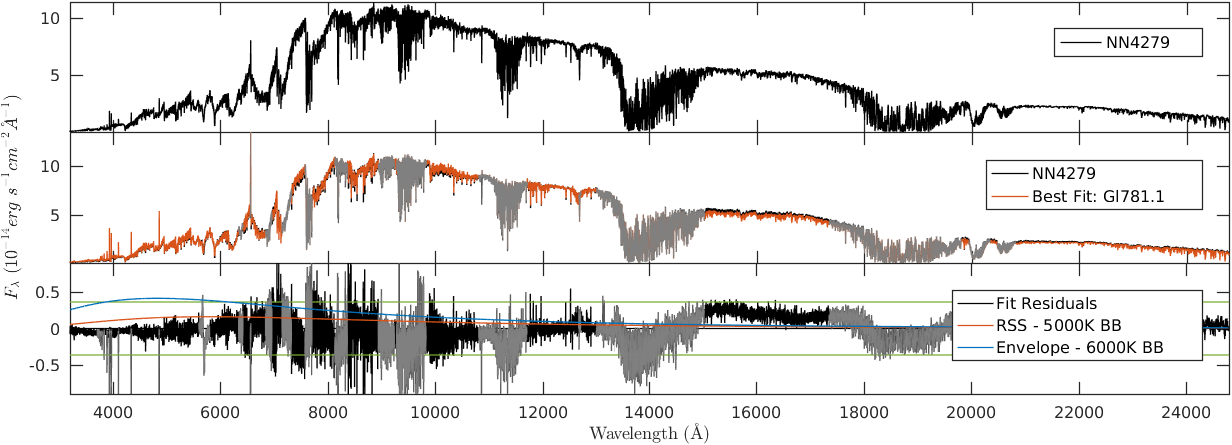}
\centering
\caption{NN4279\label{fig:Results_NN4279}}
\end{figure}

\begin{figure}[h]
\includegraphics[width=\textwidth]{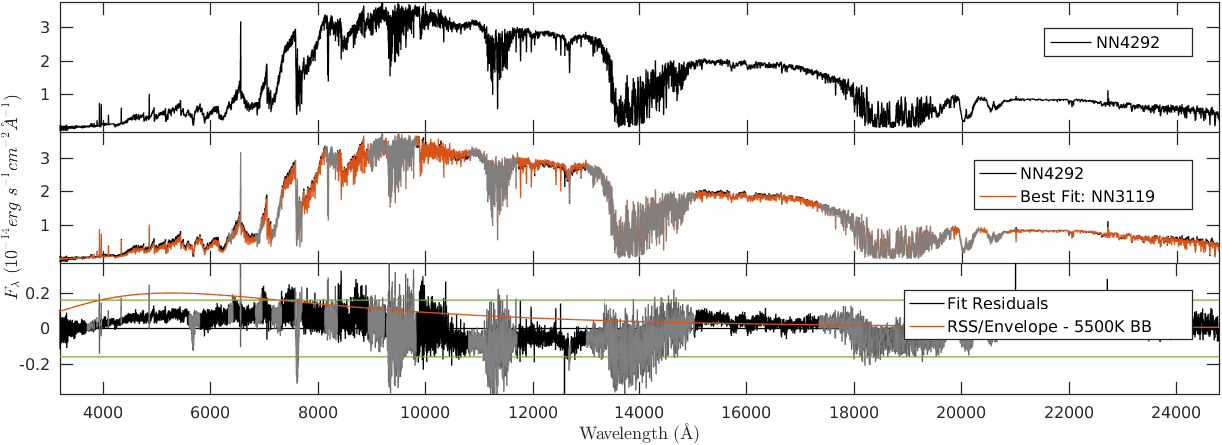}
\centering
\caption{NN4292\label{fig:Results_NN4292}}
\end{figure}

\begin{figure}[h]
\includegraphics[width=\textwidth]{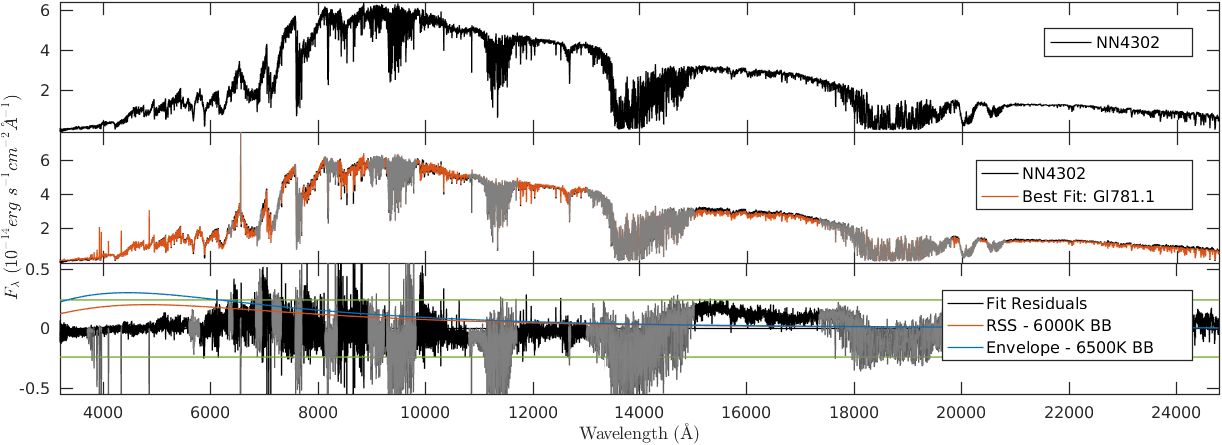}
\centering
\caption{NN4302\label{fig:Results_NN4302}}
\end{figure}

\clearpage

\begin{figure}[h]
\includegraphics[width=\textwidth]{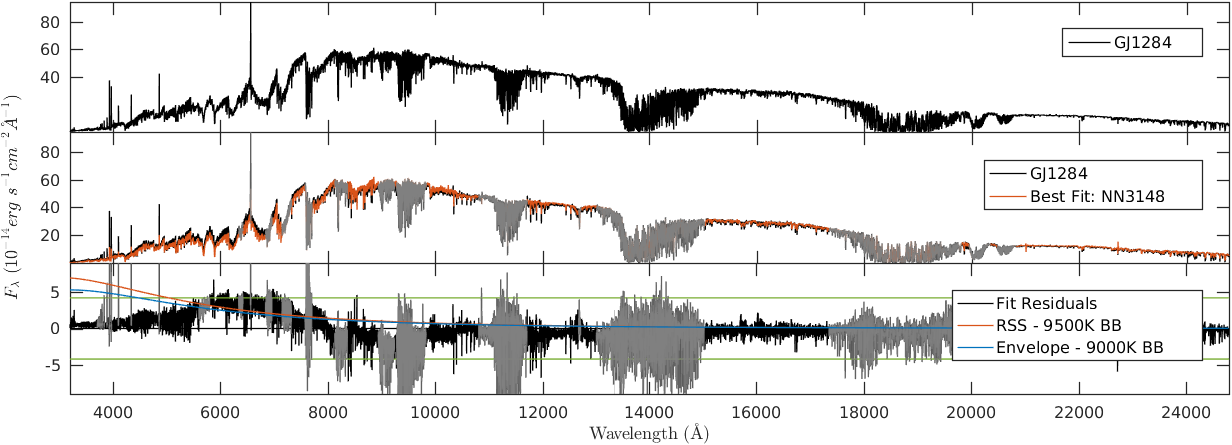}
\centering
\caption{GJ1284\label{fig:Results_GJ1284}}
\end{figure}

\begin{figure}[h]
\includegraphics[width=\textwidth]{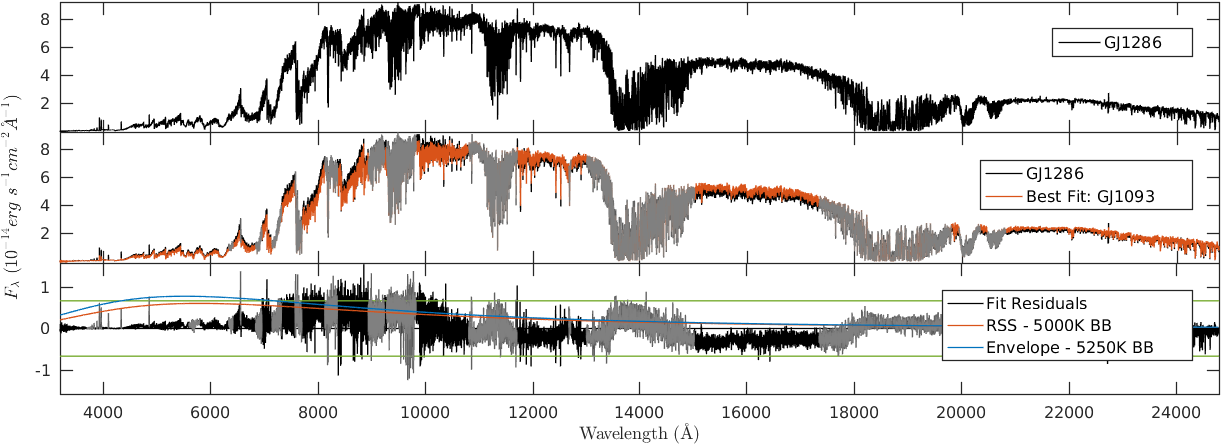}
\centering
\caption{GJ1286\label{fig:Results_GJ1286}}
\end{figure}

\begin{figure}[h]
\includegraphics[width=\textwidth]{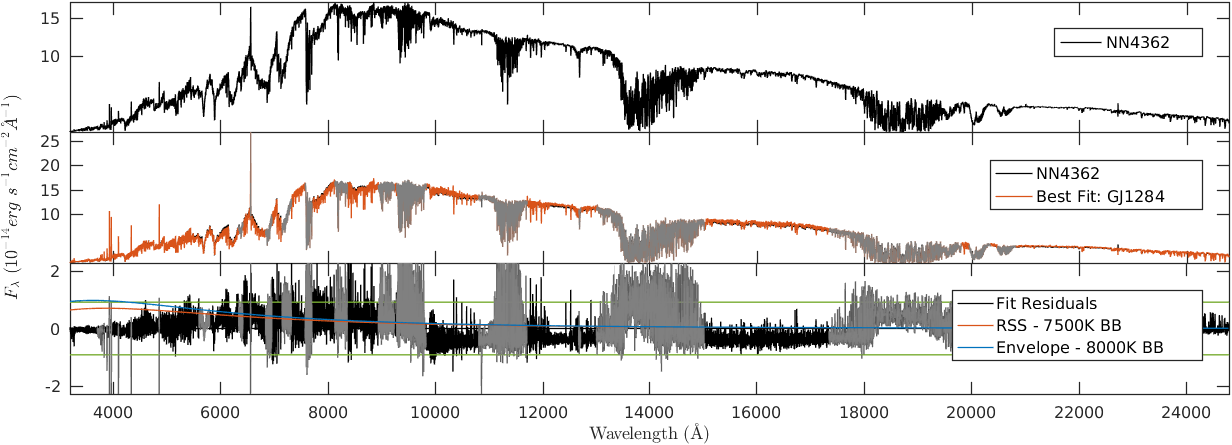}
\centering
\caption{NN4362\label{fig:Results_NN4362}}
\end{figure}

\clearpage

%% file: nonlimit_file_list.tex
\begin{figure}[h]
\includegraphics[height=1.37in]{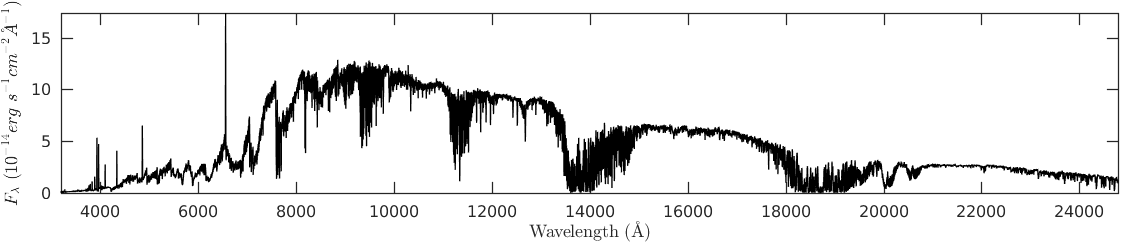}
\centering
\caption{NN3010\label{fig:Results_NN3010}}
\end{figure}

\begin{figure}[h]
\includegraphics[height=1.37in]{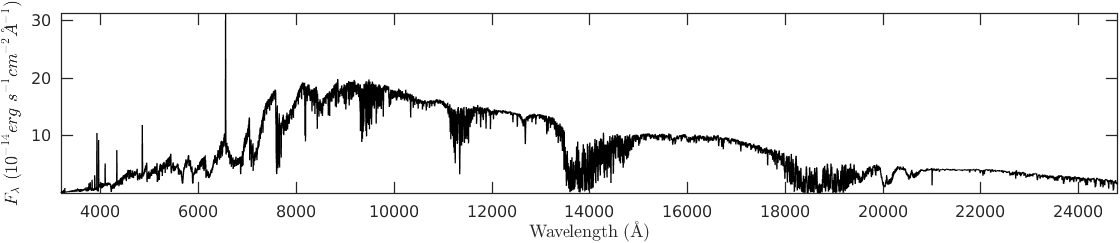}
\centering
\caption{NN3039\label{fig:Results_NN3039}}
\end{figure}

\begin{figure}[h]
\includegraphics[height=1.37in]{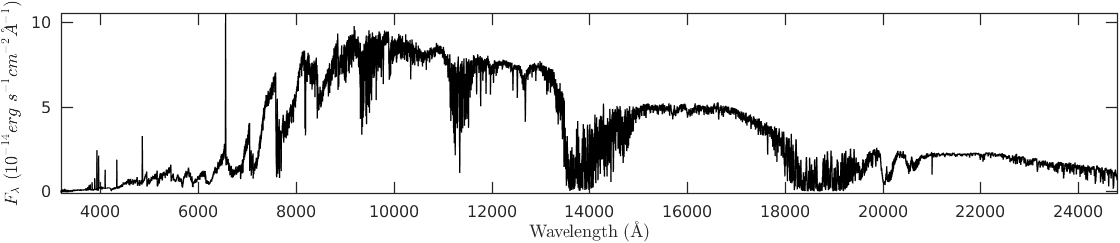}
\centering
\caption{NN3076\label{fig:Results_NN3076}}
\end{figure}

\begin{figure}[h]
\includegraphics[height=1.37in]{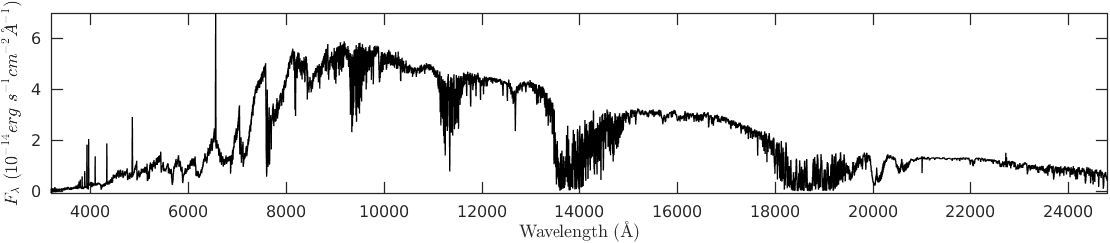}
\centering
\caption{NN3129\label{fig:Results_NN3129}}
\end{figure}

\begin{figure}[h]
\includegraphics[height=1.37in]{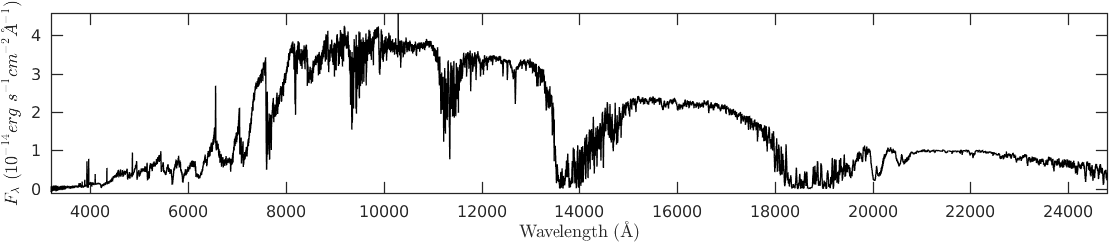}
\centering
\caption{NN3224\label{fig:Results_NN3224}}
\end{figure}

\clearpage

\begin{figure}[h]
\includegraphics[height=1.37in]{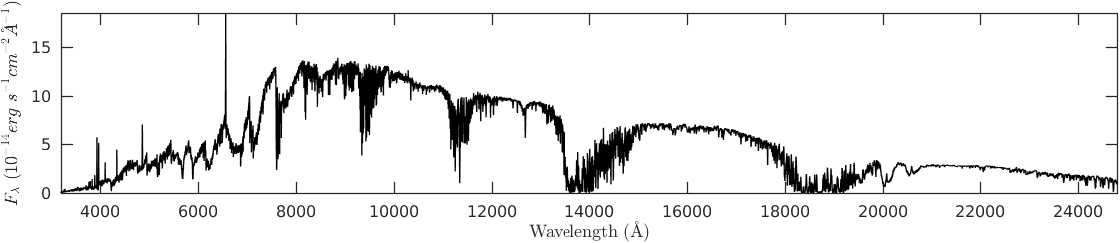}
\centering
\caption{NN3261\label{fig:Results_NN3261}}
\end{figure}

\begin{figure}[h]
\includegraphics[height=1.37in]{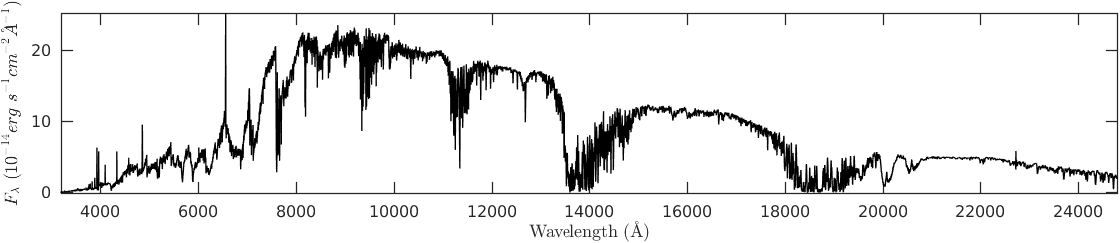}
\centering
\caption{NN3304\label{fig:Results_NN3304}}
\end{figure}

\begin{figure}[h]
\includegraphics[height=1.37in]{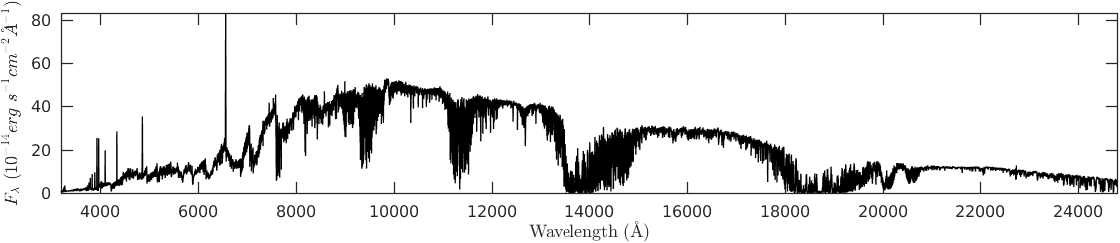}
\centering
\caption{NN3322\label{fig:Results_NN3322}}
\end{figure}

\begin{figure}[h]
\includegraphics[height=1.37in]{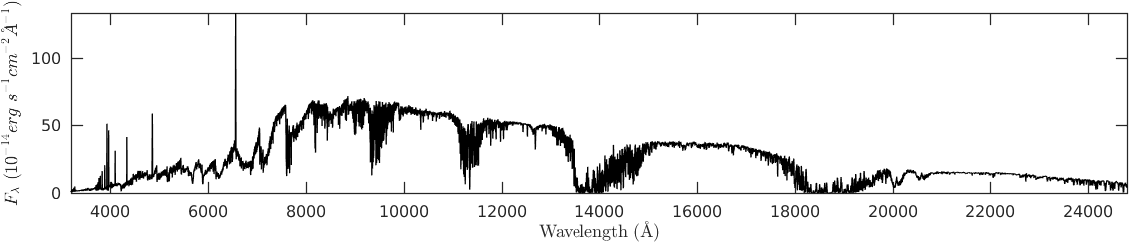}
\centering
\caption{NN3332\label{fig:Results_NN3332}}
\end{figure}

\begin{figure}[h]
\includegraphics[height=1.37in]{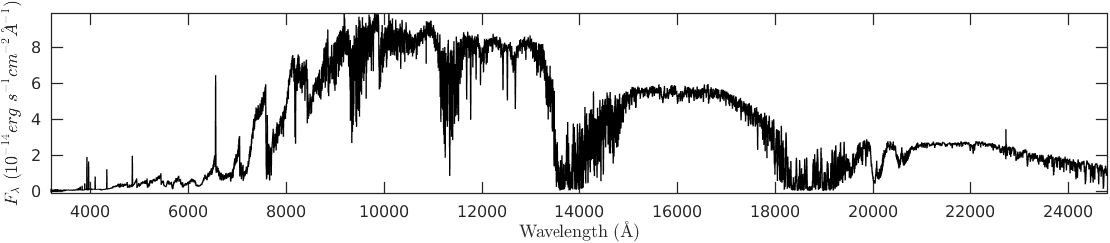}
\centering
\caption{GJ1083\label{fig:Results_GJ1083}}
\end{figure}

\clearpage

\begin{figure}[h]
\includegraphics[height=1.37in]{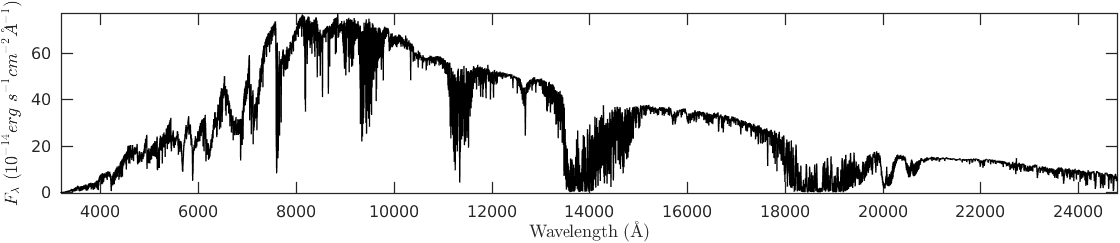}
\centering
\caption{Gl268.3\label{fig:Results_Gl268.3}}
\end{figure}

\begin{figure}[h]
\includegraphics[height=1.37in]{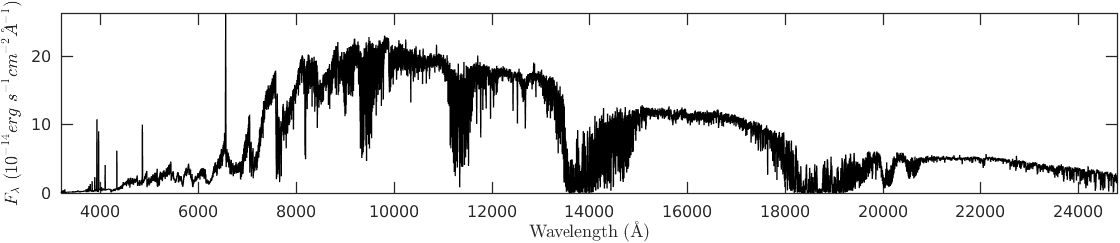}
\centering
\caption{NN3454\label{fig:Results_NN3454}}
\end{figure}

\begin{figure}[h]
\includegraphics[height=1.37in]{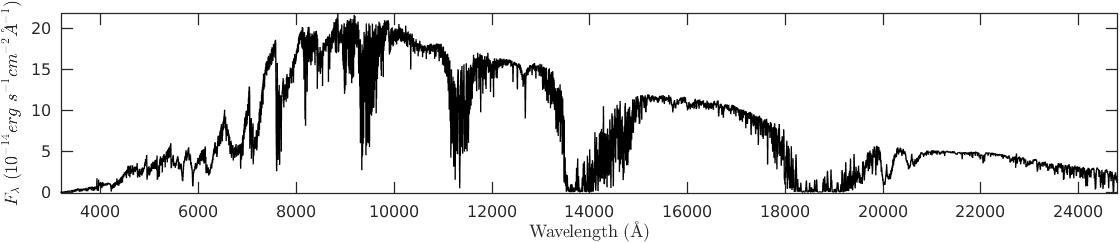}
\centering
\caption{NN3466\label{fig:Results_NN3466}}
\end{figure}

\begin{figure}[h]
\includegraphics[height=1.37in]{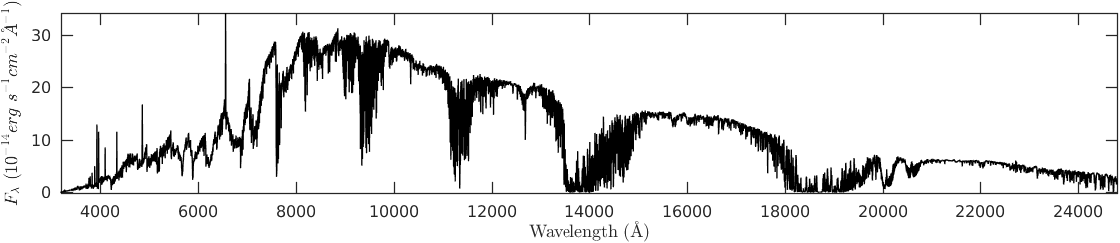}
\centering
\caption{GJ1108\label{fig:Results_GJ1108}}
\end{figure}

\begin{figure}[h]
\includegraphics[height=1.37in]{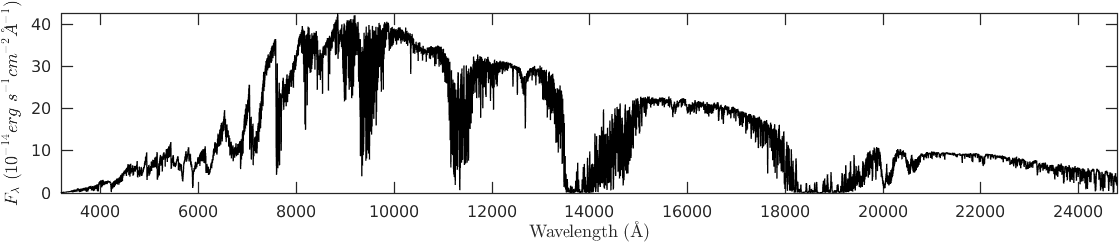}
\centering
\caption{Gl300\label{fig:Results_Gl300}}
\end{figure}

\clearpage

\begin{figure}[h]
\includegraphics[height=1.37in]{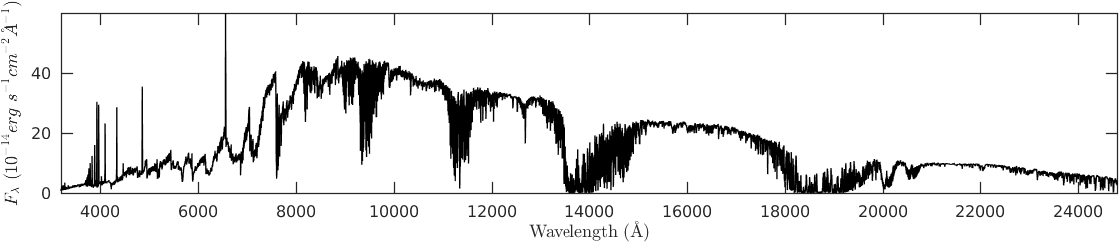}
\centering
\caption{GJ2069\label{fig:Results_GJ2069}}
\end{figure}

\begin{figure}[h]
\includegraphics[height=1.37in]{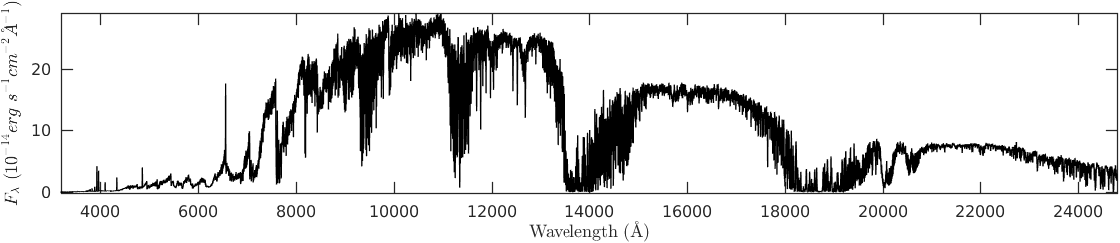}
\centering
\caption{GJ1116\label{fig:Results_GJ1116}}
\end{figure}

\begin{figure}[h]
\includegraphics[height=1.37in]{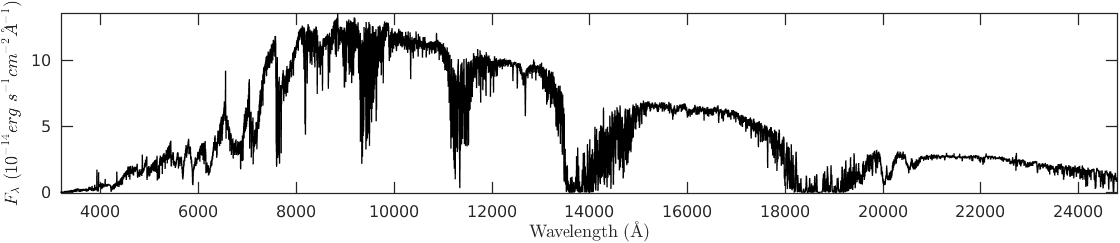}
\centering
\caption{NN3981\label{fig:Results_NN3981}}
\end{figure}

\begin{figure}[h]
\includegraphics[height=1.37in]{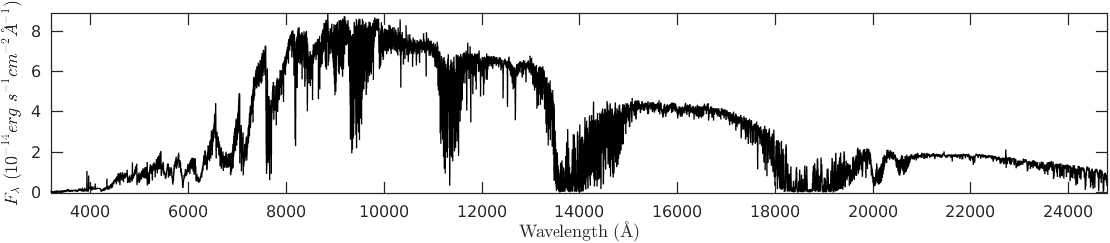}
\centering
\caption{GJ1210\label{fig:Results_GJ1210}}
\end{figure}

\begin{figure}[h]
\includegraphics[height=1.37in]{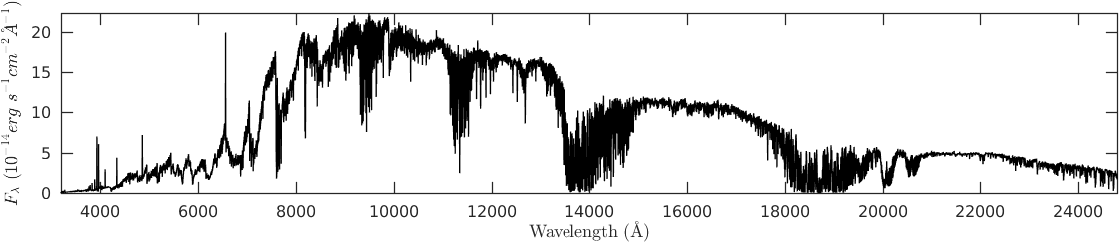}
\centering
\caption{Gl791.2\label{fig:Results_Gl791.2}}
\end{figure}

\clearpage

\begin{figure}[h]
\includegraphics[height=1.37in]{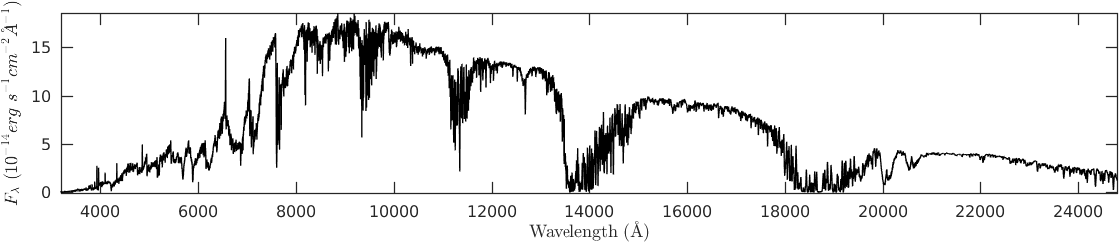}
\centering
\caption{NN4201\label{fig:Results_NN4201}}
\end{figure}

\begin{figure}[h]
\includegraphics[height=1.37in]{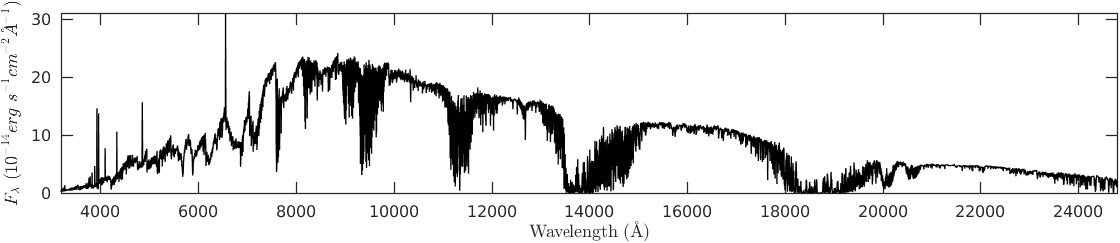}
\centering
\caption{NN4231\label{fig:Results_NN4231}}
\end{figure}

\begin{figure}[h]
\includegraphics[height=1.37in]{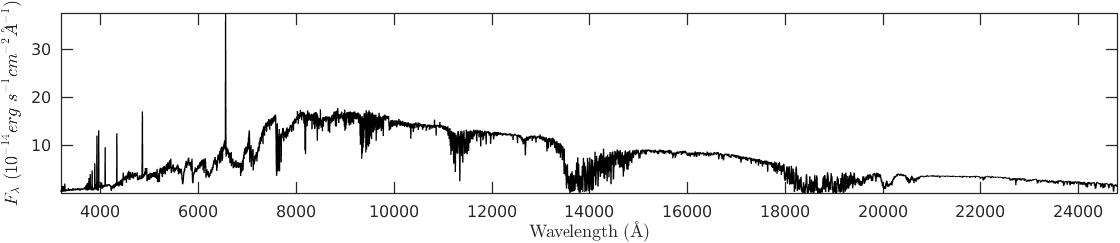}
\centering
\caption{NN4282\label{fig:Results_NN4282}}
\end{figure}

\begin{figure}[h]
\includegraphics[height=1.37in]{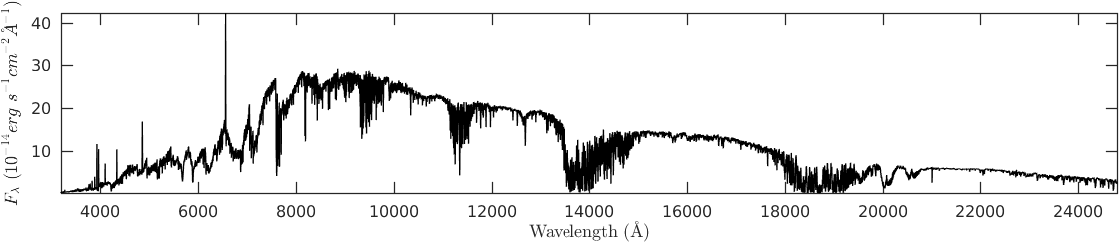}
\centering
\caption{NN4326\label{fig:Results_NN4326}}
\end{figure}

\begin{figure}[h]
\includegraphics[height=1.37in]{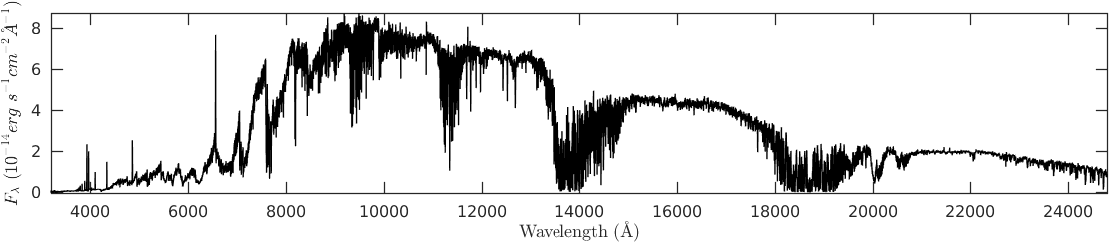}
\centering
\caption{NN4360\label{fig:Results_NN4360}}
\end{figure}

\clearpage

\begin{figure}[h]
\includegraphics[height=1.37in]{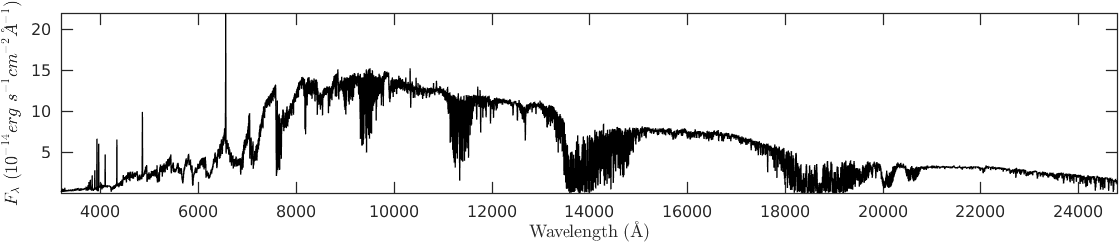}
\centering
\caption{NN4378\label{fig:Results_NN4378}}
\end{figure}